\documentclass[11pt]{article}%
\usepackage{amssymb}
\usepackage{amsfonts}
\usepackage{amsmath}
\usepackage{graphicx}%
\usepackage{authblk}
\usepackage{mwe}
\usepackage{subcaption}
\usepackage{url}
\usepackage{hyperref}
\usepackage{wrapfig}
\usepackage{xcolor}
\usepackage{soul}
\usepackage[normalem]{ulem}
\usepackage[british]{babel}
\providecommand{\U}[1]{\protect\rule{.1in}{.1in}}
\newtheorem{theorem}{Theorem}

\begin{document}

\title{Time Dynamics of the Dutch Municipality Network}
\author[1]{\small Ivan Joki\'{c}\thanks{\emph{ivan.jokic.93@gmail.com}}}
\author[1, 3]{\small Edgar van Boven}
\author[1]{\small Ioannis Manolopoulos}
\author[2]{\small Trivik Verma}
\author[4]{\small Gert Buiten}
\author[4,5]{\small Frank Pijpers}
\author[4]{\small Hans van Hooff}
\author[1]{\small Piet Van Mieghem}

\affil[1]{\footnotesize Faculty of Electrical Engineering, Mathematics, and Computer Science,
Delft University of Technology, P.O. Box 5031, 2600 GA Delft, The Netherlands}
\affil[2]{\footnotesize Faculty of Technology Policy and Management, Delft University of Technology, P.O. Box 5031, 2600 GA Delft, The Netherlands}
\affil[3]{\footnotesize KPN Royal, P.O. Box 30.000, 2500 GA The Hague, The Netherlands}
\affil[4]{\footnotesize Statistics Netherlands, P.O. Box 24.500, 2490 HA, The Hague, The Netherlands}
\affil[5]{\footnotesize Kortweg-de Vries Institute for Mathematics, University of Amsterdam, Science Park 105-107,1098 XG Amsterdam{\tiny }, The Netherlands}
\date{\today}
\maketitle 

\begin{abstract}
Based on data sets provided by Statistics Netherlands and the International Institute of Social History, we investigate the Dutch municipality merging process and the survivability of municipalities over the period $1830-2019$. We examine the dynamics of the population and area per municipality and how their distributions evolved during the researched period.
We apply a Network Science approach, where each node represents a municipality and the links represent the geographical interconnections between adjacent municipalities via roads, railways, bridges or tunnels which were available in each specific yearly network instance. 
Over the researched period, we find that the distributions of the logarithm of both the population and area size closely follow a normal and a logistic distribution respectively. The tails of the population distributions follow a power-law distribution, a phenomenon observed in community structures of many real-world networks.
The dynamics of the area distribution are mainly determined by the merging process, while the population distribution is also driven by the natural population growth and migration across the municipality network.
Finally, we propose a model of the Dutch Municipality Network that captures population increase, population migration between municipalities and the process of municipality merging. Our model allows for predictions of the population and area distributions over time.
\end{abstract}

\section{Introduction}\label{Sec:Introduction}

The process of urbanization by which large numbers of people permanently resided in cities\footnote{\href{https://www.britannica.com/topic/urbanization}{Urbanization|Britannica}.} has marked our history. The technological changes that enabled the urbanization and industrialization of our society took centuries to shape our cities, villages and rural areas. Identifying relevant governing factors and understanding the influence of different processes, such as population evolution, people migration, and urban growth, is essential for urban planning and policy-making.

Marchetti revealed in \cite{Marchetti1994AnthropologicalBehavior} how people's movements and commuting time depend on technological innovations in transport. In addition, Gonzales \textit{et al.} \cite{Barabasi2005HumanMobilityPatterns} observed that human motion is cha\-ra\-cterised by both temporal and spatial regularity while obeying simple, reproducible patterns. Human movement is often modelled as a random walk, known as a Levy flight \cite{Mantegna1994StochasticFlight}, where the distribution of travelling distance follows a power law \cite{Brockmann2006TheTravel}. In contrast to human mobility patterns, job-related and socioeconomic well-being variables govern migration flows of people \cite{Rayer2001Geographic19801995}, whose trends can be age-specific, as observed by Johnson and Fuguitt in \cite{Johnson2000Continuity19501995}. Migration patterns further shape the development of urban and rural areas. Makse \textit{et al.} \cite{Stanley1995UrbanGrowthPatterns} proposed a percolation-based model for city growth, following the principle that urban area development leads to further development. In addition, they found that the area distribution of towns surrounding a city follows a power law in the case of Berlin and London in years $1920,1945$ and $1981$ respectively. Schlapfer \textit{et al.} \cite{Schlapfer2014TheSize} empirically confirmed the scale-invariant increase of interactions between humans with city size.

When researching phenomena related to geographical urban areas, most often cities are considered \cite{Stanley1995UrbanGrowthPatterns, Verbavatz2020TheCities, Bettencourt2007GrowthCities, Bettencourt2013TheCities, Schneider2014Expansion19782010} as a basic unit, thus limiting the analysis to only a part of the entire urbanization spectrum of a country. Consequently, the geographical influence between neighbouring areas cannot be adequately considered \cite{Bergs2021SpatialNegligible}. In this research, a set of municipal\footnote{A municipality is a city or a town or a set of localities having a dedicated local government (\href{https://dictionary.cambridge.org/dictionary/english/municipality?q=Municipality}{Municipality|Cambridge Dictionary}).} units is chosen rather than cities, for two reasons:
\begin{itemize}
	\item [1] City boundaries are unofficial and often ambiguously defined compared to municipalities that enclose their localities and rural area situated on a particular part of national land and account for their particular part of the total national population. 
	\item [2] All cities belonging to one country do not cover together the entire national surface and do not comprise the entire national population. In contrast, municipalities together constitute an entire country in terms of land surface and population, allowing for analysis of a country as a network of interconnected municipalities. To the best of our knowledge, the evolution of municipalities over time has not been analysed from a network perspective before.
\end{itemize}

A large system of elements (nodes) and their interactions or relations (links) can be represented by a network. The characterization of networks has been extensively investigated for classification purposes and for understanding the effects of the network structure on its functioning \cite{Newman2003TheNetworks,Barabasi2016NetworkIntroduction,Newman2018Networks}.
From a network science perspective, this research focuses on understanding the underlying processes that influence the evolution and survivability of geographical areas of a country.

This paper concentrates on the population and area size distributions at the municipality level and the processes that change their characteristics over time. We argue that the population and area size, together with the underlying topology, sufficiently correlate with the probability of a municipality being annexed by a neighbouring municipality. We demonstrate that the municipality merging process changes the area size distribution. The population distribution is also influenced by the continuous (inter)national migration of people across the municipality network.  Regarding migration we distinguish two different types of migration flows:
\begin{itemize}
	\item  People moving from small(er) to large(r) municipalities in terms of population size. This migration flow occurs due to more attractive characteristics such as urban infrastructure, better facilities, employment and economic opportunities available in large(r) municipalities \cite{Bettencourt2007GrowthCities,Schlapfer2014TheSize},
	\item  People moving from municipalities with a large(r) population to municipalities with small(er) population. This migration flow, enabled by mass-commuting\footnote{The increase of the number of privately owned cars in The Netherlands is described in the $2019$ publication of Statistics Netherlands; De groei van het Nederlandse personenautopark \cite{Molnar-intVeld2019DePersonenautopark}.} since the $1960$s, occurs due to a more attractive cost of living, more space per person and affordable housing, thus avoiding the drawbacks of densely populated urban municipalities.
\end{itemize}
These two opposite migration flows take place simultaneously and shape the migration patterns across the municipality network.
These migration flows can be regarded as an optimisation process in which people aim to obtain advantages of both small(er) and large(r) municipalities as much as they can. People tend to live close enough to large urban areas to enjoy the benefits but, on the contrary, distant enough to also enjoy the additional living space and nature in smaller localities. 
Consequently, individual citizens decide about the trade-off between their commuting time \cite{Marchetti1994AnthropologicalBehavior} and geographical distance\footnote{While in $1947$ only $15 \%$ of working population in The Netherlands worked outside the municipality where they lived, in $2006$ that percentage reached $56 \%$ \cite{Molnar-intVeld2019DePersonenautopark}.} between their specific household situation and their work locations. As a result of all these individual decisions, the two migration flows directly govern the population distribution while indirectly influencing the merging process, the topology change and area distribution over time. 

\begin{figure}[!h]
	\begin{center}
		\includegraphics[angle=0, scale=0.6]{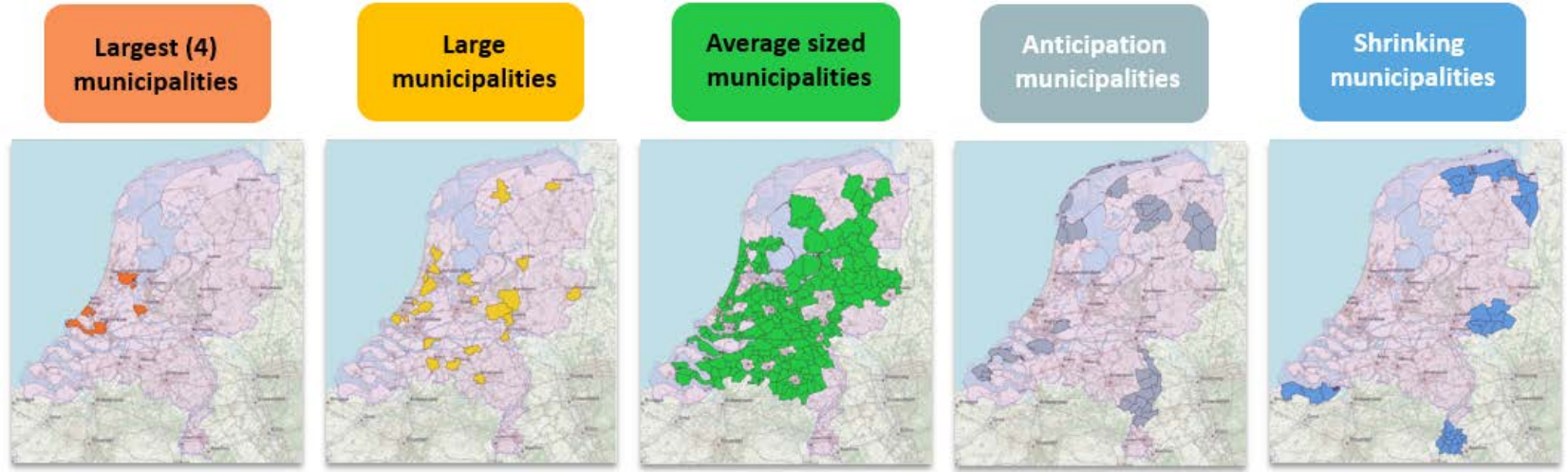}
		\caption{ Classification of the Dutch municipalities in five categories of population.}
		\label{Fig_Classification_Dutch_Municipalities}
	\end{center}
\end{figure}

After collecting and combining approximately $200$ years of Dutch statistical data into one large multi-layer network where each layer contains both the population and area per municipality per year, we focus on the Netherlands and we design a method and model allowing for quantitative network analysis. However, our research approach can be applied to any urbanized country if lengthy time series of statistical data are available from the respective national statistical offices.

From densely populated urban areas to the smallest villages in The Netherlands, more than $2000$ localities are grouped into municipalities. Continuing today, a major development observed from recorded population-related time series\footnote{\href{https://opendata.cbs.nl/statline/\#/CBS/nl/dataset/37259ned/table?ts=1638870604154}{Population development; live births, deaths and migration by region| CBS}} is the municipality merging process, also referred to as municipal restructuring \cite{Hoekveld2014UrbanSouthern-Limburg}. Figure \ref{Fig_Classification_Dutch_Municipalities} shows a classification of all Dutch municipalities in five\footnote{\label{Footnote_DMN}\href{https://opendata.cbs.nl/statline/\#/CBS/nl/dataset/37259ned/table?ts=1638870604154}{Areas of shrinkage and anticipation areas|CBS}} population categories. Although sometimes newly established municipalities occur in municipal restructuring, due to coalitions, renaming and in a few cases creating land from water, in this paper, we use the term \textit{municipality merging process} for all processes influencing CBS-codes (explained in Appendix \ref{App_Merger_Types}). 
The number of Dutch municipalities decreased since the beginning of industrialization (the end of the first half of the $19$th century), while the population steadily increased. 
For example, at the beginning of the industrial revolution, The Netherlands consisted of 1228  mainly rural municipalities, gradually decreasing to 1016 in 1947 towards 355 mainly urbanized municipalities in 2019. 
In the meantime, the national population density tripled between $1905$ and $2010$. 
According to the Ministry of Interior Affairs, The Netherlands has 9 shrinking areas\footnote{In Dutch: krimpgebieden or krimpregio’s} and 11 anticipation areas\footnote{In Dutch: anticipeergebieden or anticipeerregio’s} as shown in Figure \ref{Fig_Classification_Dutch_Municipalities}. A shrinking area \cite{Hoekveld2014UrbanSouthern-Limburg} is defined as an area where the population is expected to decrease by at least 12.5\% until 2040, while the decrease in the number of households is expected at least 5\%. Areas, where the population is declining less rapidly, are called anticipation areas. In anticipation areas, the population is forecast to decrease by at least 2.5\% until 2040.

In Section \ref{Sec_DMN} we define the Dutch Municipality Network. We analyse over time its topology changes: how its population and area sizes evolved at the national, province and municipality level. Section \ref{Sec_DMN_Governing_Processes} examines the governing processes behind population and area distribution changes over time from which we propose a model for the Dutch Municipality Network in Section \ref{Sec_DMN_model}. In Section \ref{Sec_Conclusions} we conclude.

\section{Dutch Municipality Network}\label{Sec_DMN}

\begin{figure}[!h]
	\begin{center}
		\includegraphics[angle=0, scale=0.67]{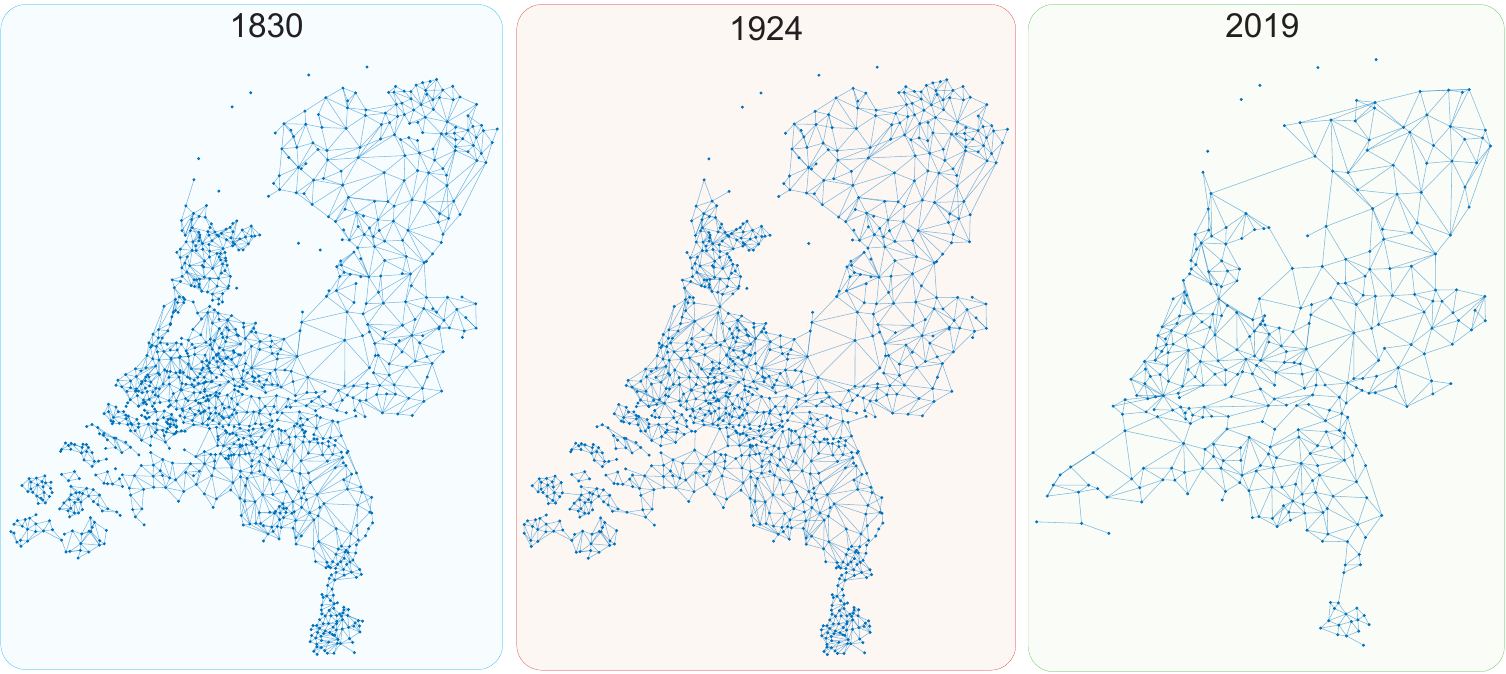}
		\caption{Dutch Municipality Network topology in the years $1830$, $1924$ and $2019$.}
		\label{Fig_Topology_1830_vs_2019}
	\end{center}
\end{figure}

We construct the Dutch Municipality Network (DMN) from a dataset consisting of geographical municipality-related polygons for each year between $1830$ and $2019$. 
As a result, Figure \ref{Fig_Topology_1830_vs_2019} shows examples of the planar graphs for the years $1830$, $1924$ and $2019$, in which the position of a node is determined by the geographic coordinates of the town hall of the corresponding municipality. The set of Dutch municipalities in year $k$ constitutes a  temporal network $G\left(\mathcal{N}[k], \mathcal{L}[k]\right)$, defined by the set $\mathcal{N}[k]$ of $N[k] = |\mathcal{N}[k]|$ nodes and set $\mathcal{L}[k]$ of $L[k] = |\mathcal{L}[k]|$ links. Each municipality in year $k$ is represented by a unique node and the underlying topology is defined by the $N[k] \times N[k]$ symmetric adjacency matrix $A[k]$. Nodes $i$ and $j$ share a link (i.e. $a_{ij}[k] =1$) if there are geographical interconnections between adjacent municipalities $i$ and $j$ via roads, railways, bridges or tunnels\footnote{If a pair of adjacent municipalities is exclusively connected in year $k$ via water, we record $a_{ij}=0$ in $A[k]$. Although there can be a ferry service connecting two adjacent municipalities, we record $a_{ij}=0$ because a ferry service can connect more than two municipality nodes in contrast to one link exclusively connecting two nodes. Another characteristic that complicates analysis is the fact that some ferry services are not available during an entire year $k$.}, which were available in year $k$, otherwise $a_{ij}[k]=0$. 

In addition to the set of DMN graphs, a DMN research construct\footnote{The DMN research construct comprises: (I) from 1830 on, the municipality area and merging data for each year k, (II)  from 1851 on, the population vectors for each year k and (III) two population vectors derived from the 1809 and 1830 censuses.} was setup, containing the municipality area size, the population size and the merging data.
Appendix \ref{App_Datasets_Overview} describes the datasets used in this research.  We applied two complementary municipality identification coding schemes to connect the yearly instances, as explained in Appendix \ref{App_A}. 
The time series of data containing the population and area per municipality were collected from the International Institute of Social History\footnote{In Dutch: Internationaal Instituut voor Sociale Geschiedenis.} recorded in the Historical Database of Dutch Municipalities \cite{BoonstraHistoricalDruid} and from Statistics Netherlands\footnote{In Dutch: Centraal Bureau voor de Statistiek (CBS).}.

To better understand the survivability of Dutch municipalities, we analyse different underlying governing processes of the Dutch Municipality Network over time. Subsection \ref{Sec_DMN_topology_in_time} analyses the municipality network topology evolution per year, while the time dynamics of the area and population distribution per municipality are inspected in subsections \ref{Sec_Area} and \ref{Sec_Pop}, respectively.

\subsection{Topology evolution over time }\label{Sec_DMN_topology_in_time}

\begin{figure}[!h]
	\begin{center}
		\includegraphics[angle =0, scale = 0.8]{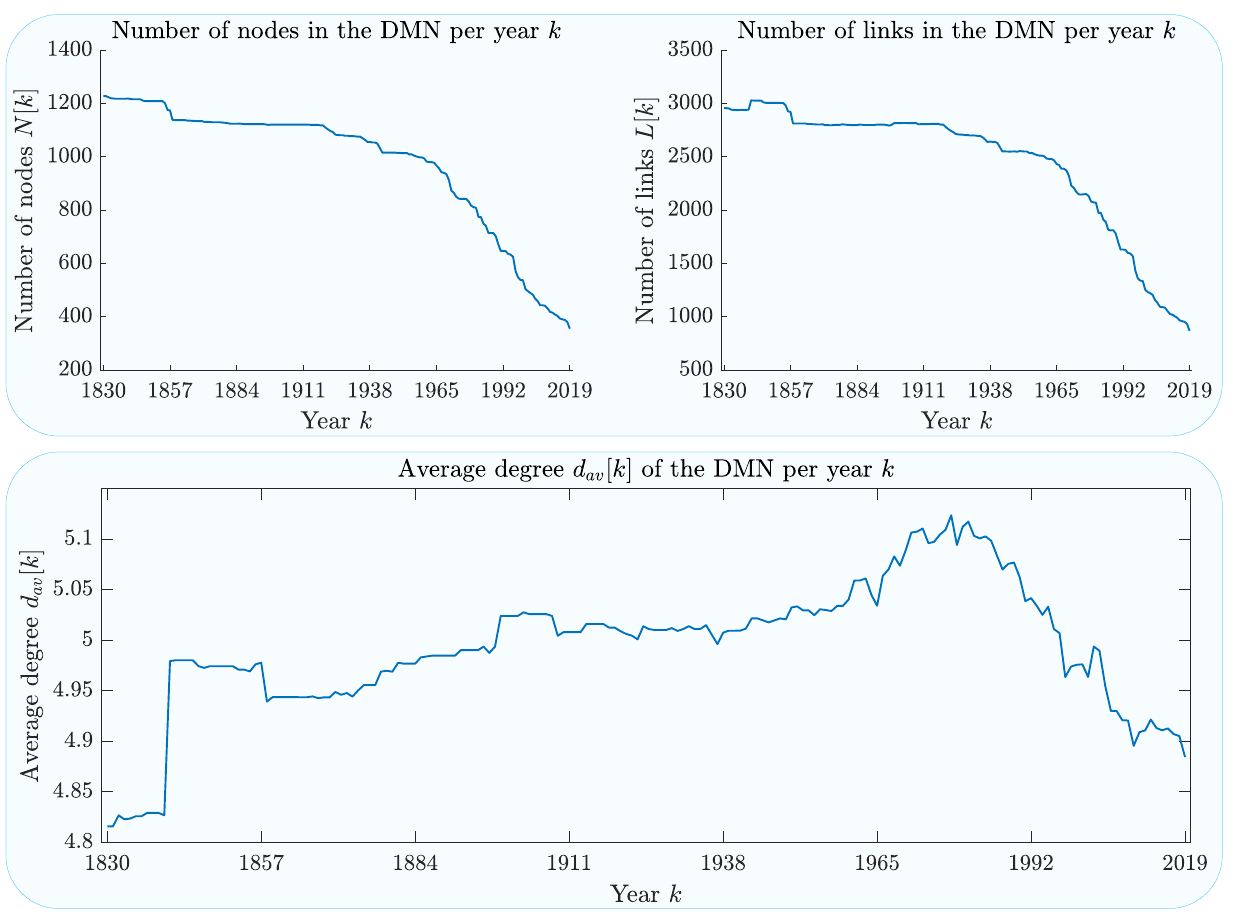}
		\caption{Number of nodes $N[k]$ (upper left-hand side), number of links $L[k]$ (upper right-hand side) and average degree $d_{av}[k]$ (lower part) in the DMN during the period $1830 - 2019$.}
		\label{Fig_Nodes_number_f1}
	\end{center}
\end{figure}
When municipality $i$ merges into an adjacent municipality $j$ in year $k$,  municipality node $i$ disappears and becomes inactive in the $k+1$ instance of the DMN. When a new municipality is created in year $k$, an additional municipality node appears and becomes active in the $k+1$ instance of the DMN. The upper left-hand side of Figure \ref{Fig_Nodes_number_f1}
shows a decrease of $873$ municipality nodes $N[k]$ as a results of municipality merging process between $1830$ and $2019$. In addition, the right-hand side of Figure \ref{Fig_Nodes_number_f1} depicts the number of links $L[k]$ evolution over the researched period.
The average degree $d_{av}[k] = \frac{1}{N[k]} \sum_{j=1}^{N[k]} d_j[k] = 2 \frac{L[k]}{N[k]}$,
with $d_{j}[k] = \sum_{i=1}^{N[k]}a_{ij}[k]$ denoting the degree of the $j$-th node in year $k$, remained almost unchanged during the research period, as depicted in the lower part of Figure \ref{Fig_Nodes_number_f1}
\begin{equation}\label{Average_Degree_eq_1}
	d_{av}[k] \approx 5, \, k\in \{1830,\dots ,2019\}.
\end{equation}
In other words, a typical Dutch municipality in the period $1830-2019$ was surrounded by five neighbouring municipalities on average. Appendix \ref{App_Geographical_Network_Merger} provides a conservation law for the average degree $d_{av}[k]$ on a planar geographical network.
The conservation equation (\ref{Eq_App_d_av}) explains the changes in $d_{av}[k]$ over the year $k$ as shown in the lower part of Figure \ref{Fig_Nodes_number_f1}. When pairs of municipalities merge, the average degree $d_{av}[k]$ slightly increases. Upward spikes in $d_{av}[k]$ occur in the DMN when newly built infrastructure\footnote{Due to road and railway infrastructure development the number of nodes in disconnected components of the DMN decreased from $191$ in the year $1830$ to only $5$ disconnected island municipalities in the Waddenzee in the year $2019$.} connects pairs of municipalities which were previously separated by water. However, in the period after $1960$, the merging process intensified and often took place in waves which involved multiple municipalities per merger. 
As shown in Appendix \ref{App_Geographical_Network_Merger}, the conservation relation (\ref{Eq_App_d_av_3_merger}) indicates a decreasing trend in $d_{av}[k]$, when mergers of clusters of three or more municipalities occur.
As a result, $d_{av}[k]$ started decreasing after $1975$.

\subsection{Area per Dutch municipality}\label{Sec_Area}
In this subsection, we consider area measurements per municipality in the period $1830-2019$ as realisations of the area random variable $S$ of a municipality and examine how the area distribution per municipality changed over time. We show that the random area per municipality on a logarithmic scale, denoted by $Y = \log S$, allows for better insight into the underlying governing processes compared to a linear scale.

The area of each Dutch municipality in year $k$ is a component of the $N[k]\times 1$ vector $s[k]$, where $s_{i}[k]$ denotes the area of municipality $i$ in year $k$. The average area per Dutch municipality in year $k$ is denoted as $s_{av}[k]$
\begin{equation}\label{Eq_S3_S2_S1_E1}
	s_{av}[k] = \frac{1}{N[k]} \cdot\sum\limits_{i=1}^{N[k]} s_{i}[k] = \frac{1}{N[k]}\cdot u^{T}\cdot s[k],
\end{equation}
where $u^{T}=\left[1,1,\cdots, 1\right]$ is the all-one vector. The $N[k]\times 1$ vector $y[k]$ contains the logarithm of area per municipality:
\begin{equation}\label{Eq_Log_Area}
	y[k] = \begin{bmatrix}
		\log (s_{1}[k]) & \log (s_{2}[k]) & \hdots & \log (s_{N[k]}[k]) \\
	\end{bmatrix}^{T}.
\end{equation}

\begin{figure}[!h]
	\begin{center}
		\includegraphics[ angle =0, scale= 0.7]{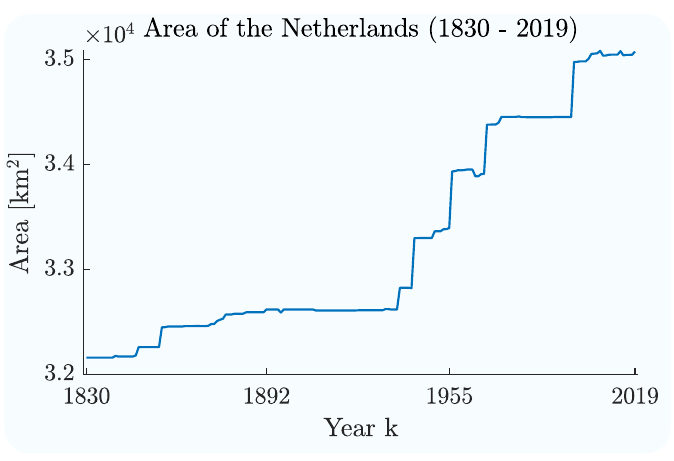}
		\caption{The total land surface of The Netherlands as the summation of the square kilometres from all municipalities over the period $1830-2019$.}
		\label{Fig_Total_Surface_Area_f1}
	\end{center}
\end{figure}

The total land surface of The Netherlands increased due to the process of building dikes, creating polders and draining land from the North sea and (after $1932$) the IJsselmeer\footnote{The Flevoland province, established in $1986$, has been created almost entirely from water and includes the municipalities of Almere, Zeewolde, Dronten, Lelystad, Noordoostpolder and the former island Urk.}. Figure \ref{Fig_Total_Surface_Area_f1} shows that the national area size has increased by $9\%$ between $1830$ and $2019$.

\subsubsection{Area distribution}\label{Sec_Area_Distribution}

The logarithm $Y$ of the area of a typical Dutch municipality closely follows a Gaussian or normal distribution and a logistic or Fermi-Dirac\footnote{The Fermi-Dirac distribution was introduced to describe energy states of particles.} distribution \cite{Tadikamalla1982SystemsVariables}, which are reviewed in Appendix \ref{App_Distribution_Models}.
Instead of applying lognormal and log-logistic distributions on the area random variable $S[k]$ in year $k$, the fitting accuracy with a normal and logistic distribution of the {\em logarithm} of the area random variable $Y[k]=\log(S[k])$ is higher.  
In Appendix \ref{App_Area_Goodness_of_Fit}, we apply the Anderson-Darling (AD) and the Kolmogorov-Smirnov (KS) statistical tests to examine to which extent the hypothesis, that the random variable $Y[k]$ follows a Gaussian or a logistic distribution, holds.
\begin{figure}[!h]
	\begin{center}
		\includegraphics[ angle =0, scale= 0.82]{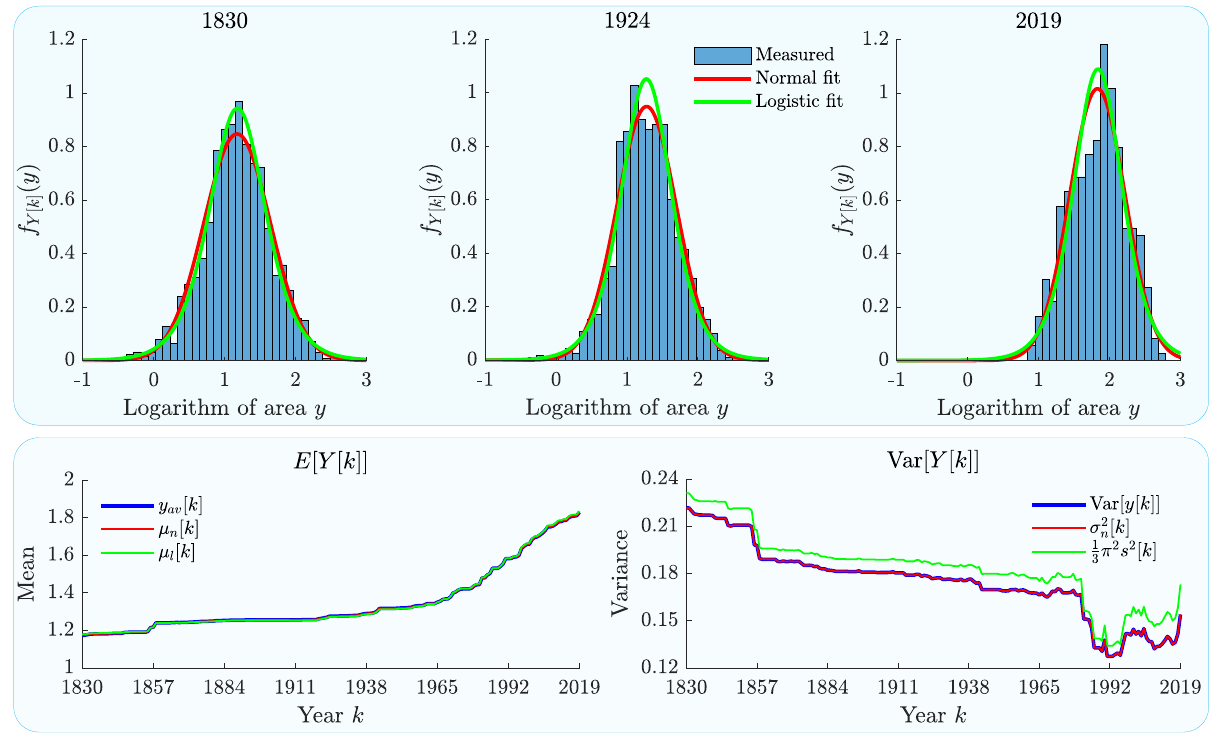}
		\caption{The probability density function $f_{Y[k]}(y)$ of the logarithm of the measured area $Y[k]$ per Dutch municipality (blue bars), fitted with a normal distribution  (defined in (\ref{Eq_Normal_Distribution_f}); red colour) and a logistic distribution  (defined in (\ref{Eq_Logistic_Distribution_f}); green colour) for the years $1830$, $1924$ and $2019$ (upper part). The mean $y_{av}[k]$ (lower left-hand side) and variance $\text{Var}[y[k]]$ (lower right-hand side) of the measured logarithm of area vector $y[k]$ versus the mean and the variance by the  normal fit (red colour) and the logistic fit (green colour) in the period $1830-2019$.}
		\label{Fig_Surface_Distribution_1830_2019_f1}
	\end{center}
\end{figure}

The upper part of Figure \ref{Fig_Surface_Distribution_1830_2019_f1} illustrates the probability density function $f_{Y[k]}(y)$ of the logarithm of measured area per municipality (blue bars), fitted with a normal (red) and a logistic (green) distribution, for the years $1830$, $1924$ and $2019$. 
The mean $\mu_n[k]$ and the variance $\sigma_n^2[k]$ of the normal distribution (defined in Section \ref{App_Normal_Distribution}), together with the mean $\mu_l[k]$ and variance $\sigma_l^2[k] = \frac{1}{3}\pi^2s^2[k]$ of the logistic distribution  (defined in Section \ref{App_Logistic_Distribution}), per year $k$, are compared with the measured mean $y_{av}[k]$ and the variance $\text{Var}(y[k])$ in the lower part of Figure \ref{Fig_Surface_Distribution_1830_2019_f1}. The mean $E[Y[k]]$ of the logarithm of area $Y[k]$ is estimated equally precisely with a normal and logistic distribution. On the contrary, the variance $\text{Var}[Y[k]]$ is better fitted with a normal distribution. The lognormal distribution  (defined in Sec \ref{App_Lognormal_Distribution}) possesses a weaker right tail than a log-logistic distribution  (defined in Sec \ref{App_Log_Logistic_Distribution}), which follows more realistically the geographical boundary that areas of municipalities obey. 
In general, the area of a municipality can increase only at the cost of another municipality annexation, because the total area is almost\footnote{Total area of the mainland of The Netherlands is constant over time, except for the newly built land, as presented in Figure \ref{Fig_Total_Surface_Area_f1}.} constant over time.

\begin{figure}[!h]
	\begin{center}
		\includegraphics[ angle =0, scale= 0.67]{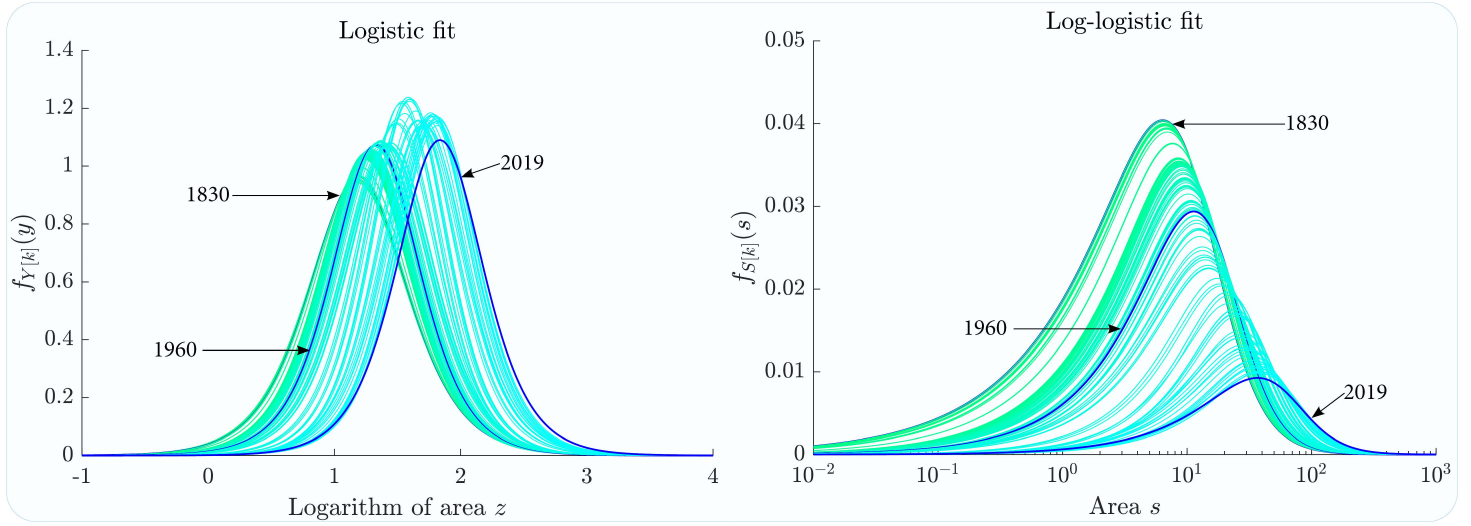}
		\caption{Probability density functions $f_{Y[k](y)}$ of the logistic fit of the logarithm of the area distribution in the period $1830-2019$ (left-hand part). Probability density functions $f_{S[k]}(s)$ of the log-logistic fit of the area distribution in the period $1830-2019$ (right-hand part).}
		\label{Fig_Surface_area_distribution_agregated_fit_f1}
	\end{center}
\end{figure}

As will be derived in (\ref{Eq_Mean_Log_Area}) in Section \ref{Sec_Area_Governing_Processes}, the mean\footnote{The mean $y_{av}[k] = \frac{1}{N[k]}\cdot \sum_{i=1}^{N[k]}\log(s_{i}[k]) = \log\left(\prod_{i=1}^{N[k]}(s_{i}[k])^{\frac{1}{N[k]}}\right)$ represents the logarithm of the geometric mean of the $N[k]\times 1$ area vector $s[k]$.} $y_{av}[k]$ monotonically increases over time due to the merging process, with a pace depending on the merging rate and the area of abolished municipalities. The variance $\text{Var}(y[k])$ mainly decreases over time, except in the last two decades, where fluctuations occur. The decreasing trend of $\text{Var}(y[k])$ with time $k$ reveals the nature of the merging process. In order to visualise how the merging process affected the distribution of the logarithm of the area, the probability density function $f_{Y[k]}(y)$ of the logistic distribution fit (left-hand side of Figure  \ref{Fig_Surface_area_distribution_agregated_fit_f1}) of the logarithm of random area $Y[k]$ and the probability density function $f_{S[k]}(s)$ of the log-logistic distribution fit (right-hand side of Figure  \ref{Fig_Surface_area_distribution_agregated_fit_f1}) of the area random variable $S[k]$ for each year $k$ in the period $1830-2019$.

The left-hand side of Figure \ref{Fig_Surface_area_distribution_agregated_fit_f1} reveals that due to the merging process, municipalities are predominantly abolished from the left-hand side of the distribution curve and were annexed by a neighbouring municipality with a larger area. As a result, the left-hand side of the distribution curve is constantly shifting towards the right-hand side at a faster pace than the right-hand side of the distribution\footnote{While the left distribution tail is shifted towards the right-hand side over time because municipalities with the relatively small area are abolished, the right distribution tail is shifted due to the increase in the area of municipalities that absorbed the abolished ones.}. Therefore, the variance $\text{Var}(y[k])$ decreases, while the mean $y_{av}[k]$ increases over time. The merging process reduces the diversity of municipalities in area size, while the fluctuations in $\text{Var}(y[k])$ indicate the outliers (such as island municipalities) on the left tail. 

\subsection{Population per Dutch municipality}\label{Sec_Pop}

In this subsection, we analyse the population distribution per municipality over time, where the collected population values per municipality are considered a realisation of the population random variable. Similar to the area size in Section \ref{Sec_Area}, we find that the population random variable $P$ reveals less information about underlying governing processes than its logarithm $Z = \log P$.

The population of Dutch municipalities in year $k$ is represented by the $N[k]\times 1$ vector $p[k]$, where the population of municipality $i$ in year $k$ is denoted by $p_{i}[k]$. The total Dutch population $T[k]$ in year $k$ is obtained by summing the population of each active municipality
\begin{equation}\label{Eq_Total_Population}
	T[k] = \sum\limits_{i = 1}^{N[k]} p_{i}[k],
\end{equation}
or $T[k]=u^T p[k]$.
The $N[k]\times 1$ vector $z[k]$ contains the logarithm  $z_{i}[k] = \log (p_{i}[k])$ of the population of municipality $i$ in year $k$, 
\begin{equation}\label{Eq_Pop_Log}
	z[k] = \begin{bmatrix}
	\log (p_{1}[k]) & \log (p_{2}[k]) & \hdots & \log (p_{N[k]}[k]) \\
	\end{bmatrix}^{T},
\end{equation}

\begin{figure}[!h]
	\begin{center}
		\includegraphics[ angle =0, scale= 0.6]{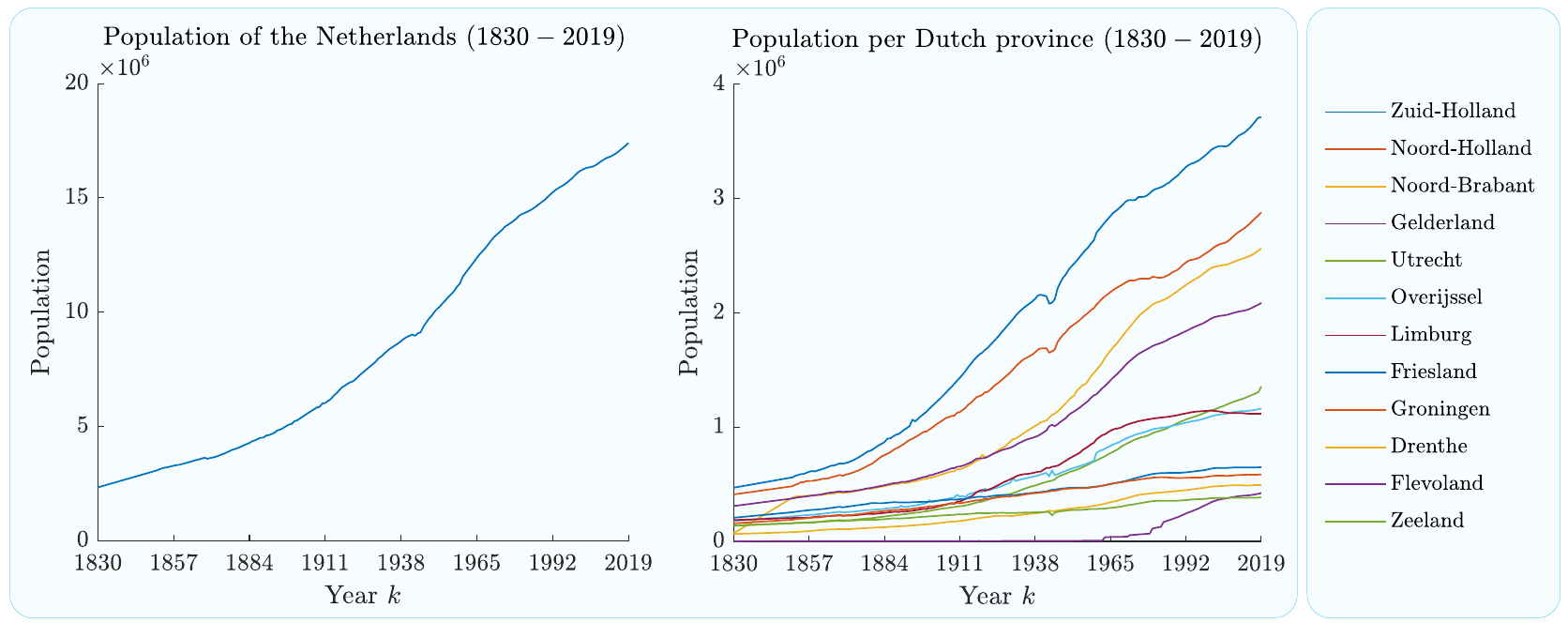}
		\caption{Population development of The Netherlands (left-hand side figure) and population development per Dutch province (right-hand side figure) in the period $1830 - 2019$.}
		\label{Fig_Total_Population_f1}
	\end{center}
\end{figure}
In $1830$, The Netherlands had a population of $2.33$ million people living in $1048$ municipalities, which increased to $17.41$ million citizens in $2019$. 
Although the total population of The Netherlands, shown in Figure \ref{Fig_Total_Population_f1}, has steadily increased during the period $1830 - 2019$, the population increase per province significantly varies (right-hand side of Figure \ref{Fig_Total_Population_f1}). The impact of the Second World War on the population per Dutch province also varies in intensity: the population of the
provinces South Holland\footnote{In Dutch: Zuid-Holland} and North Holland\footnote{In Dutch: Noord-Holland} temporarily decreased most significantly.

\subsubsection{Population distribution}\label{Sec_Pop_Distribution}

\begin{figure}[!h]
	\begin{center}
		\includegraphics[ angle =0, scale= 0.82]{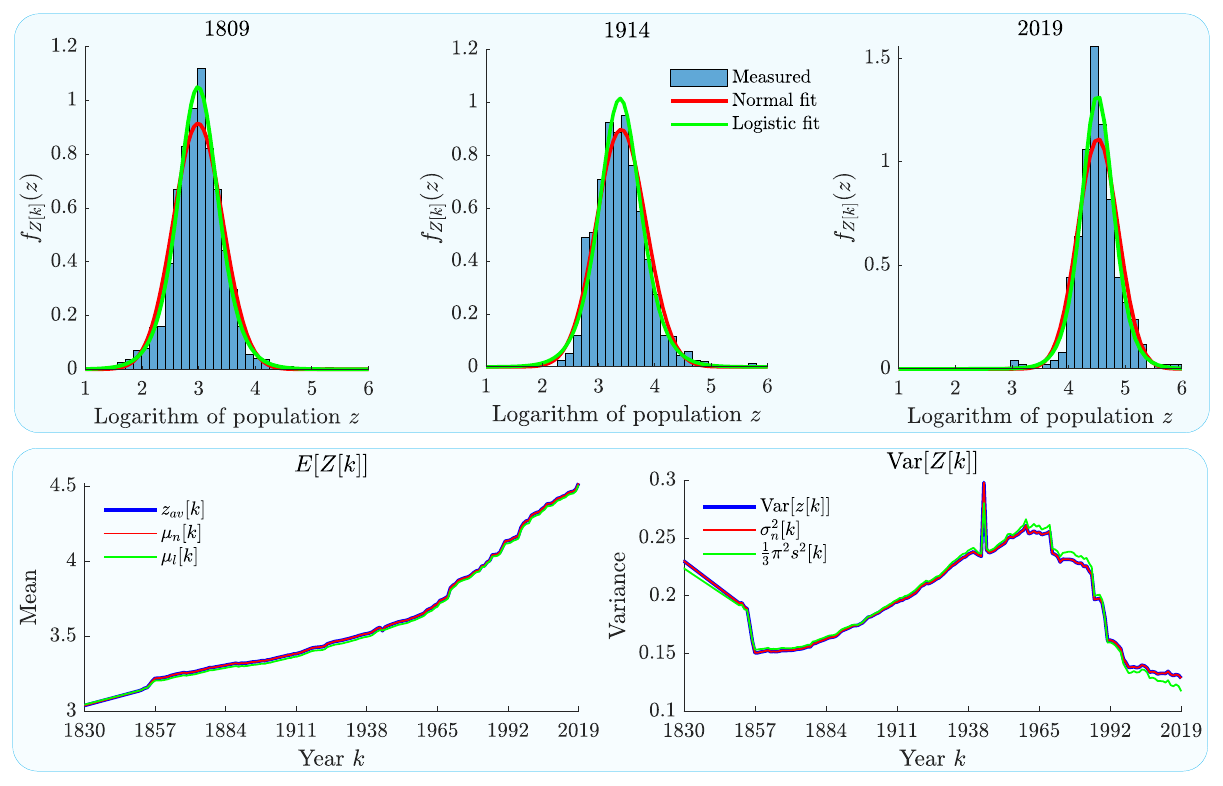}
		\caption{The probability density function of the logarithm of the population vector $Z[k]$ per Dutch municipality (blue bars), fitted with a normal distribution (defined in (\ref{Eq_Normal_Distribution_f}); red colour) and a logistic distribution (defined in (\ref{Eq_Logistic_Distribution_f}); green colour) for the years $1830$, $1914$ and $2019$ (upper part). Mean $z_{av}[k]$ (left-lower part) and variance $\text{Var}[z[k]]$ (right-lower part) versus the mean and the variance by the  normal fit (red colour) and the logistic fit (green colour) in the period $1830-2019$.}
		\label{Fig_Population_Distribution_1809_2019_f1}
	\end{center}
\end{figure}

The logarithm of the Dutch municipality population random variable 
$Z[k]$ closely follows a normal and a logistic distribution in the period $1830 - 2019$. Similarly, the population random variable $P[k]$ follows a lognormal and a log-logistic distribution.
The upper part of Figure \ref{Fig_Population_Distribution_1809_2019_f1} depicts the probability density function $f_{Z[k]}(z)$ of the logarithm of the population per Dutch municipality $Z[k]$ (blue bars), fitted with a normal (red) and a logistic (green) distribution, for the years $1830$, $1914$ and $2019$. 
In Appendix \ref{App_Population_Goodness_of_Fit}, we apply the Anderson-Darling (AD) and the Kolmogorov-Smirnov (KS) statistical tests to examine to which extent the hypothesis of the random variable $Z[k]$ following a normal or a logistic distribution holds.

To understand how the population distribution evolved over the researched period, we analyse the mean $E[Z[k]]$ and the variance $E\left[\left(Z[k] - E[Z[k]]\right)^{2}\right]$ trends over time.
The lower left-hand side of Figure \ref{Fig_Population_Distribution_1809_2019_f1} illustrates the average logarithm of the population $z_{av}[k] = \frac{1}{N[k]}\sum_{i=1}^{N[k]}z_{i}[k]$ in the period $1830-2019$, together with the mean $E[Z[k]]$ of the normal fit $\mu_{n}[k]$ (red colour) and the logistic fit $\mu_{l}[k]$ (green colour). Both considered distributions precisely fit the measured mean $z_{av}[k]$ over time. The monotonically increasing mean $z_{av}[k]$ reveals the national population growth, but also comprises the effects of the migration and merging process, as will be discussed in Section \ref{Sec_DMN_Governing_Processes}.

The variance $\text{Var(z[k])} = \frac{1}{N[k]}\cdot \sum_{i=1}^{N[k]}\left(z_{av}[k] - z_{i}[k]\right)^{2}$ over time is compared in the lower right-hand side of Figure \ref{Fig_Population_Distribution_1809_2019_f1} with the expected variance $E\left[\left(Z[k] - E[Z[k]]\right)^{2}\right]$ of the normal fit $\sigma_{n}^{2}[k]$ (red colour) and of the logistic fit $\frac{1}{3} \pi^{2} \sigma_{l}^{2}[k]$ (green colour). Different trends of the variance $\text{Var(z[k])}$ over time reveal opposite migration patterns\footnote{We refer here to two opposite migration flows, from small(er) to large(r) municipalities and from large(r) to small(er) municipalities, as defined in the Introduction.}
across the geographical network of municipalities, which is derived in Section \ref{Sec_DMN_Governing_Processes}.

\begin{figure}[!h]
	\begin{center}
		\includegraphics[ angle =0, scale= 0.55]{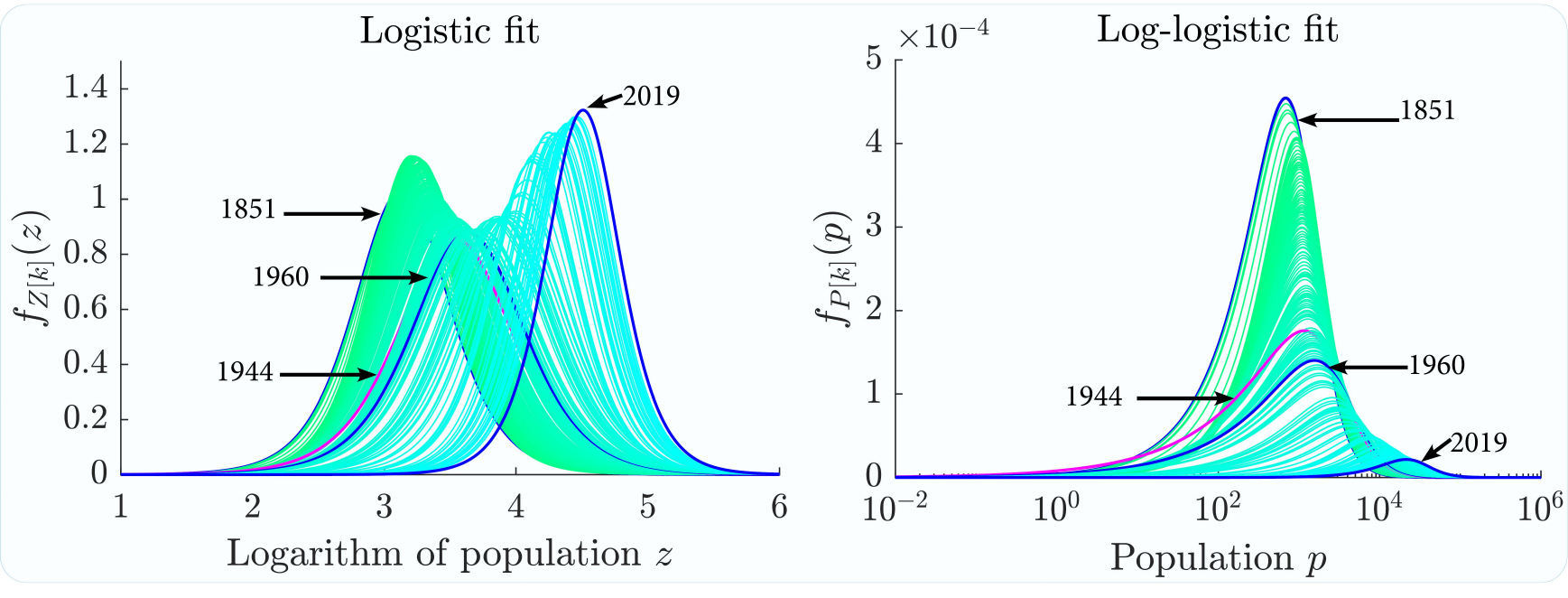}
		\caption{Probability density functions $f_{Z[k]}(z)$ of the logistic fit of the logarithm of the population distribution in the period $1830-2019$ (left-hand part). Probability density functions $f_{P[k]}(p)$ of the log-logistic fit of the population distribution in the period $1830-2019$ (right-hand part).}
		\label{Fig_Population_Distribution_Aggregated_fit_f1}
	\end{center}
\end{figure}

The left-hand side of Figure \ref{Fig_Population_Distribution_Aggregated_fit_f1} illustrates the probability density functions $f_{Z[k]}(z)$, from annual logistic fits of the logarithm of the population $Z[k]$ in the period $1830 - 2019$. On the right-hand side of Figure \ref{Fig_Population_Distribution_Aggregated_fit_f1}, we provide the probability density functions $f_{P[k]}(p)$ of the log-logistic annual fit of the population random variable $P[k]$. Figure \ref{Fig_Population_Distribution_Aggregated_fit_f1} reveals the following general trends:
\begin{itemize}
	\item The mode of the probability density function $\text{mode}(f_{Z[k]}(z)) = \frac{1}{\sigma_{l}[k]}$ is inversely proportional to the square root of variance $\sqrt{\text{Var}(Z[k])}$, as the enclosed surface under an bell-shaped curve obeys $\int_{-\infty}^{\infty}f_{Z[k]}(z) dz = 1$. Therefore, the increasing $\text{mode}(f_{Z[k]}(z))$ over time reflects a decreasing diversity in population on a logarithmic scale.
	\item Both probability density functions $f_{Z[k]}(z)$ and $f_{P[k]}(p)$ are continuously shifted to the right-hand side during the entire researched period, reflecting an almost $8$ times increase in the total population of The Netherlands between $1830$ and $2019$ (see left-hand side of Figure \ref{Fig_Total_Population_f1}). 
\end{itemize}

Since the mean $E[Z[k]]$ and the variance $\text{Var}[Z[k]]$ of the logarithm of the random population $Z[k] = \log P[k]$ are fitted precisely by both distributions, the difference lies in the deep tails, where normal and logistic distributions behave considerably different. 
As derived in (\ref{Eq_Logistic_Tail}), the probability density function $f_{Z[k]}(z)$ of a logistic distribution on a double logarithmic scale decays linearly with population $p$, while the probability density function $f_{Z[k]}(z)$ of a normal distribution decreases as a square function of the population $p$, as derived in (\ref{Eq_Normal_Tail}).
Linear decay in the probability density function  $f_{Z[k]}(z)$ of the logistic distribution indicates that the population distribution in the deep tail follows a power-law distribution.

\subsubsection{Power-Law Fitting}\label{Sec_Power_Law}

The population per random municipality $P[k]$ in year $k$ follows a power law if it obeys the distribution function $F_{P[k]}(p) = \text{Pr}(P[k]\geq p)$ obeys
\begin{equation}\label{Eq_Power_Law_P}
	F_{P[k]}(p) = \left(\frac{p}{p_{min}[k]}\right)^{-\tau[k] + 1},
\end{equation}
where the distribution parameter $\tau[k]$ in year $k$ is known as the exponent or scaling parameter, while $p_{min}[k]$ denotes the minimum population value in year $k$ that obeys the power law. In empirical datasets, a power law often fits only a subset of a vector, explaining the rare occurrence of large outcomes \cite{Newman2000}. The corresponding probability density function $f_{P[k]}(p)=\frac{dF_{P[k]}(p)}{dp}$ of the power-law distribution in (\ref{Eq_Power_Law_P}) is as follows
\begin{equation}\label{Eq_Power_Law_p}
	f_{P[k]}(p) = C[k]\cdot p^{-\tau[k]},
\end{equation}
where $C[k] = (\tau[k] - 1)\cdot p_{min}[k]^{(\tau[k] - 1)}$ denotes the normalisation constant in year $k$. 
\begin{figure}[!h]
	\begin{center}
		\includegraphics[angle=0, scale=0.6]{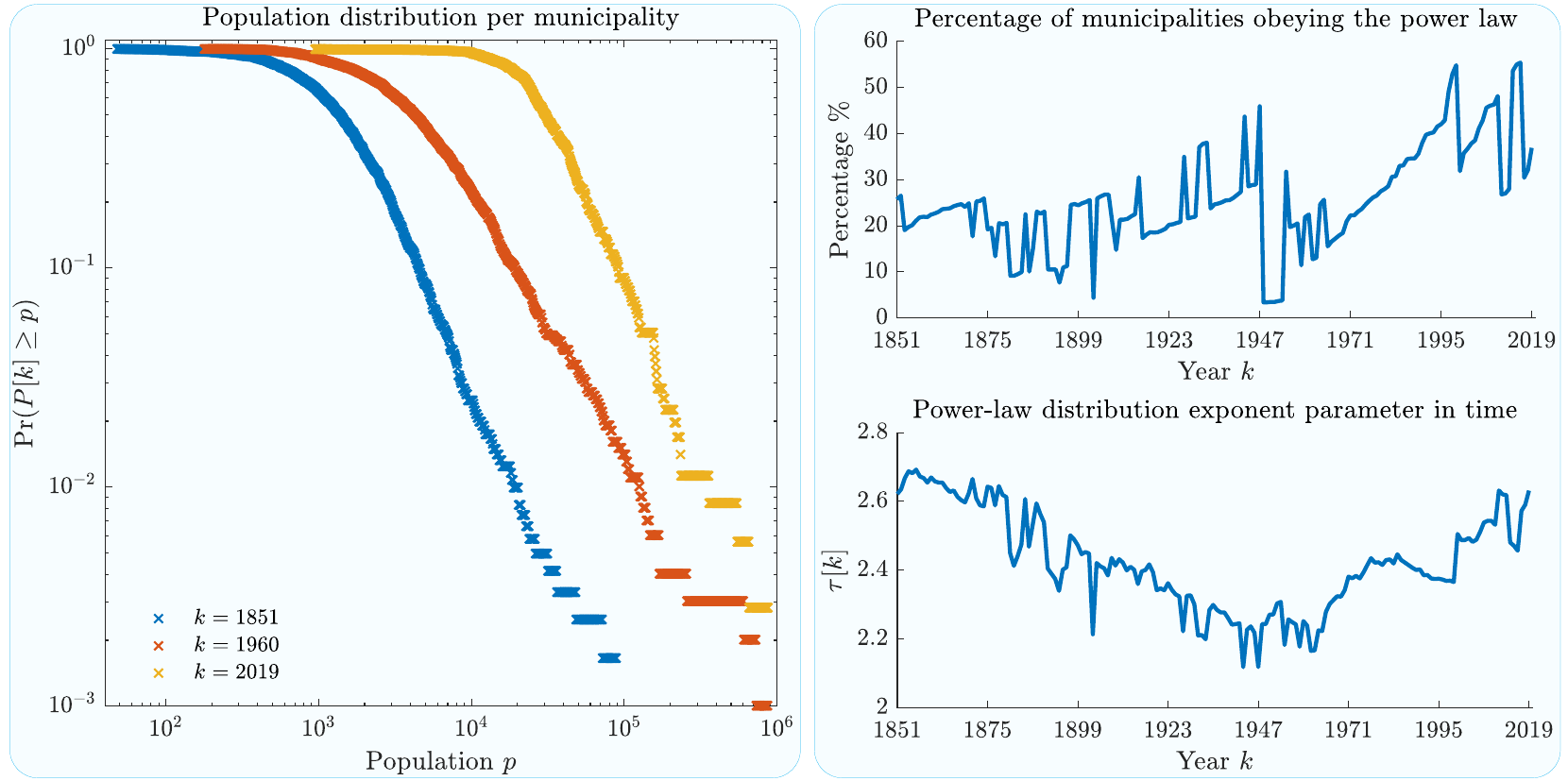}
		\caption{Distribution function $\text{Pr}(P[k]\geq p)$ of the population per municipality in years $1851$, $1960$ and $2019$ (left-hand side). The percentage of the $N[k]\times 1$ vector $p[k]$ that is fitted by a power-law fit in the period $1851 - 2019$ (upper right-hand part). Estimated exponent $\tau [k]$ of the power-law fit of population distribution per municipality in the period $1851 - 2019$ (lower right-hand part).}
		\label{Fig_Population_Power_Law}
	\end{center}
\end{figure}
For each year $k$ in the period $1830-2019$, the distribution function $F_{P[k]}(p) = \text{Pr}(P[k]\geq p)$ is fitted by a power-law distribution for a subset of medium and larger municipalities and the power law exponent $\tau[k]$ is estimated. Figure \ref{Fig_Population_Power_Law} shows the distribution function $F_{P[k]}(p) = \text{Pr}(P[k]\geq p)$ for the years $1851$, $1960$ and $2019$ on the left-hand side, while the percentage of the total number of municipalities, that approximately follow a power law over time, is drawn on the upper right-hand part. In the lower right-hand part of Figure \ref{Fig_Population_Power_Law}, the power law exponent $\tau[k] \in (2.1,2.7)$ roughly decreases up to $1960$ and increases after $1960$. 
Apparently, during the Dutch urbanisation period, featured by a dominant migration flow of people towards large(r) municipalities, the power law exponent $\tau[k]$ decreased towards about $\tau[1961]=2,17$ at $1961$. Subsequently, the opposite migration flow dominated after $1960$ and caused an increasing trend in the power law exponent $\tau[k]$. Probability functions in years $1851$, $1960$ and $2019$, presented on the left-hand side of Figure \ref{Fig_Population_Power_Law}, depict the effect of different migration flows on the population distribution per municipality.


In this section, we showed that the average degree $d_{av}[k]$ slightly increases when two municipalities are merged while $d_{av}[k]$ decreases due to a merger of more than two municipalities. Irrespective of the merger type, the average area size per municipality on a logarithmic scale $y_{av}[k]$ monotonically increased. In contrast, the variability in the area size $\text{Var}(y[k])$ decreased over the entire research period. Multiple underlying processes govern the time dynamics of the population. The increasing trend of the average logarithm $z_{av}[k]$ of population per municipality reveals a national population increase, but also the opposite trends of variability in population size $\text{Var}(z[k])$.

\section{Dynamic Processes on the Dutch Municipality Network}\label{Sec_DMN_Governing_Processes}

In this section, we identify underlying processes in the Dutch municipality network and how they impacted population and area distribution in the period $1830-2019$. Overall, we show that taking the logarithmic of the relevant quantities simplifies the analysis of the governing processes, which is a quite remarkable observation\footnote{Often human behaviour seems to follow a lognormal distribution as in Twitter \cite{Doerr2013LognormalSpread} and online social platforms like Digg \cite{VanMieghem2011LognormalNetwork}. To the best of our knowledge, there is no rigorous theory of why the {\em logarithm} of quantities related to human behaviour (as here population and area) appears so often.}. While both population and area distributions are heavy-tailed on a linear scale, they are bell-shaped on a logarithmic scale. Thus, we find that the mean and variance on a logarithmic scale describe the population and area distribution more precisely than on a linear scale.

\subsection{Municipality merging process}\label{Sec_Merging_Process}

\begin{figure}[!h]
	\begin{center}
		\includegraphics[angle=0, scale=0.81]{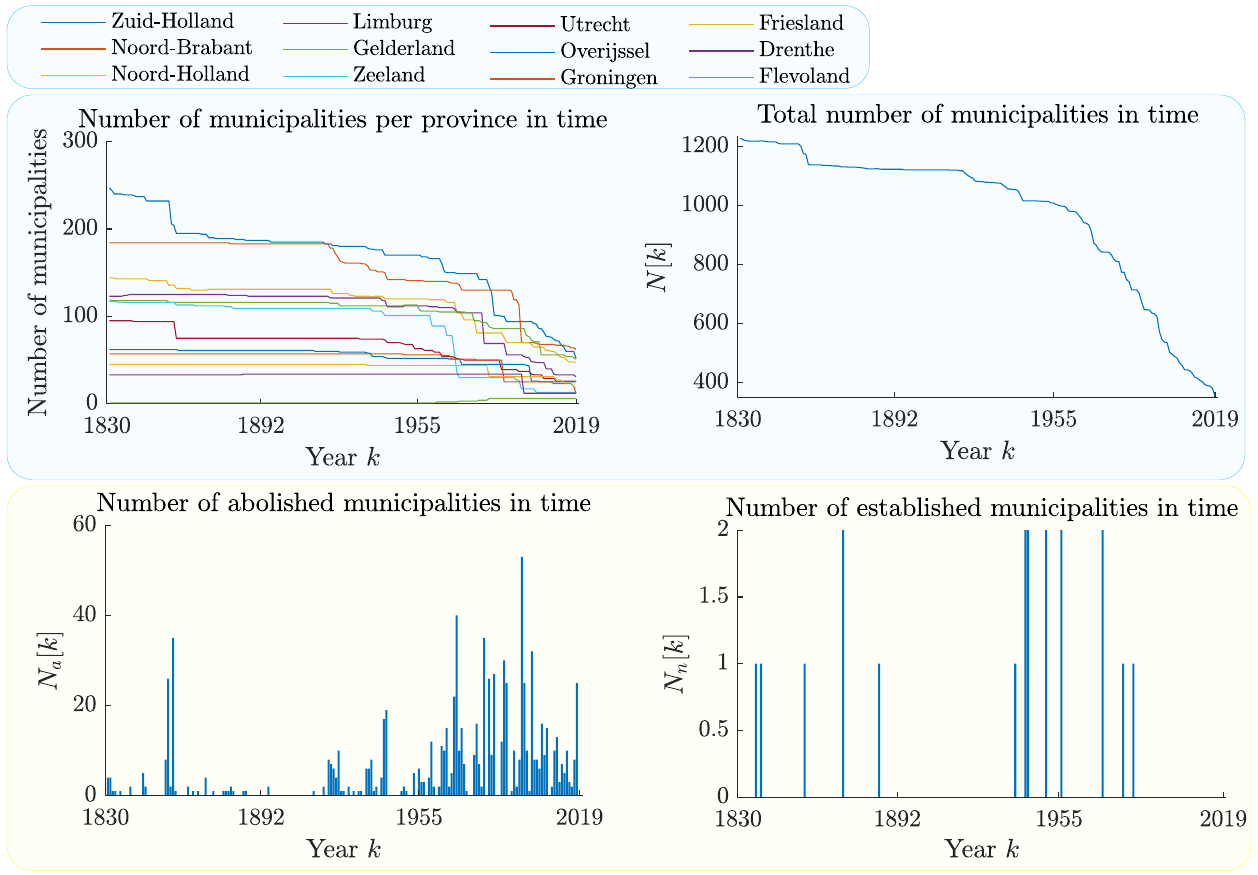}
		\caption{Number of municipalities per each of the $12$ Dutch provinces in the period $1830 - 2019$ (upper left-hand side). The total number of Dutch municipalities $N[k]$ in the period $1830 - 2019$ (upper right-hand side). Number of abolished $N_{a}[k]$ (lower left-hand side) and newly established municipalities $N_{n}[k]$ (lower right-hand side) in the period $ 1830 - 2019$.}
		\label{Fig_Number_of_Municipalities_abo_est}
	\end{center}
\end{figure}

At the end of each year $k$, a number -- possibly none -- of municipalities is abolished and annexed by one or more neighbouring municipalities. We denote the set of abolished municipalities at the end of year $k$ as $\mathcal{A}[k]$, with the number of abolished municipalities denoted by $N_{a}[k] = |\mathcal{A}[k]|$. The evolution of the number of abolished municipalities $N_a[k]$ over time is depicted in the lower left-hand side of Figure \ref{Fig_Number_of_Municipalities_abo_est}. Appendix \ref{App_Merger_Types} provides an overview of the merger types in the Dutch Municipality Network during the research period. The merging process became most intensive in the second part of the 20th century,  decreasing population and area diversity while increasing the average size per municipality. 

Newly established municipalities additionally modify the Dutch Municipality Network topology over time. 
The total area of The Netherlands increased since $1830$ due to the reclaimed land from the sea on which new municipalities have been established.
We denote the number of newly established municipalities at the end of year $k$ as $N_{n}[k]$. The lower right-hand side of Figure \ref{Fig_Number_of_Municipalities_abo_est} depicts how often new municipalities were established in the period $1830-2019$. 
The evolution of the number of municipalities $N[k]$ over time, as presented in the upper left-hand side of Figure \ref{Fig_Number_of_Municipalities_abo_est}, obeys the following conservation law
\begin{equation}\label{Eq_Number_of_municipalities}
	N[k+1] = N[k] + N_{n}[k] - N_{a}[k].
\end{equation}
However, since very few new municipalities\footnote{As presented in the lower right-hand side of Figure \ref{Fig_Number_of_Municipalities_abo_est}, since $1830$ until $2019$ in total $\sum_{i=1830}^{2019}N_n[i] = 19$ new municipalities have been established.} have been established during $1830-2019$, the municipality merging process predominantly drives the changes in the DMN topology. Thus, for the following analysis, we approximate (\ref{Eq_Number_of_municipalities}) as
\begin{equation}\label{Eq_Number_of_Nodes_Approx}
	N[k+1] \approx N[k] - N_{a}[k].
\end{equation}
The difference equation (\ref{Eq_Number_of_Nodes_Approx}) appears commonly in literature and can be solved by iteration\footnote{By iteration over $k$, we obtain $N[k]=N[m]-\sum_{j=p}^{m-1} N_a[j]$, where we assume that year $m < k$ is known or is the initial condition.} over $k$. The general exact solution is found via generating functions (see, e.g. \cite[p. 123]{VanMieghem2006DataNetworking}).

\subsection{Governing processes of the area dynamics}\label{Sec_Area_Governing_Processes}

The area distribution per municipality evolves due to merging and establishing new municipalities. Since the latter occurs relatively rarely, we focus on how the merging process impacts the area distribution. We show that the area dynamics on a linear scale depend solely on the number of abolished municipalities $N_a[k]$, while the analysis on a logarithmic scale reveals additional information about the merging process. 

\begin{figure}[!h]
	\begin{center}
		\includegraphics[ angle =0, scale= 0.8]{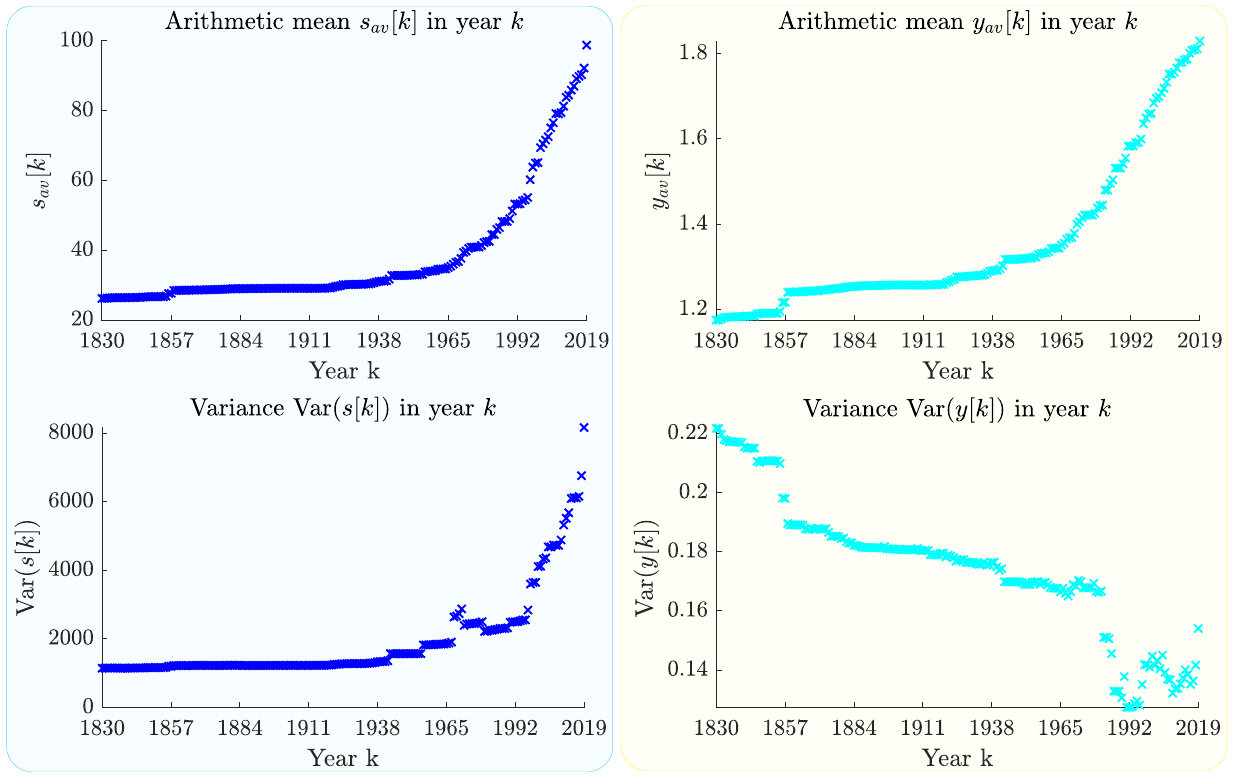}
		\caption{Mean $s_{av}[k]$ (upper left-hand part) and Variance $\text{Var}(s[k])$ (lower left-hand part) of the $N[k]\times 1$ area vector $s[k]$ in the period $1830-2019$. Mean $y_{av}[k]$ (upper right-hand part) and Variance $\text{Var}(y[k])$ (lower right-hand part) of the $N[k]\times 1$ logarithm of area vector $y[k]$ in the period $1830-2019$.}
		\label{Fig_Mean_Variance_Area}
	\end{center}
\end{figure}

In Figure \ref{Fig_Mean_Variance_Area}, we provide the mean $s_{av}[k]$ (upper left-hand side) and the variance $\text{Var}(y[k])$ (lower left-hand side) of the $N[k]\times 1$ area vector $s[k]$, as well as the mean $y_{av}$ (upper right-hand side) and the variance $\text{Var}(y[k])$ (lower right-hand side) of the $N[k] \times 1$ logarithm of area vector $y[k]$ in the period $1830-2019$.

We consider a merger case where $N_{a}[k] = |\mathcal{A}[k]|$ abolished municipalities in year $k$ are annexed by an existing municipality $\eta \in \mathcal{N}[k]$. The set of municipalities in the following year becomes $\mathcal{N}[k+1] = \mathcal{N}[k]\setminus \mathcal{A}[k]$, where $\setminus$ denotes the set difference. As a result of a municipality merger, the area of the annexing municipality $\eta$ grows as $s_{\eta}[k+1]=s_{\eta}[k] + \sum_{i\in \mathcal{A}[k]} s_{i}[k]$ in the next year $k+1$. The average area $s_{av}[k+1]$ in the following year $k+1$ evolves as follows
\begin{equation}\label{Eq_Area_Average_Law}
	s_{av}[k+1] = \left(1 + \frac{N_{a}[k]}{N[k]-N_{a}[k]}\right)\cdot s_{av}[k].
\end{equation}
The mean $s_{av}[k]$ over time is presented in the upper right-hand side of Figure \ref{Fig_Mean_Variance_Area}. From combining (\ref{Eq_Number_of_Nodes_Approx}) and (\ref{Eq_Area_Average_Law}), we observe that the mean $s_{av}[k]$ in year $k$ is inversely proportional to the number of municipalities $N[k]$ 
\[
\frac{s_{av}[k+1]}{s_{av}[k]}=\frac{N[k]}{N[k+1]},
\]
thus revealing solely the information about the intensity of the merging process over time. On the contrary, the conservation law for the average $y_{av}[k]$ of the $N[k]\times 1$ vector $y[k]=\log (s[k])$ of the area $s[k]$ per Dutch municipality in year $k$ is
\begin{equation}\label{Eq_Mean_Log_Area}
y_{av}[k+1] = \left(1+\frac{N_{a}[k]}{N[k]-N_{a}[k]}\right)\cdot y_{av}[k] + \frac{1}{N[k]-N_{a}[k]}\log \left(\frac{\sum\limits_{i\in \eta \cup \mathcal{A}[k]}s_{i}[k]}{\prod\limits_{j\in \eta \cup \mathcal{A}[k]}s_{j}[k]}\right),
\end{equation}
as derived in Appendix \ref{App_Area_Evolution}. The second term in (\ref{Eq_Mean_Log_Area}) reveals additional information about the mergers, compared to the conservation law in (\ref{Eq_Area_Average_Law}).
The increase of the mean $y_{av}[k]$ over time, as depicted in the upper right-hand side of Figure \ref{Fig_Mean_Variance_Area}, is bounded by the second term in (\ref{Eq_Mean_Log_Area}). From the left-hand side of Figure \ref{Fig_Surface_area_distribution_agregated_fit_f1}, we observe that the left distribution tail of the logarithm of the area $Y[k]$ is impacted mostly after $1960$, indicating that the municipalities with the smallest areas were often abolished. Mergers involving municipalities from the left distribution tail cause the second term in (\ref{Eq_Mean_Log_Area}) to increase in value and consequently increase the mean $y_{av}[k]$ at a larger pace than before $1960$. From the upper right-hand side of Figure \ref{Fig_Mean_Variance_Area}, we clearly distinguish two linear patterns over time, until and after $1960$.

From the decreasing trend of the variance $\text{Var}(y[k])$ over time, presented in the lower right-hand side of Figure \ref{Fig_Mean_Variance_Area}, we observe that the merging process continuously lowered the area size diversity of Dutch municipalities on a logarithmic scale. In other words, the area size of a municipality negatively correlates with the probability of its abolishment, taking into account the increasing trend of the mean $y_{av}[k]$. Therefore, the municipality area could be considered a predictor of the probability of municipality abolishment.

Since the area and population distribution per Dutch municipality follow the same distribution model, the insights into how the merging process impacted the area distribution also hold for the population, allowing for recognising the impact of other governing processes, such as population growth and people migration.

\subsection{Governing processes of the population dynamics}\label{Sec_Pop_Governing_Processes}

In this section, we analyse how the population increase, population migration between municipalities and the process of municipality merging determined the population distribution per municipality.

\begin{figure}[!h]
	\begin{center}
		\includegraphics[ angle =0, scale= 0.8]{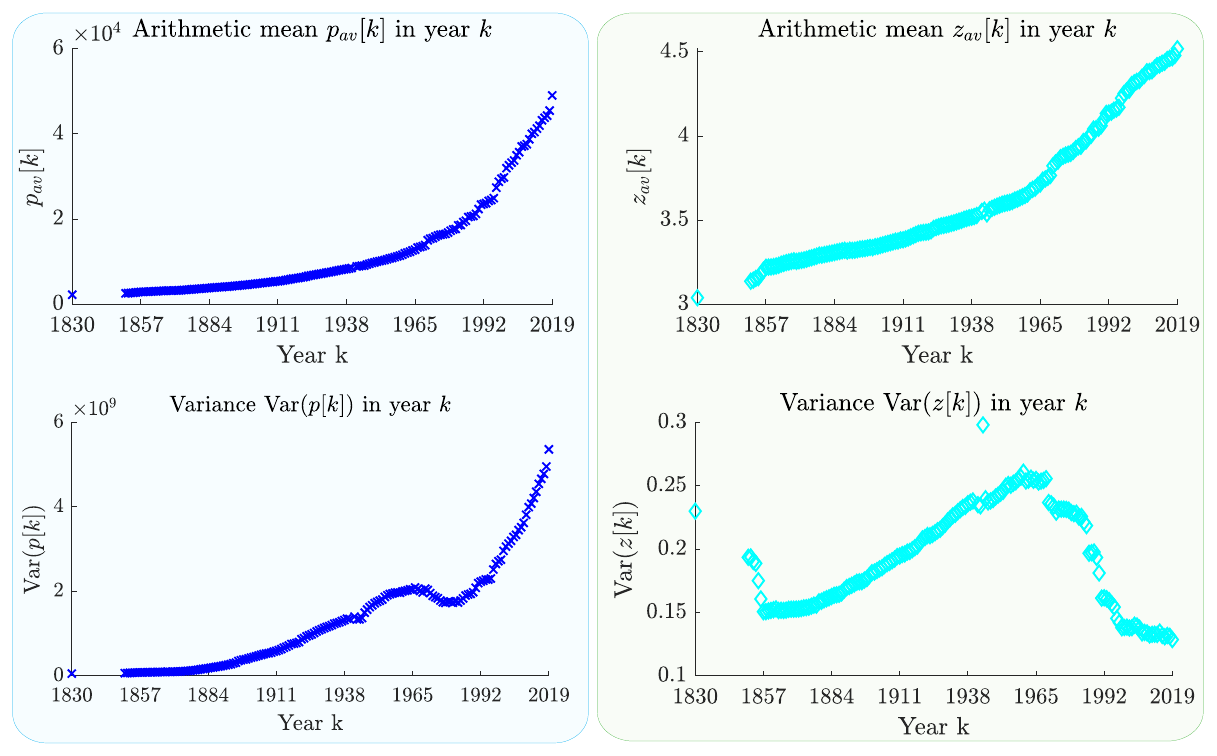}
		\caption{Mean $x_{av}[k]$ (upper left-hand part) and Variance $\text{Var}(x[k])$ (lower left-hand part) of the $N[k]\times 1$ population vector $x[k]$ in the period $1830-2019$. Mean $z_{av}[k]$ (upper right-hand part) and Variance $\text{Var}(z[k])$ (lower right-hand part) of the $N[k]\times 1$ logarithm of population vector $z[k]$ in the period $1830-2019$.}
		\label{Fig_Mean_Variance_Skewness_Population_Measured_f1}
	\end{center}
\end{figure}

In Figure \ref{Fig_Mean_Variance_Skewness_Population_Measured_f1}, we present the mean $p_{av}[k]$ (upper left-hand side) and the variance $\text{Var}(p[k])$ (lower left-hand part) of the $N[k]\times 1$ population vector $p[k]$. In addition, we depict the mean $z_{av}[k]$ (upper right-hand side) and the variance $\text{Var}(p[k])$ (lower right-hand side) of the $N[k]\times 1$ logarithm of the population vector $z[k]$ in the period $1830-2019$.

The trends of both the mean $z_{av}[k]$ and the variance $\text{Var}(z[k])$ in Figure \ref{Fig_Mean_Variance_Skewness_Population_Measured_f1} can be approximated by a two-segment linear function of time $k$, before and after $1960$. 
Further, the variance\footnote{The variance $\text{Var}(z[k])$ spikes in the year $k=1944$ due to the Second World War, and this year represents an outlier in the time dynamics of the DMN population distribution.} of the logarithm of the population vector $\text{Var}(z[k])$ peaks around $1960$ and starts decreasing afterwards, revealing a change in the dynamic pattern of the underlying processes. A decreasing trend of the variance $\text{Var}(z[k])$ coincides with the intensified municipality merging process that took place after $1960$, as presented in Figure \ref{Fig_Number_of_Municipalities_abo_est}.
Another underlying process governing both the mean $z_{av}[k]$ and the variance $\text{Var}(z[k])$ over time is the population evolution per municipality.

\subsubsection{Population rank-size distribution}\label{Sec_Pop_Ranking_Distribution}
\begin{figure}[!h]
	\begin{center}
		\includegraphics[angle=0, scale=0.77]{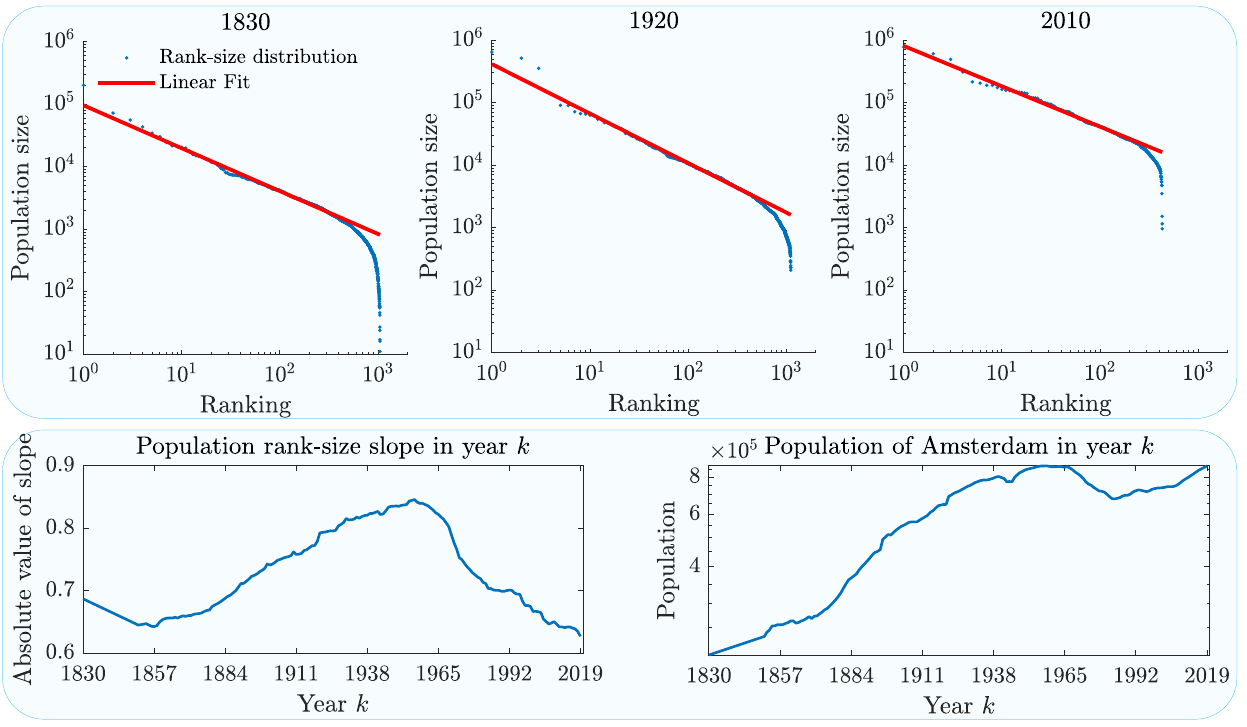}
		\caption{Population rank-size distribution per municipality and the linear fit in double logarithmic scale, in years $1830$, $1920$ and $2010$ (upper part). Estimated slope of the population rank-size distribution in the period $1809 - 2019$ (lower left-hand part). Population of Amsterdam in the period $1830-2019$ (lower right-hand part).}
		\label{Fig_Population_Rank_Size_1}
	\end{center}
\end{figure}

The population distribution of a country's large(r) cities often follows Zipf's Law, which reveals a relationship between the frequency and size of a set \cite{Cristelli2012ThereZipf}. We analyse the rank-size distribution of the population per Dutch municipality in the period $1830 - 2019$. In each year $k$, the population vector $p[k]$ rank-size distribution is fitted with a linear function on a double logarithmic scale. 
The absolute value of the slope of the population rank-size distribution we denote as the population rank-size slope $\beta[k]$.
In the upper part of Figure \ref{Fig_Population_Rank_Size_1}, we provide the population rank-size distribution per municipality, together with the fitted line on a double logarithmic scale for the years $1830$, $1920$ and $2010$.
In addition, we depict the slope $\beta[k]$ of the linear fit (lower left-hand part) and the population of Amsterdam $p_{A}[k]$ in the period $1830 - 2019$.

A general trend of vertical movement\footnote{A vertical movement of the dots in the upper part of Figure \ref{Fig_Population_Rank_Size_1} is similar to the horizontal movement of the logistic probability density function $f_{Z[k]}(z)$ over time, presented in Figure \ref{Fig_Population_Distribution_Aggregated_fit_f1}.} of the point cloud in the upper part of Figure \ref{Fig_Population_Rank_Size_1} reflects the population increase over time, as we explain in Section \ref{Sec_DMN_Model_Step_1}.
The rank-size distribution slope $\beta[k]$ over time reveals two opposite dynamic trends of the population evolution per Dutch municipality in the period $1851-2019$. Since $1851$, the rank-size slope $\beta[k]$ continuously increased and peaked at $\beta[1960] = 0.846$ in $1960$, from when it started decreasing. The population rank-size slope $\beta[k]$ in year $k$ can be approximated as 
\begin{equation}\label{Eq_slope_linear_fit}
	\beta[k] \approx b_{1}\cdot k + b_{2},
\end{equation}
where the parameters $b_{1}$ and $b_{2}$ are estimated for the periods $1851-1960$ and $1960-2019$ as follows
\begin{alignat}{3}\label{Eq_Slope_b_coeffs}
	b_{1} =&  2\cdot 10^{-3}, \,&& b_{2} = -3.23 , \, & & k\in \{1851,1959\} \\
	b_{1} =&  -3.4\cdot 10^{-3}, \,&& b_{2} = 7.47 , \, & & k\in \{1960,2019\}. \nonumber
\end{alignat}
A positive increase in the population rank-size slope $\beta[k+1] - \beta[k]$ between two consecutive years $k$ and $k+1$ in the period $1851 - 1960$ informs us that larger municipalities increased in population size faster than smaller municipalities on a logarithmic scale. 
Here we introduce an assumption important for the following analysis. We assume that the population of each municipality approximately follows Zipf's Law. Therefore, from the rank-size distribution in Figure \ref{Fig_Population_Rank_Size_1}, the logarithm of municipality $i$ population $z_{i}[k]$ in year $k$ can be approximated as
\begin{equation}\label{Eq_Pop_Zipf_Law}
	z_{i}[k] \approx z_{A}[k] -\beta[k]\cdot \log r_{i}[k],
\end{equation}
where $i \in \mathcal{N}[k]$ and municipality $i$ ranking in the $N[k]\times 1$ population vector $p[k]$ in year $k$ is denoted as $r_{i}[k]$, while the logarithm of the population in Amsterdam (i.e. the largest Dutch municipality by population) in year $k$ is denoted as $z_{A}[k] = \log(p_{A}[k])$.
When assuming that the ranking of municipality $i$ in the $N[k]\times 1$ population vector $p[k]$ does not change $r_{i}[k+1] = r_{i}[k]$ in two consecutive years $k$ and $k+1$, we obtain the following governing equation
\begin{equation}\label{Eq_Pop_Zipf_Period_1}
\frac{p_i[k+1]}{p_i[k]} = \left(r_{i}[k]\right)^{-b_{1}}\cdot \frac{p_{A}[k+1]}{p_{A}[k]},
\end{equation}
as derived in Appendix \ref{App_Pop_Evolution}.
The governing equation (\ref{Eq_Pop_Zipf_Period_1}) of the population increase per municipality, for different values of the linear fit parameter $b_{1}$ in (\ref{Eq_Slope_b_coeffs}), reveal two opposite trends of people migration. Until $1960$, people predominantly migrated from small(er) to large(r) municipalities. 
Consequently, municipalities with a large(r) population grew faster. 
In contrast, in the period after $1960$, the largest municipalities in population no longer grew at a dominant pace, revealing the migration flow towards small(er) municipalities in population size. The two dynamic trends are also observable from the population of Amsterdam $p_{A}[k]$, presented on the lower right-hand side of Figure \ref{Fig_Population_Rank_Size_1}. 
Based on the governing equation (\ref{Eq_Pop_Zipf_Law}), we derive the impact of the rank-size slope $\beta[k]$ over time onto the mean $z_{av}[k]$ of the logarithm of population vector $z[k]$
\begin{equation}\label{Eq_Pop_Rank_Mean_Variance}
z_{av}[k] = z_{A}[k] -\beta[k]\cdot \log\left(N[k]!^{\frac{1}{N[k]}}\right).
\end{equation}
which can be further simplified using Stirling's approximation \cite[p. 257]{Abramowitz1968HandbookFunctions}  of the Gamma function
\begin{equation}\label{Eq_Pop_Rank_Mean_Variance_Approx}
z_{av}[k] \approx z_{A}[k] -\beta[k]\cdot \left(\log\left(N[k]\right) - \log(e) + \frac{1}{2N[k]}\cdot\log\left(2\pi N[k]\right)\right).
\end{equation}
Relation (\ref{Eq_Pop_Rank_Mean_Variance}) explains two linear patterns in the mean $z_{av}[k]$ evolution over time, presented in the upper right-hand side of Figure \ref{Fig_Mean_Variance_Skewness_Population_Measured_f1}.
In the period $1830-1960$, the increase in Amsterdam population dominantly impacted the mean $z_{av}$. In the next period $1960-2019$, Amsterdam population saturated on a logarithm scale $z_{A}[k]$. However, both the rank-size distribution slope $\beta[k]$ and the number of municipalities $N[k]$ monotonically decreased in value, keeping the increasing trend of the mean $z_{av}[k]$.

The assumption introduced in (\ref{Eq_Pop_Zipf_Law}) allows to derive the variance $\text{Var}(z[k])$ as provided in Appendix \ref{App_Pop_Evolution}
\begin{equation}\label{Eq_Var_z_via_Beta}
\text{Var}(z[k]) = \beta^{2}[k]\cdot g(N[k]),
\end{equation}
where 
\[
g(N[k])  = \frac{1}{N[k]}\sum_{i=1}^{N[k]}\left(\log \frac{N[k]!^{\frac{1}{N[k]}}}{i}\right)^{2},
\]
explaining the behaviour of the Variance $\text{Var}(z[k])$ over time. Since $1830$ until $1960$, the slope $\beta[k]$ monotonically increased, causing an increase of the variance $\text{Var}(z[k])$. On the contrary, the decreasing trend of the slope $\beta[k]$ after $1960$, together with the decreasing number of municipalities $N[k]$ due to the merging process, caused the $\text{Var}(z[k])$ to decrease. Moreover, the aggressive merging process that took place after $1960$ amplified the decreasing rate of the variance $\text{Var}(z[k])$.

In Appendix \ref{Sec_App_Rank_Size_Power_Law}, we derive an explicit relation between the exponent of the power-law probability density function (depicted in Figure \ref{Fig_Population_Power_Law}) and the rank-size distribution slope $\beta[k]$ (provided in the lower left-hand side of Figure \ref{Fig_Population_Rank_Size_1})
\begin{equation}\label{Eq_Beta_vs_Tau}
\beta[k] = \frac{1}{\tau[k] - 1}.
\end{equation}

\subsubsection{Evolution of municipalities across fixed population size categories}\label{Sec_trajecotires}

\begin{figure}[!h]
	\begin{center}
		\includegraphics[angle=0, scale=0.47]{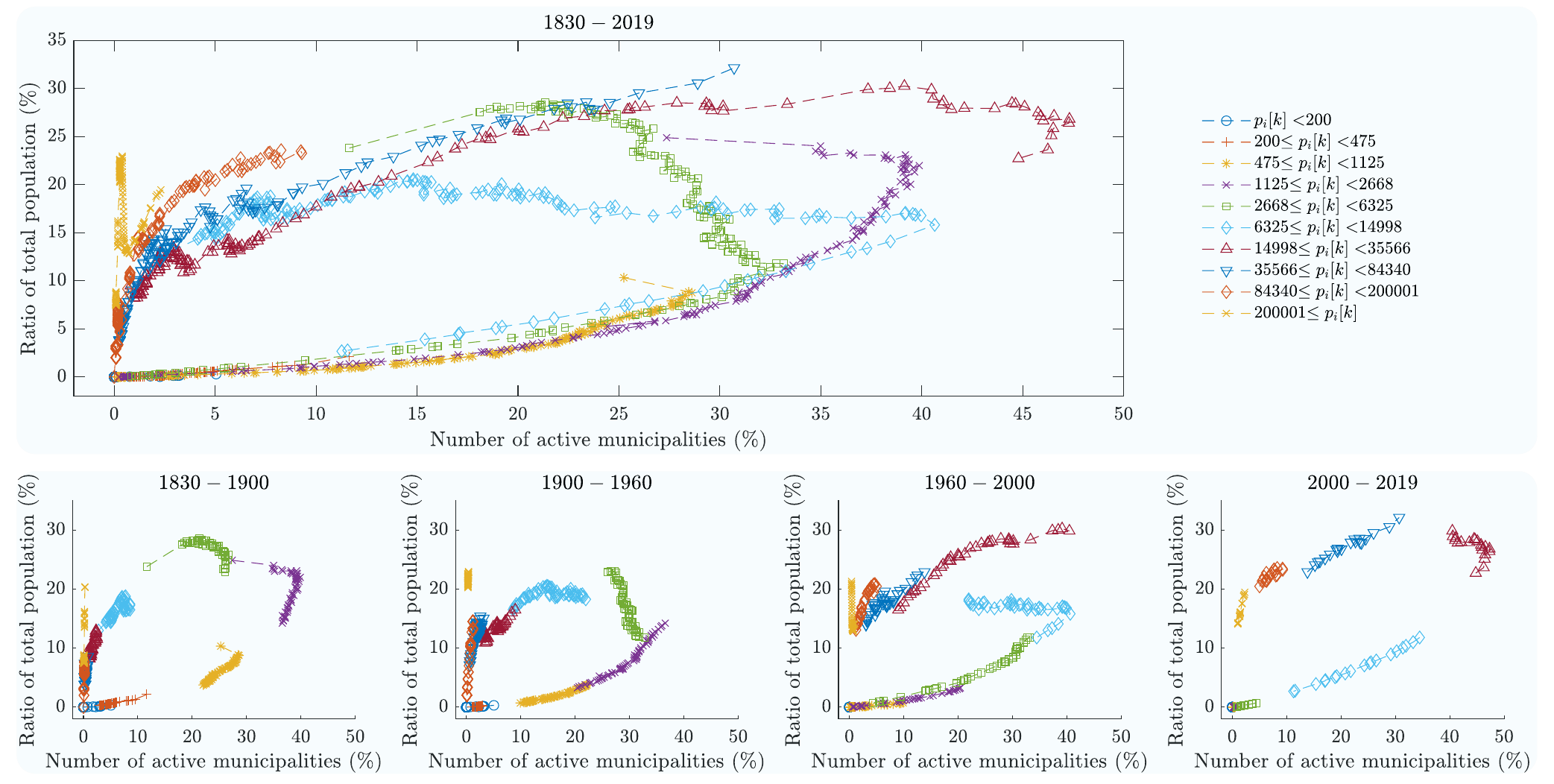}
		\caption{Correlation between the relative number of municipalities and the relative population per logarithmic equidistant population size category in the periods (lower part, from left to right-hand side) $1830 - 1900$, $1900 - 1960$, $1960-2000$ and $200-2019$, and during the entire researched period $1830-2019$  (upper part).}
		\label{Fig_Population_Trajectories_1}
	\end{center}
\end{figure}

To better understand how the population distribution changed over time, we analyse the evolution of municipalities over time across fixed population size categories. The first size category contains municipalities with less than $200$ inhabitants. On the other side, the last category includes municipalities with more than $200.000$ inhabitants. In between, we define equidistant intervals of the population size on a logarithmic scale for a given number of intervals.
For each year $k$ in the period $1830-2019$, we correlate the percentage of the total population in The Netherlands with the ratio of the total number of municipalities per population size category. The correlation is presented in the upper part of Figure \ref{Fig_Population_Trajectories_1}.

Except for the largest municipalities  (i.e. those with more than $200.000$ inhabitants), we observe a consistent correlation pattern in the two-dimensional space. A fixed population size category initially\footnote{Under the assumption, we find an adequate time instant. However, for a given time instant, different categories are in different regions of their respective paths.} contains no municipalities until their population sizes reach values determining the size category. Each category draws a path in the clockwise direction, as presented in Figure \ref{Fig_Population_Trajectories_1}, and eventually tends back to the coordinate origin.
On the contrary, a municipality advances upwards through adjacent categories as population size increases. When advancing, a municipality is the largest in the current category while becoming the new smallest element in the next larger category. Therefore, the impact of a municipality leaving the current category is negative and of considerably higher intensity than a municipality's positive impact upon entering the adjacent larger category. Consequently, the correlation patterns in Figure \ref{Fig_Population_Trajectories_1} must always describe paths in the clockwise direction. 

In the lower part of Figure \ref{Fig_Population_Trajectories_1}, we present correlation patterns in different periods to analyse the impact of different underlying processes. The merging process negatively affects the abolished municipality's category, as its annexation is equivalent to removing that municipality from the corresponding group. On the contrary, the annexing municipality either positively influences the path of its size category or advances to a (possibly non-adjacent)  larger-size group of municipalities. Indeed, the trajectory of smaller-size categories in the period $1830-1900$ (first plot from the left) is considerably shorter than the trajectory in the following periods $1900-1960$ (second plot) and $1960-2000$ (third plot), respectively, which coincides with the merging intensification over time, as presented in Figure \ref{Fig_Number_of_Municipalities_abo_est}.

The dominant increase in the population ratio of the largest municipalities in the period $1830-1900$ indicates an intensive migration of people from small(er) to large(r) municipalities, as presented in the lower left-hand side of Figure \ref{Fig_Population_Trajectories_1}. On the contrary, the largest municipalities significantly decreased in population size in the period $1960-2000$. The migration flow from large(r) to small(er) municipalities became dominant in this period. In addition, an intensive merging process took place, degrading small(er) size categories while further reinforcing the population increase of municipalities of large(r) sizes. 

Finally, trajectories in Figure \ref{Fig_Population_Trajectories_1} enclose an area. Such phenomena can be explained by the fact that the distribution of the logarithm of population per municipality follows Gaussian distribution consistently over time, as depicted in Figure \ref{Fig_Population_Distribution_1809_2019_f1}. In other words, the probability density function $f_{Z[k]}(z)$ defines a bell-shaped curve, being horizontally shifted over time. Therefore, a population-size category of municipalities firstly increases in the number of municipalities (and thus population ratio), peaks and starts decreasing, explaining the trajectories in Figure \ref{Fig_Population_Trajectories_1}.

\section{Model of the Dutch Municipality Network}\label{Sec_DMN_model}
This section proposes a model which captures the time dynamics of the DMN. The purpose of the model is not to explain the evolution of a single Dutch municipality over time but rather to describe the evolution of the population and area distribution per Dutch municipality. The DMN model consists of three sequential sub-models:
\begin{enumerate}
	\item Population increase model per municipality,
	\item Inter-municipal migration model, and
	\item Merging model,
\end{enumerate}

\subsection{Population increase  model}\label{Sec_DMN_Model_Step_1}
\begin{figure}[!h]
	\begin{center}
		\includegraphics[angle=0, scale=0.57]{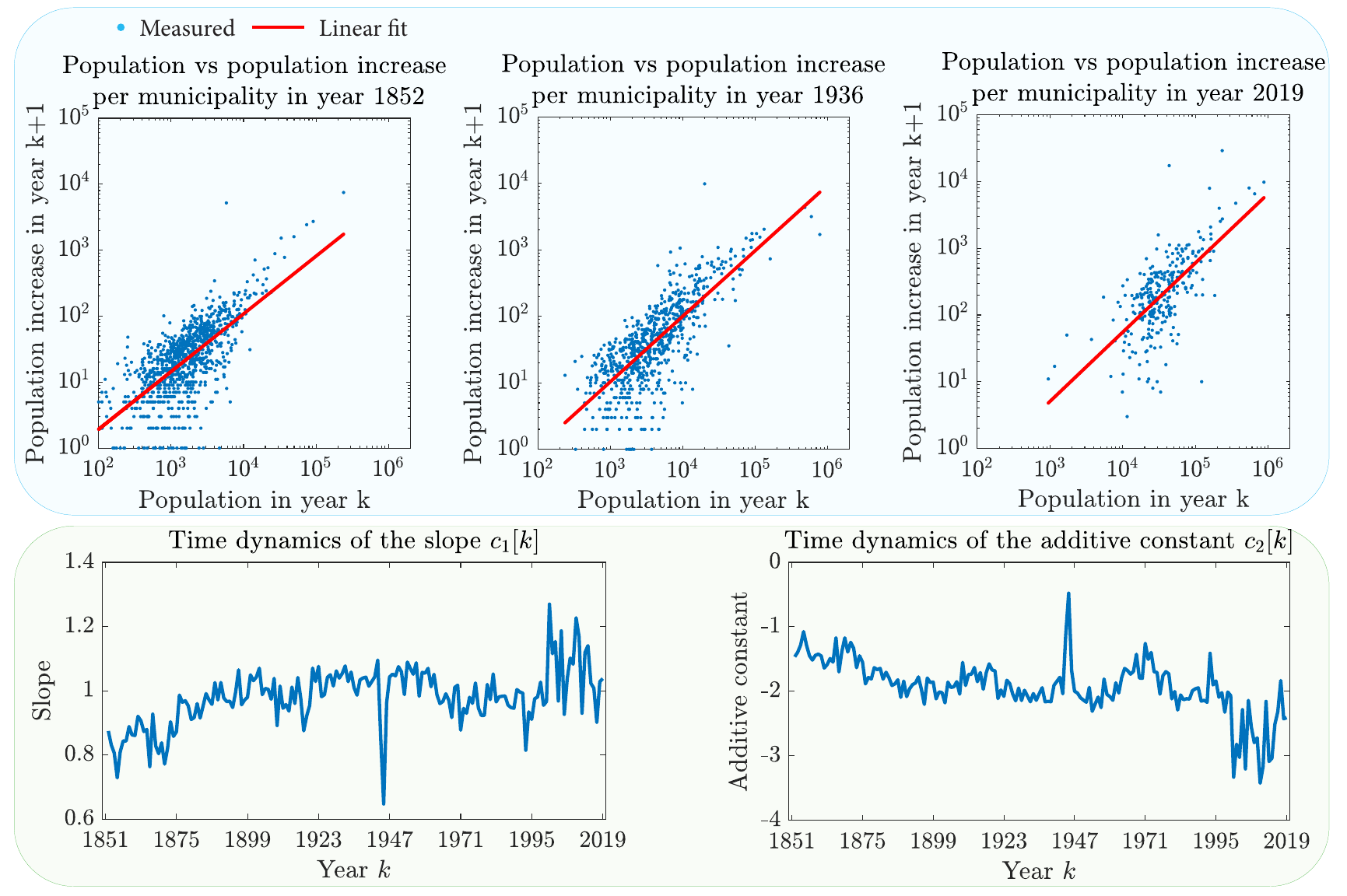}
		\caption{Population per municipality $p[k]$ in year $k$, versus population increase $(p[k+1]-p[k])$ in the following year $k+1$, on a double logarithmic scale for the years $1852$, $1936$ and $2019$, together with a fitted linear function (upper-part). Time dynamics of the slope $c_{1}[k]$ and the additive constant $c_{2}[k]$ of a fitted line (lower part).}
		\label{Fig_Topology_Log_Log_1829_vs_2019}
	\end{center}
\end{figure}

Available measurements of the population per municipality in the period $1830-2019$ reveal a consistent correlation pattern between the population $p_i[k]$ of municipality $i$ in year $k$ and its increase $p_i[k+1] - p_i[k]$, which a linear function on a double logarithmic scale can approximate. 
In the upper part of Figure \ref{Fig_Topology_Log_Log_1829_vs_2019}, we present the correlation\footnote{A minor percentage of municipalities with negative population increase is not presented in Figure \ref{Fig_Topology_Log_Log_1829_vs_2019}.} between values $p_i[k+1]-p_i[k]$ and $p_i[k]$, where $i\in \mathcal{N}[k]$, on a double logarithmic scale in years $1852$, $1936$ and $2019$, respectively. 
A positive correlation can be approximated as follows
\begin{equation}\label{Eq_Step_1_Correlation}
	E\left[\log \left(P[k+1] - P[k]\right)\right] = c_{1}[k]\cdot E\left[Z[k]\right] + c_{2}[k],
\end{equation}
where coefficients $c_{1}[k]$ and $c_{2}[k]$ represent the slope and additive constant of the linear fit in year $k$, as presented in the lower part of Figure \ref{Fig_Topology_Log_Log_1829_vs_2019}. While the slope $c_{1}[k]$ slightly oscillates around $1$ for a period of $130$ years, the additive constant $c_{2}[k]$ decreases over time, allowing us to introduce the following approximations
\begin{equation}\label{Eq_Step_1_Linear_Fit}
	\begin{split}
		c_{1}[k] & \approx 1 \\
		c_{2}[k] & \approx 9.27 - 5.8 \cdot 10^{-4}\cdot k.
	\end{split}
\end{equation}
By choosing $c_{1}[k] = 1$ and adopting the stronger assumption that the difference equation (\ref{Eq_Step_1_Correlation}) holds not only for the mean, but also for the random variables themselves, i.e. $\log \left(P[k+1] - P[k]\right)=Z[k]$, we deduce that $\text{Cov}\left[\log \left(P[k+1] - P[k]\right),Z[k]\right] = \text{Var}\left[Z[k]\right]$, meaning that the variability in the difference (an increase of population in a random municipality $P[k]$) equals that of $P[k]$. The governing equation for the population increase model per municipality is obtained by importing (\ref{Eq_Step_1_Linear_Fit}) into (\ref{Eq_Step_1_Correlation}):
\begin{equation}\label{Eq_Step_1_Governing_Law}
	E[P[k+1]] \approx  \left(1 + e^{c_{2}[k]}\right)\cdot E[P[k]].
\end{equation}
The slope $c_1[k]$ and the additive constant $c_2[k]$ values considerably oscillate in the last two decades of the researched time series. A reason that causes such behaviour is an intensified merging process (see lower left-hand part of Fig \ref{Fig_Number_of_Municipalities_abo_est}) that took place in the mentioned period, as a result of which certain municipalities (in most cases with a relatively small population) are being abolished and annexed by a neighbour municipality with a larger population. Consequently, the population increase of annexing municipalities in the following year spikes. Indeed, on a closer look at Figure \ref{Fig_Topology_Log_Log_1829_vs_2019}, the slope $c_1[k]$ (additive constant $c_2[k]$) exhibit only positive (negative) spikes during the last $20$ years of the researched period, respectively, and return to the previous state in the subsequent year.

Under the assumption that the number of municipalities remains unchanged within two consecutive years, $N[k+1]=N[k]$, the mean $z_{av}[k]$ evolves due to the proposed population increase model in (\ref{Eq_Step_1_Governing_Law}) as follows
\[
z_{av}[k+1] = \frac{1}{N[k]}\sum\limits_{i=1}^{N[k]}\left(\log \left(1 + e^{c_{2}[k]}\right) + z_{i}[k]\right) = \log \left(1 + e^{c_{2}[k]}\right) + z_{av}[k],
\]
because a multiplicative increase on a linear scale is equivalent to an additive increase on a logarithmic scale. On the contrary, the variance $\text{Var}(z[k])$ 
\[
\text{Var}(z[k+1]) = \frac{1}{N[k]}\cdot \sum\limits_{i=1}^{N[k]}\left(z_{av}[k+1] - \log \left(1 + e^{c_{2}[k]}\right) - z_i[k]\right)^{2} = \text{Var}(z[k]),
\]
is invariant to the population increase model in (\ref{Eq_Step_1_Governing_Law}). An argument behind neglecting the slope $c_1[k]$ deviations around value $1$ and adopting (\ref{Eq_Step_1_Linear_Fit}) is the idea of decoupling two population processes that occur on the DMN, namely the population increase and population migration. We argue that the variability in multiplicative population increases per municipality is a consequence of the people migrating between municipalities. Thus, in the following subsection, we introduce the migration model on a network that complements the population increase model (\ref{Eq_Step_1_Governing_Law}).

\subsection{Inter-municipal migration model}\label{Sec_DMN_Model_Step_2}
The population increase model in (\ref{Eq_Step_1_Governing_Law}) reveals a common trend in population increase per Dutch municipality.
Other simultaneous processes on the DMN are immigration/emigration and internal migration of people. In this subsection, we introduce a migration model of people on a geographical network and apply it to the Dutch Municipality Network. 

We propose a linear model that captures the migration of people across a geographical network of municipalities and complements the population increase model. The proposed migration model is a diffusion-like process founded on the assumption that there are two opposite migration flows taking place simultaneously on a network:
\begin{itemize}
	\item \textbf{Forward migration:} People moving from small(er) to large(r) municipalities in terms of population size, denoted as the forward migration flow with forward migration rate $\alpha$. This migration flow became dominant during the urbanisation period, from approximately $1850$ until $1960$ (see Figure \ref{Fig_Migration_Strenghts}).
	\item \textbf{Backward migration:} People moving from  large(r) to small(er) municipalities, denoted as the backward migration flow with backward migration rate $\delta$. This migration flow became dominant after $1960$.
\end{itemize}
We define the $N[k] \times N[k]$ migration matrix $M[k]$, with elements $m_{ij}[k]$
\begin{equation}\label{Migration_Matrix_Y_eq}
	m_{ij}[k] = a_{ij}[k] 1_{\left\{E[p_i[k]] < E[p_j[k]]\right\}},
\end{equation}
with the indicator function denoted as $1_{x}$, which is defined as $1$ if the statement $x$ is true, otherwise equals $0$. Relation (\ref{Migration_Matrix_Y_eq}) transforms the undirected DMN into a directed network, in which each link points to the {\em adjacent} municipality with a larger population, from where we conclude
\[
A[k] = M[k] + M^{T}[k].
\]
The $N[k]\times N[k]$ migration matrix $M[k]$ allows for introducing a model of people migrating on a municipality network
\begin{equation}\label{Eq_Migration_Process_Law}
	E[P[k+1]] = \left(I + \alpha\cdot M^{T}[k] + \delta\cdot M[k] - \alpha\cdot \text{diag}\left(M[k]\cdot u\right) - \delta\cdot \text{diag}\left(M^{T}[k]\cdot u\right) \right)\cdot E[P[k]],
\end{equation}
where the $N[k]\times N[k]$ matrix $I$ denotes the identity matrix. Each matrix term in (\ref{Eq_Migration_Process_Law}) allows for a physical interpretation. The $N\times N$ identity matrix $I$ indicates that the proposed migration model describes an additive process. The second term $\alpha\cdot M^{T}[k]$ calculates arrivals of people per municipality due to the forward migration flow (i.e. from smaller to larger adjacent municipality). The same migration flow, away from a smaller adjacent municipality, is accounted for in the matrix term $\alpha\cdot \text{diag}\left(M[k]^{T}\cdot u\right)$. The third matrix term $\delta\cdot M[k]$ computes the arrivals of people per municipality due to backward migration (i.e. from large(r) to small(er) adjacent municipality). As each migration flow has an origin and a destination municipality, the number of departures due to the backward migration is accounted for by $\delta\cdot \text{diag}\left(M[k]\cdot u\right)$. 
The sum of the opposite forward and backward migration flows provides the resulting migration flow from municipality $i$ to a larger adjacent municipality $j$ (i.e. $p_j[k]>p_i[k]$) $\alpha\cdot p_i[k] - \delta \cdot p_j[k]$. In the particular case when $\alpha = \delta$, the governing equation (\ref{Eq_Migration_Process_Law}) describes a diffusion process
\[
E\left[P[k+1]\right] = \left(I - \alpha\cdot Q[k]\right)\cdot E\left[P[k]\right],
\]
where the $N\times N$ Laplacian $Q = \text{diag}(d) - A$.
Properties of the proposed migration model in (\ref{Eq_Migration_Process_Law})
are provided in the Appendix \ref{App_Migration_Model}.

\subsection{Merging model}\label{Sec_Merging_Model}

The merging dynamics of the Dutch Municipality Network is a complex, government-controlled process that depends on numerous factors, such as economics, politics and social aspects, to name a few. Instead, we argue that all these aspects correlate with the population and area of municipalities. Thus, we model the Dutch municipal merging process by considering the population and area measurements per municipality and the network effect.

\begin{figure}[!h]
	\begin{center}
		\includegraphics[angle=0, scale=0.65]{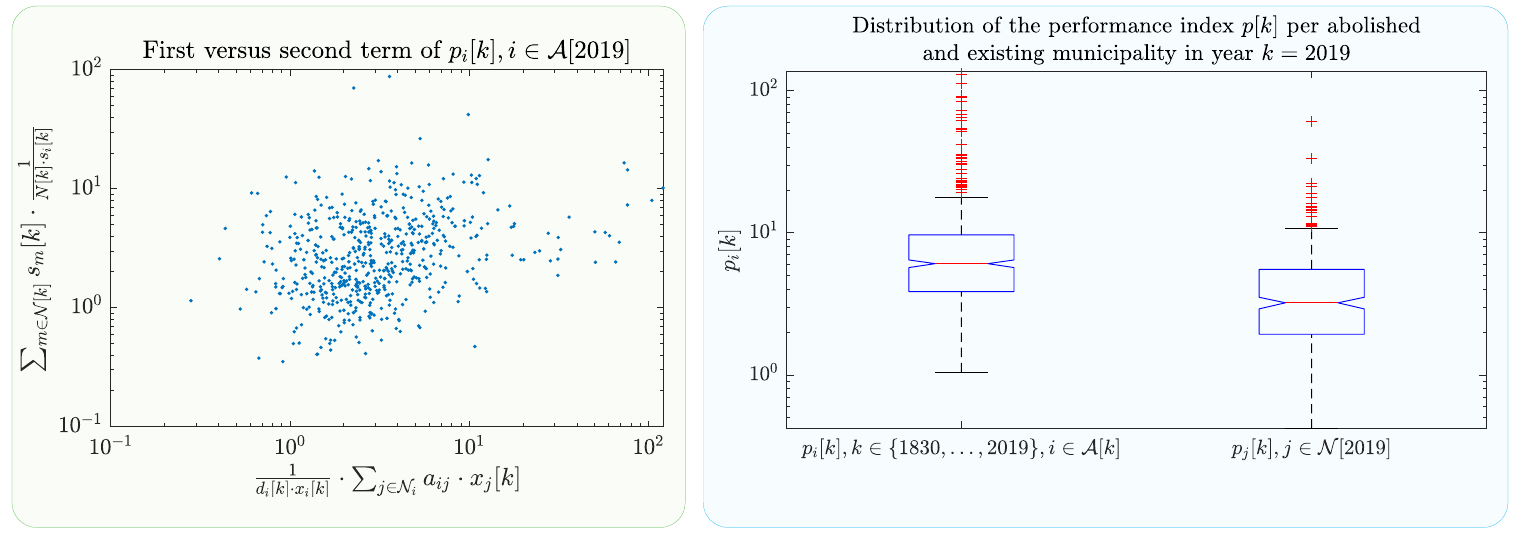}
		\caption{Correlation between the first and the second term of the Abolishment Likelihood index $x_i[k]$ per abolished municipality (i.e. $i\in \mathcal{A}[k]$) in the year of their abolishment (left-hand side figure). Distribution of the Abolishment Likelihood index $p[k]$ per abolished municipality in the year of their abolishment versus distribution of the Abolishment Likelihood index vector $x_i[k]$ per each municipality in year $k = 2019$ (right-hand side figure).}
		\label{Fig_Step_3_of_Model_f1}
	\end{center}
\end{figure}

We propose an Abolishment Likelihood index per municipality that estimates the set of abolished municipalities $\mathcal{A}[k]$ in year $k$. The Abolishment Likelihood index of municipality $i$ in year $k$, denoted as $x_i[k]$, is defined as
\begin{equation}\label{Eq_Abolishment_Index}
	x_i[k] = \frac{1}{d_{i}[k] \cdot p_i[k]} \sum\limits_{j\in \mathcal{N}_{i}[k]}a_{ij}[k]\cdot p_j[k] + \frac{1}{3} \cdot \frac{s_{av}[k]}{s_{i}[k]},
\end{equation}
with $\mathcal{N}_{i}[k]$ denoting the set of the node $i$ neighbours in year $k$ (i.e. $\mathcal{N}_{i}[k] = \{\, j \mid j \in \mathcal{N}[k],\,a_{ij}[k] = 1 \,\}$). The first term of the Abolishment Likelihood index $x_i[k]$ in (\ref{Eq_Abolishment_Index}) compares the population $p_i[k]$ of municipality $i$ with the mean population of its direct neighbours $\frac{1}{d_{i}[k] \cdot p_i[k]}\cdot \sum_{j\in \mathcal{N}_{i}[k]}a_{ij}[k]\cdot p_j[k]$, while the second term in (\ref{Eq_Abolishment_Index})  compares the area $s_{i}[k]$ of municipality $i$ with the mean area $s_{av}[k]$ per Dutch municipality in year $k$. The set of abolished municipalities in year $k$, denoted by $\mathcal{A}[k]$, is determined as $N_{a}[k] = |\mathcal{A}[k]|$ municipalities with highest ranking in the $N[k]\times 1$ Abolishment Likelihood index  vector $x[k]$.

On the left-hand side of Figure \ref{Fig_Step_3_of_Model_f1}, we correlate terms of the Abolishment Likelihood index $x_i[k]$ in (\ref{Eq_Abolishment_Index}) per each abolished municipality $i$, in the year of its abolishment. The absence of correlation confirms the validity of our choice to consider both population and area size per municipality. In addition, the two box-whisker plots (abolished versus existing municipalities) on the right-hand side of Figure \ref{Fig_Step_3_of_Model_f1} are clearly shifted with respect to each other, confirming that the Abolishment Likelihood index indeed captures the abolishment likelihood.

\subsection{Model validation}\label{Sec_Model_Validation}

\begin{figure}[!h]
	\begin{center}
		\includegraphics[angle=0, scale=0.74]{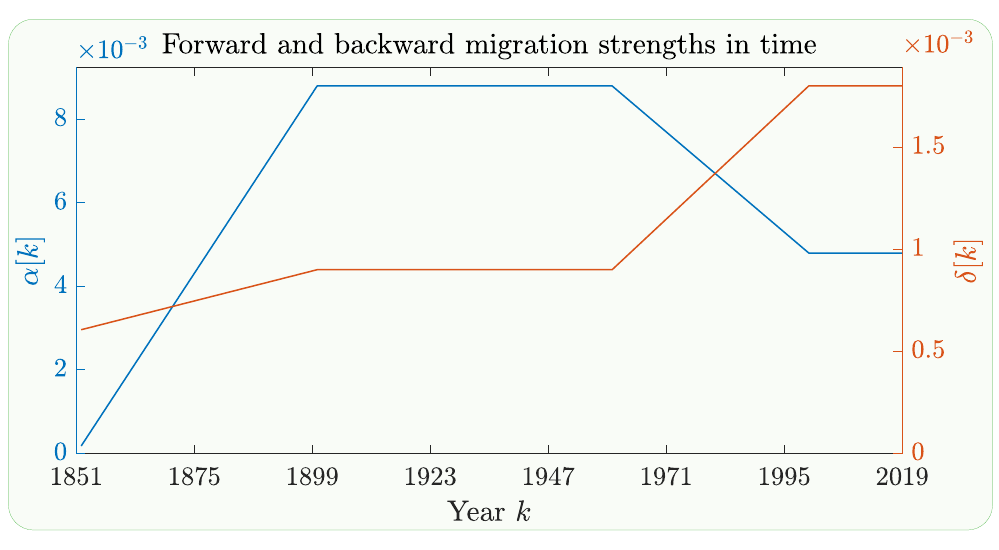}
		\caption{Forward migration rate $\alpha[k]$ (blue colour) and the backward migration rate $\delta[k]$ (red colour) over time $k$.}
		\label{Fig_Migration_Strenghts}
	\end{center}
\end{figure}

By combining the assumption from (\ref{Eq_Step_1_Linear_Fit}), the proposed migration model (\ref{Eq_Migration_Process_Law}) and the merging model (\ref{Eq_Abolishment_Index}), we obtain a complete model for the time dynamics of the Dutch Municipality Network. 
The model is initialized by the $\left(N[k]\times N[k]\right)$ adjacency matrix $A[k]$ of the DMN, the $\left(N[k]\times 1\right)$ population vector $p[k]$ and the $\left(N[k]\times 1\right)$ area vector $s[k]$ from year $k = 1851$. Starting from $k = 1852$, the input to the model is the total population of The Netherlands $T[k]$ and the number of abolished municipalities $N_{a}[k]$ in each year $k$ in the period ($1851-2019$). The DMN model is iteratively applied for each year $k$ in the following order
\begin{itemize}
	\item Based on the assumption in (\ref{Eq_Step_1_Linear_Fit}), update the population vector as $p[k+1] = \frac{T[k+1]}{T[k]}\cdot p[k]$.
	\item Apply the migration model, defined in (\ref{Eq_Migration_Process_Law}).
	\item Compute the Likelihood Abolishment index $x[k]$ per municipality, as in (\ref{Eq_Abolishment_Index}). Determine $N_{a}[k]$ municipalities with the highest ranking in the sorted index vector $x[k]$. Assign the population and area of each abolished municipality to an adjacent municipality closest in the ranking in the sorted vector $x[k]$. 
	\item Update the $N[k+1]\times N[k+1]$ adjacency matrix $A[k+1]$ of the DMN, the $N[k+1]\times 1$ population vector $p[k+1]$ and the $N[k+1]\times 1$ area vector $s[k+1]$.
\end{itemize}
For the migration model in (\ref{Eq_Migration_Process_Law}), the used values of the forward migration rate $\alpha[k]$ and the backward migration $\delta[k]$ rate per year $k$ are provided in Figure \ref{Fig_Migration_Strenghts}. The migration rates are determined heuristically, motivated by observations in Section \ref{Sec_trajecotires}.
In the following subsection, we analyse the DMN model prediction accuracy.

\subsection{Prediction Accuracy of the DMN Model}\label{Sec_DMN_Model_Prediction}
\begin{figure}[!h]
	\begin{center}
		\includegraphics[angle=0, scale=0.8]{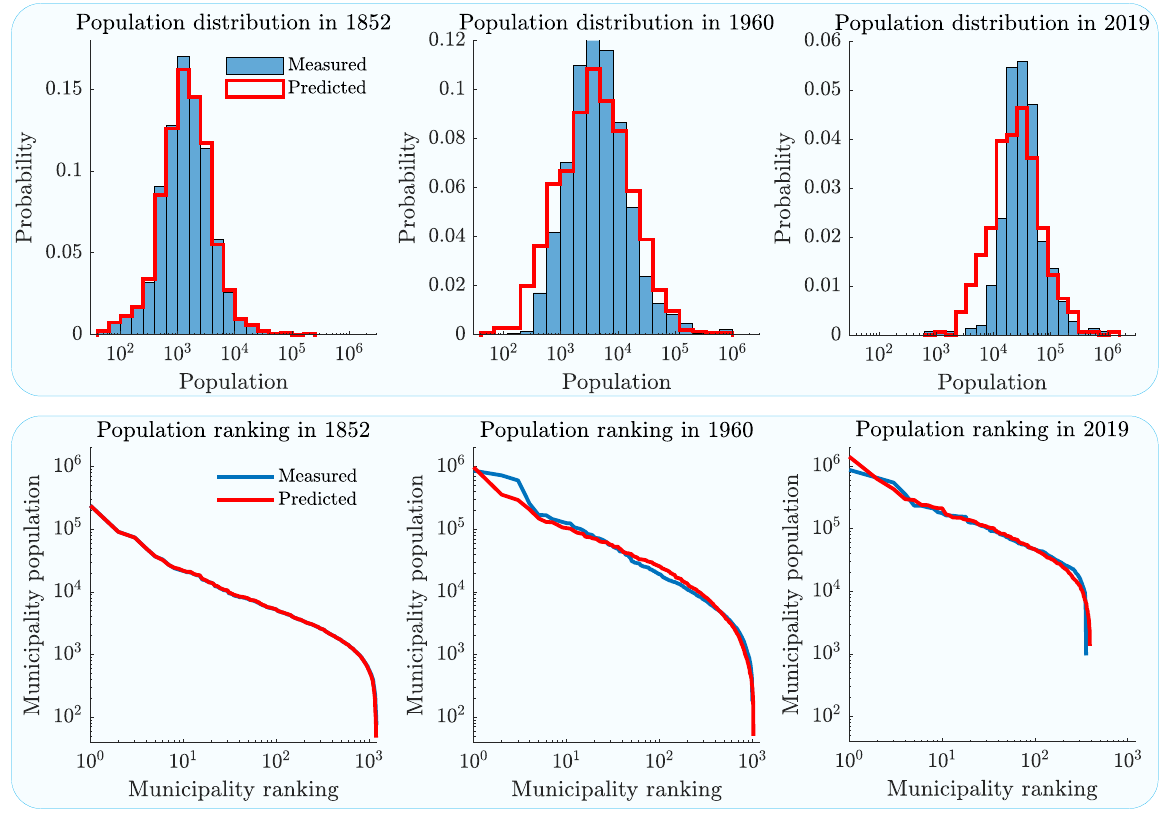}
		\caption{Measured versus predicted population distribution per municipality in years $1852$, $1960$ and $2019$ (upper part). Measured versus predicted population rank-size distribution per municipality in years $1852$, $1960$ and $2019$ (lower part).}
		\label{Fig_Model_Output_Population}
	\end{center}
\end{figure}

The measured population distribution per municipality is compared with the predicted population distribution by the DMN model. The measured and predicted population distributions are compared for the years $1852$, $1960$ and $2019$ in the upper part of Figure \ref{Fig_Model_Output_Population}. Further, the lower part of the Figure provides the rank-size distribution of both the measured and predicted population vector. 

The distribution of the predicted population vector per Dutch municipality closely follows the distribution of the measured population vector over time during the entire research period. With the proposed decoupling of the population dynamics into an equal increase per municipality in (\ref{Eq_Step_1_Linear_Fit}) and the migration process in (\ref{Eq_Migration_Process_Law}), we can explain how the Dutch population distribution evolved in the last 170 years.

\begin{figure}[!h]
	\begin{center}
		\includegraphics[angle=0, scale=0.6]{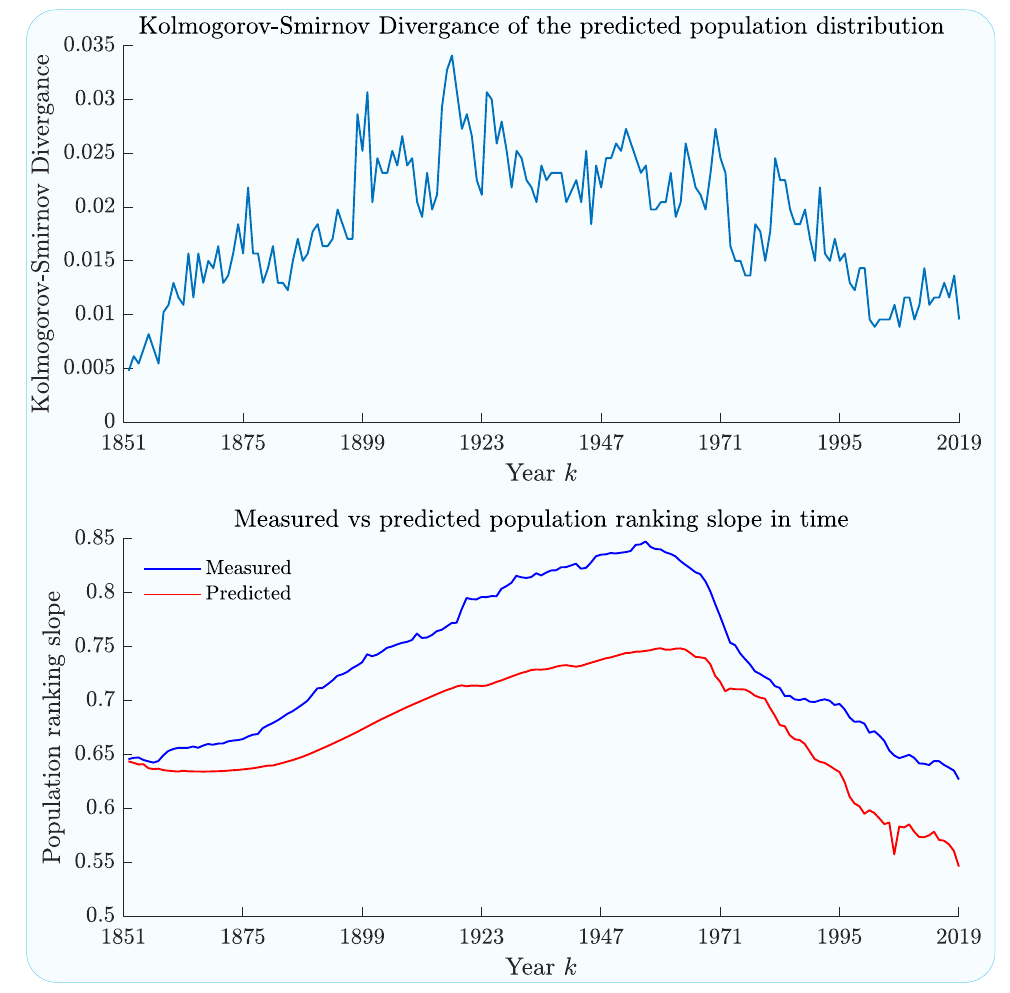}
		\caption{Kolmogorov-Smirnov divergence between the measured and predicted population distribution in the period $(1852 - 2019)$ (upper figure). Predicted versus measured population rank-size slope in the period $(1852 - 2019)$ (lower figure).}
		\label{Fig_Model_Output_TVD}
	\end{center}
\end{figure}

To quantify the precision of the predicted population distribution over time, we compute the Kolmogorov-Smirnov divergence between the predicted and measured population distribution and provide the results in the upper part of Figure \ref{Fig_Model_Output_TVD}. The divergence value remains relatively low during the entire researched period, indicating that the adopted values for the forward migration rate $\alpha[k]$ and the backward migration rate $\delta[k]$ indeed reveal the migration flows in The Netherlands.

The predicted versus the measured population rank-size distribution slope $\beta[k]$ is provided in the lower part of Figure \ref{Fig_Model_Output_TVD}. The rank-size slope of the predicted population vector depends solely on the migration process in (\ref{Eq_Migration_Process_Law}) and, thus, on the forward $\alpha$ and the backward $\delta$ migration rate, provided in Figure \ref{Fig_Migration_Strenghts}. Opposite trends in migration rates $\alpha[k]$ and $\delta[k]$ until and after $1960$ marked a dynamic transition in the rank-size slope $\beta[k]$ of the population vector. 

\begin{figure}[!h]
	\begin{center}
		\includegraphics[angle=0, scale=0.9]{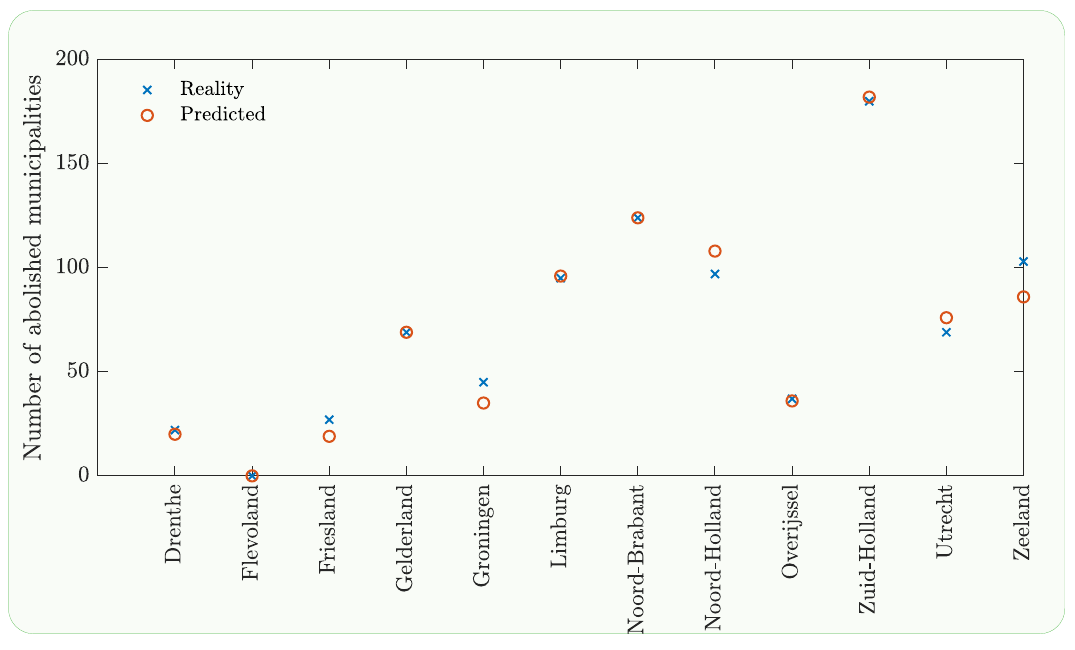}
		\caption{Predicted versus the measured number of abolished municipalities per Dutch province in the period $1851-2019$.}
		\label{Fig_Model_Output_Mergers}
	\end{center}
\end{figure}

In Figure \ref{Fig_Model_Output_Mergers}, we compare the distribution of all abolished municipalities in the period ($1851-2019$) per Dutch province with predicted mergers by the DMN model. The proposed DMN model achieves a fantastic precision of $91,7 \%$. The DMN topology in $1851$ initialises the DMN model. However, the road bridges and dikes built after $1851$ connected many isolated components of the Dutch Municipality Network to the mainland, as discussed in Appendix \ref{App_Geographical_Network_Merger}. Consequently, the number of isolated components in the DMN monotonically decreased over time, which is not taken into account in the DMN model. For example, the entire province of Zeeland remains disconnected from the mainland in the DMN model, which is not the case in reality. We argue that the model precision could be even higher if the topology changes of the DMN after $1851$ were taken into account in the proposed DMN model.

\section{Conclusion}\label{Sec_Conclusions}

Linking the data sets collected by Statistics Netherlands and the International Institute of Social History enabled us to investigate the Dutch municipality merging process and the survivability of municipalities over the period 1830-2019. In a geographic sense,  a municipality can vary from a
densely populated urbanized (city) area to a set of sparsely populated localities\footnote{In 2019, 2168 population localities were grouped into 355 Dutch domestic municipalities.} in a rural area. 
In a governmental sense, a municipality is a highly autonomous administrative unit serving its population at the local government level and interacting with the overarching levels of the provincial and national government.
All 1467 Dutch domestic municipalities that have ever existed between 1830 and 2019 are captured in a research construct referred to as the Dutch Municipality Network (DMN).

Compared to the first 160 years of the researched time series, the last three decades reveal considerable fluctuations, such as in the observed values of the slope of the population increase per municipality (see section \ref{Sec_DMN_Model_Step_1}). Can these recent fluctuations be solely attributed to the accelerated decrease in the number of Dutch municipalities due to the intensified merging process? What do these fluctuations mean for a country as a whole?

Over the entire research period, we have observed the amazing property that {\em the logarithm} of population size and municipality area features an almost linear difference equation over the years. 
This underlying "log-linearity" in the evolution process resulted in a predictive accuracy of 91.7\% at province level (see section \ref{Sec_DMN_model}) in spite of all municipality mergers and dynamic population fluctuations that took place in reality.
The remarkable "log-linearity" in population size and municipality area asks for a scientific explanation.
The collected data researched here are macroscopic statistical observations derived from many individual movements. Just as for interacting particle systems in physics, we think that the macroscopic observations can be explained if the rules or laws on the microscopic level, i.e. on individual human level, are known. 
Unfortunately, human behavior is far more complicated than the already exceedingly complex interacting physical systems in nature, because the latter obey physical laws, while the governing laws -- expressed in differential equations to allow computations -- of human behavior are yet unknown. Many complex networks (flock of birds, synchronization of fire-flies or heart muscles, interacting particle systems at atomic or molecule level, epidemics, etc.) possess reasonably simple local rules at the nodal or individual level, but the multitude of the interacting local rules gives rise to a surprisingly complex emerging behavior, often characterized by phase transitions. Ab-initio calculation of a possible phase transition for the ‘packing’ of humans is therefore out of reach.

Nevertheless, we hypothesize on the observations of "log-linearity".
The scale free power law behaviour, i.e. the linear relationship of population and area of municipalities as a function of rank-order in double logarithmic plots (figure \ref{Fig_Population_Rank_Size_1}), could be a manifestation that the system of human habitation is in a self-organised critical state, typically associated with phase transitions. The population in the Netherlands over the past 200 years has remained at or near a certain phase transition. The distribution function of the population over municipalities has similar long power-law tails (figure \ref{Fig_Population_Power_Law}), although, of course,
there is a cut-off at population sizes (which are too small to justify the existence of a municipality). One might speculate that a ‘fully solid’ phase for human habitation occurs when the entire population of the Netherlands lives in households occupying a minimum acceptable amount of space. Trying to squeeze more humans will cause repulsive forces.
On the other hand, a ‘fluid/gas phase’ would be a fully dispersed population, in which inhabitants have a comfortable individual living space. However, the benefits of being close to other people for social interaction as well as the mutual exchange of goods or services constitutes an attraction for humans, that pulls them towards the ‘fully solid’ phase. 
Therefore, the sketched hypothetical human packing process shares some qualitative properties with known dynamics in complex networks in which phase transitions occur, making it worthwhile to explore the possibility of phase transitions in the habitation of people.

\section*{Acknowledgements}
This research is part of NExTWORKx, a collaboration between the Delft University of Technology and KPN Royal on future communication networks.
We cordially thank all experts of Statistics Netherlands and the Delft University of Technology for their contributions and for supporting the team of co-authors. PVM acknowledges the European
Research Council (ERC) under the European Union’s Horizon
2020 research and innovation programme (grant agreement No 101019718).

\section*{Author contributions}
IJ performed data analysis and mathematical modelling and was involved in realising the research construct. 
EvB came up with the idea to study municipalities as a network and was involved in research design, connecting output and supervision.
IM performed data exploration and found a method to connect the selected datasets. 
TV provided expertise on the confrontation of the city-related research with municipality-related research. 
GB explained long-term socio-economic dynamics from a historical perspective.
HvH explained the municipality merging process and types of mergers, and performed qualitative merger analysis.
FP was involved in mathematical modelling, explaining and connecting research results.
PVM was involved in stochastic analysis and concluding, performed supervision, end-responsible for the research.

\bibliography{references_DMN}{}

\begin{thebibliography}{10}

\bibitem{Marchetti1994AnthropologicalBehavior}
C.~Marchetti, ``{Anthropological invariants in travel behavior},'' {\em
  Technological Forecasting and Social Change}, vol.~47, pp.~75--88, 9 1994.

\bibitem{Barabasi2005HumanMobilityPatterns}
M.~C. Gonz{\'{a}}lez, C.~A. Hidalgo, and A.-L. Barab{\'{a}}si, ``{Understanding
  individual human mobility patterns},'' {\em Nature 2008 453:7196}, vol.~453,
  pp.~779--782, 6 2008.

\bibitem{Mantegna1994StochasticFlight}
R.~N. Mantegna and H.~E. Stanley, ``{Stochastic Process with Ultraslow
  Convergence to a Gaussian: The Truncated L{\'{e}}vy Flight},'' {\em Physical
  Review Letters}, vol.~73, pp.~2946--2949, 11 1994.

\bibitem{Brockmann2006TheTravel}
D.~Brockmann, L.~Hufnagel, and T.~Geisel, ``{The scaling laws of human
  travel},'' {\em Nature 2006 439:7075}, vol.~439, pp.~462--465, 1 2006.

\bibitem{Rayer2001Geographic19801995}
S.~Rayer and D.~L. Brown, ``{Geographic diversity of inter-county migration in
  the united states, 1980–1995},'' {\em Population Res. Policy Rev.},
  vol.~20, no.~3, pp.~229--252, 2001.

\bibitem{Johnson2000Continuity19501995}
K.~M. Johnson and G.~V. Fuguitt, ``{Continuity and change in rural migration
  patterns, 1950–1995},'' {\em Rural Sociol.}, vol.~65, no.~1, pp.~27--49,
  2000.

\bibitem{Stanley1995UrbanGrowthPatterns}
H.~A. Makse, S.~Havlin, and H.~E. Stanley, ``{Modelling urban growth
  patterns},'' {\em Nature 1995 377:6550}, vol.~377, pp.~608--612, 10 1995.

\bibitem{Schlapfer2014TheSize}
M.~Schl{\"{a}}pfer, L.~M. Bettencourt, S.~Grauwin, M.~Raschke, R.~Claxton,
  Z.~Smoreda, G.~B. West, and C.~Ratti, ``{The scaling of human interactions
  with city size},'' {\em Journal of The Royal Society Interface}, vol.~11, 9
  2014.

\bibitem{Verbavatz2020TheCities}
V.~Verbavatz and M.~Barthelemy, ``{The growth equation of cities},'' {\em
  Nature}, vol.~587, pp.~397--401, 11 2020.

\bibitem{Bettencourt2007GrowthCities}
L.~M. Bettencourt, J.~Lobo, D.~Helbing, C.~K{\"{u}}hnert, and G.~B. West,
  ``{Growth, innovation, scaling, and the pace of life in cities},'' {\em Proc.
  Natl Acad. Sci.}, vol.~104, pp.~7301--7306, 4 2007.

\bibitem{Bettencourt2013TheCities}
L.~M. Bettencourt, ``{The origins of scaling in cities},'' {\em Science},
  vol.~340, no.~6139, pp.~1438--1441, 2013.

\bibitem{Schneider2014Expansion19782010}
A.~Schneider and C.~M. Mertes, ``{Expansion and growth in Chinese cities,
  1978–2010},'' {\em Environ. Res. Lett.s}, vol.~9, no.~2, p.~024008, 2014.

\bibitem{Bergs2021SpatialNegligible}
R.~Bergs, ``{Spatial dependence in the rank-size distribution of cities –
  weak but not negligible},'' {\em PLOS ONE}, vol.~16, p.~e0246796, 2 2021.

\bibitem{Newman2003TheNetworks}
M.~E.~J. Newman, ``{The Structure and Function of Complex Networks},'' {\em
  SIAM Review}, vol.~45, pp.~167--256, 1 2003.

\bibitem{Barabasi2016NetworkIntroduction}
A.-L. Barab{\'{a}}si, ``{Network science introduction},'' {\em Network
  science}, pp.~1--27, 2016.

\bibitem{Newman2018Networks}
M.~Newman, ``{Networks},'' vol.~1, 10 2018.

\bibitem{Molnar-intVeld2019DePersonenautopark}
H.~Moln{\'{a}}r-in~‘t Veld, ``{De groei van het Nederlandse
  personenautopark},'' 2019.

\bibitem{Hoekveld2014UrbanSouthern-Limburg}
J.~J. Hoekveld, {\em {Urban decline within the region: Understanding the
  intra-regional differentiation in urban population development in the
  declining regions Saarland and Southern-Limburg}}.
\newblock PhD thesis, University of Amsterdam, 2014.

\bibitem{BoonstraHistoricalDruid}
O.~Boonstra and R.~Mourits, ``{Historical Database of Dutch Municipalities
  (Historische Database Nederlandse Gemeenten (HDNG)) - dataLegend - Druid}.''

\bibitem{Tadikamalla1982SystemsVariables}
P.~R. Tadikamalla and N.~L. Johnson, ``{Systems of Frequency Curves Generated
  by Transformations of Logistic Variables},'' {\em Biometrika}, vol.~69,
  p.~461, 8 1982.

\bibitem{Newman2000}
A.~Clauset, C.~R. Shalizi, and M.~E.~J. Newman, ``{Power-Law Distributions in
  Empirical Data},'' {\em SIAM Review}, vol.~51, pp.~661--703, 11 2009.

\bibitem{Doerr2013LognormalSpread}
C.~Doerr, N.~Blenn, and P.~Van~Mieghem, ``{Lognormal Infection Times of Online
  Information Spread},'' {\em PLoS ONE}, vol.~8, 5 2013.

\bibitem{VanMieghem2011LognormalNetwork}
P.~Van~Mieghem, N.~Blenn, and C.~Doerr, ``{Lognormal distribution in the digg
  online social network},'' {\em The European Physical Journal B}, vol.~83,
  pp.~251--261, 9 2011.

\bibitem{VanMieghem2006DataNetworking}
P.~Van~Mieghem, ``{Data communications networking},'' p.~414, 2010.

\bibitem{Cristelli2012ThereZipf}
M.~Cristelli, M.~Batty, and L.~Pietronero, ``{There is More than a Power Law in
  Zipf},'' {\em Scientific Reports 2012 2:1}, vol.~2, pp.~1--7, 11 2012.

\bibitem{Abramowitz1968HandbookFunctions}
M.~Abramowitz and I.~A. Stegun, {\em {Handbook of Mathematical Functions}}.
\newblock Dover Publications, revised 9~ed., 6 1968.

\bibitem{PVM2014PerformanceAnalysis}
P.~Van~Mieghem, {\em {Performance Analysis of Complex Networks and Systems}}.
\newblock Cambridge University Press, 4 2014.

\bibitem{2005TestingHypotheses}
{\em {Testing Statistical Hypotheses}}.
\newblock Springer Texts in Statistics, New York, NY: Springer New York, 2005.

\end{thebibliography}
\bibliographystyle{ieeetrans}

\appendix

\section{Datasets Description}\label{App_Datasets_Overview}

Since 1809, the nature and evolution of the Dutch society and economy were recorded in national statistics such as municipality-related population and surface measurements. From this data we selected three datasets which together describe the national year-on-year dynamics at municipality level:
\begin{itemize}
	\item [1] Population  measurements of each municipality,
	\item [2] Digital geometries representing the area of each municipality,
	\item [3] Municipality merging.
\end{itemize}
Population measurements have been collected from two different sources. In period $(1809 - 1960)$ the population data set is obtained from the Historical Database of Dutch Municipalities\footnote{Historische Database Nederlandse Gemeenten} (HDNG), collected by the International Institute of Social History, which is part of the Royal Netherlands Academy of Arts and Sciences. Further, the number of inhabitants per Dutch municipality in period $(1960 - 2019)$ is obtained from Statistics Netherlands\footnote{Centraal Bureau voor de Statistiek} (CBS).

\begin{figure}[!h]
	\begin{center}
		\includegraphics[angle=0, scale=0.83]{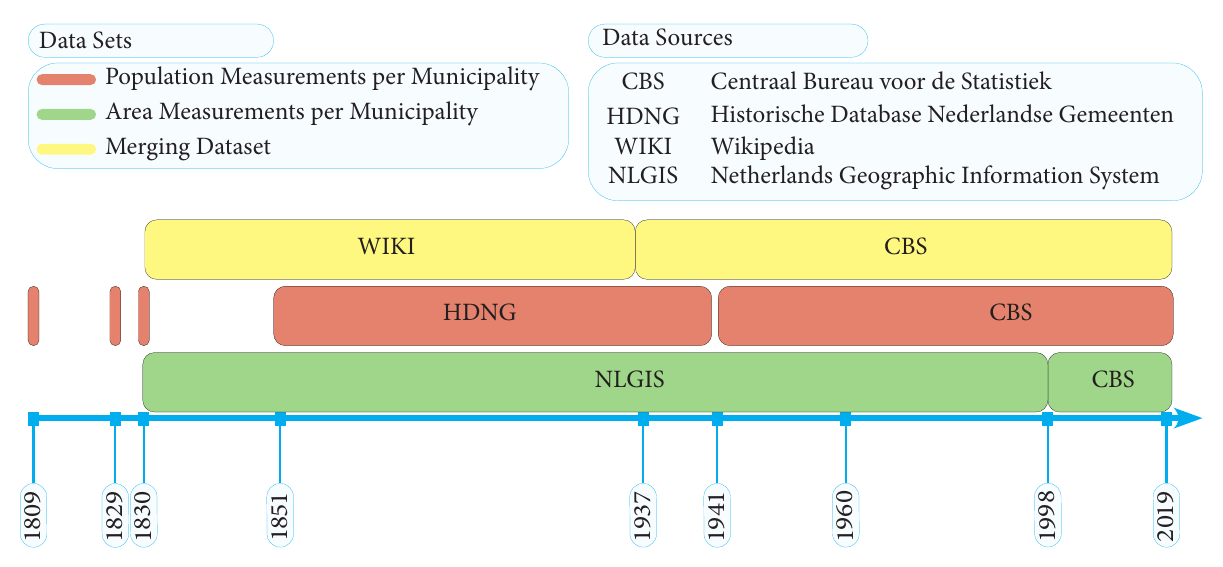}
		\caption{Datasets Overview.}
		\label{Fig_Datasets}
	\end{center}
\end{figure}

The other two datasets are collected from the online repositories of the CBS website. While digital geometries and the municipality merging datasets exist for each year in the period $(1830 - 2019)$ consistent over time, population data sets cover in total more than two centuries, but with varying time resolution. A detailed overview of the availability of data sets over time is provided in Figure \ref{Fig_Datasets}.

\section{Coding schemes identifying municipalities}\label{App_A}
In this research, two complementary coding schemes are used to identify municipalities and their geographic area, namely the four digit Central Bureau of Statistics code (CBS code) and the five digit Amsterdam code (AMS code). The CBS code identifies municipalities that existed since $1830$, while the Amsterdam code can be traced back to Dutch municipalities that existed since $1812$. The CBS code identifies specific administrative entities (municipality names), while the Amsterdam code identifies specific geographical areas on which municipalities are/were located.
\begin{figure}[!h]
	\begin{center}
		\includegraphics[angle=0, scale=0.48]{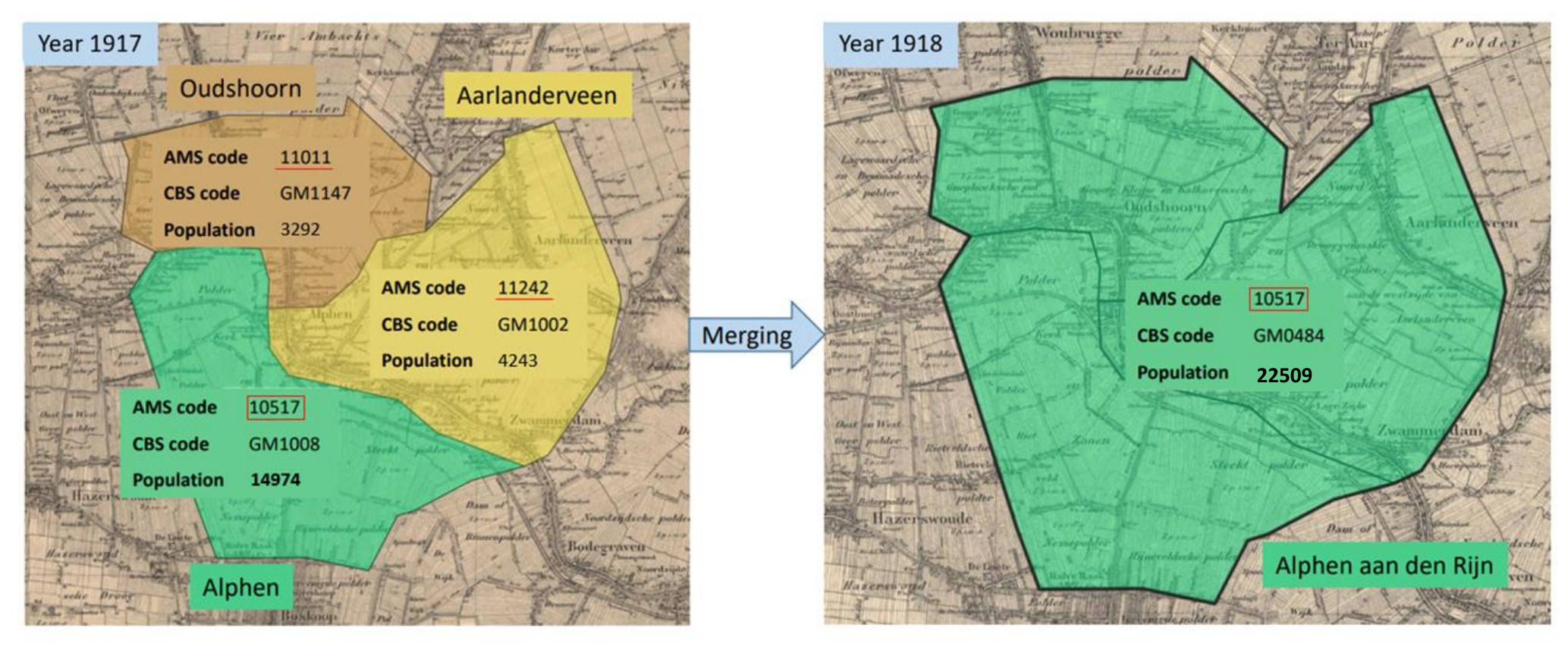}
		\caption{Municipality merging example of Alphen aan den Rijn in $1918$.}
		\label{Fig_Merging_Example}
	\end{center}
\end{figure}

Whenever municipal restructuring leads to a municipality merger or a name change, a new CBS code is generated and assigned to the new municipality. However, the new municipality is assigned an existing Amsterdam code that belonged to one of the municipalities that were involved in the merging process, in order to ensure the  historical continuity of the geographical area. As exemplified in Figure \ref{Fig_Merging_Example}, the municipality with the largest population passes on its Amsterdam code to the newly formed municipality. For example, when a municipality annexes an adjacent municipality, the Amsterdam code of the annexing municipality is preserved and the Amsterdam code of the annexed municipality is abolished, while all their CBS codes are abolished.
The CBS code can be considered a unique identifier for a municipality, because it uniquely specifies a municipal entity that exists (or has existed) for a certain defined time period. The Amsterdam code, however, is not a unique municipality identifier; it has been designed in such a way that all Dutch municipalities possess an Amsterdam code that can be traced back to an Amsterdam code of a Dutch municipality which existed in 1812.

\section{Municipality merging process}\label{App_Merger_Types}

During the researched period $(1830-2019)$ we analysed the process of municipality merging and municipality name changing. We distinguish five event types, of which each involves  discontinuation of the CBS code of the municipalities involved. In case of a merging/renaming event, the CBS code is abolished (becomes inactive) at the end of year $k$, and the administrative change takes place at the beginning of the following year $k+1$. The five event types (A-E) are explained below:
\begin{itemize}
	\item Type A (Annexation): the abolished municipality is annexed by an existing (usually
	adjacent) municipality at the end of year k. This process is officially called ‘light merger’
	(in Dutch: lichte samenvoeging) and the CBS code of the abolished municipality
	becomes inactive at the end of year $k$. This reclassification type has occurred 542 times in total
	during the studied time period (1830-2019). \\
	\item Type B (Border split): the area of the abolished municipality is split among an existing municipality and a newly formed municipality. This reclassification type is a combination of Type A and Type C, as both processes occur at the same time within the former municipality’s boundaries. This reclassification type has occurred 10 times during the
	studied time period (1830-2019). \\
	\item Type C (Coalition): the abolished municipality, along with one or more neighboring municipalities which are abolished at the end of the same year k, form a coalition by merging into one new municipality. The new municipality is assigned a new CBS code at the beginning of year k+1, and the CBS codes of the merger participants become inactive at the end of year k. This process is officially called ‘regular merger’ (in Dutch: Reguliere samenvoeging). This reclassification type has occurred 502 times during the studied time period (1830-2019). \\
	\item Type D (Dutch and/or Frisian Name-change): only the official name of a municipality is changed in Dutch or Frisian language, while its borders remain unchanged. The municipality is assigned a new CBS code at the beginning of year k+1, and the old CBS
	code of the municipality becomes inactive at the end of year k. A main difference
	between the Amsterdam and CBS coding schemes is that the municipality retains its Amsterdam code when undergoing a name-change. A municipality name-change has occurred 56 times during the studied time period (1830-2019). \\
	\item Type E (Exchanged internationally): the area of a municipality is exchanged between a neighboring country and The Netherlands. In case a municipality is allocated to a neighboring country, it is recorded in statistics to be no longer part of The Netherlands in year k+1. This reclassification type has occurred 2 times during the studied time period
	(1830-2019). The German municipalities Tudderen (Drostambt) and Elten were temporarily annexed by The Netherlands after the Second World War and were re-annexed by Germany in 1963.
\end{itemize}

\section{Municipality merging process demonstrated on a planar graph}\label{App_Geographical_Network_Merger}
\begin{figure}[!h]
	\begin{center}
		\includegraphics[ angle =0, scale= 0.4]{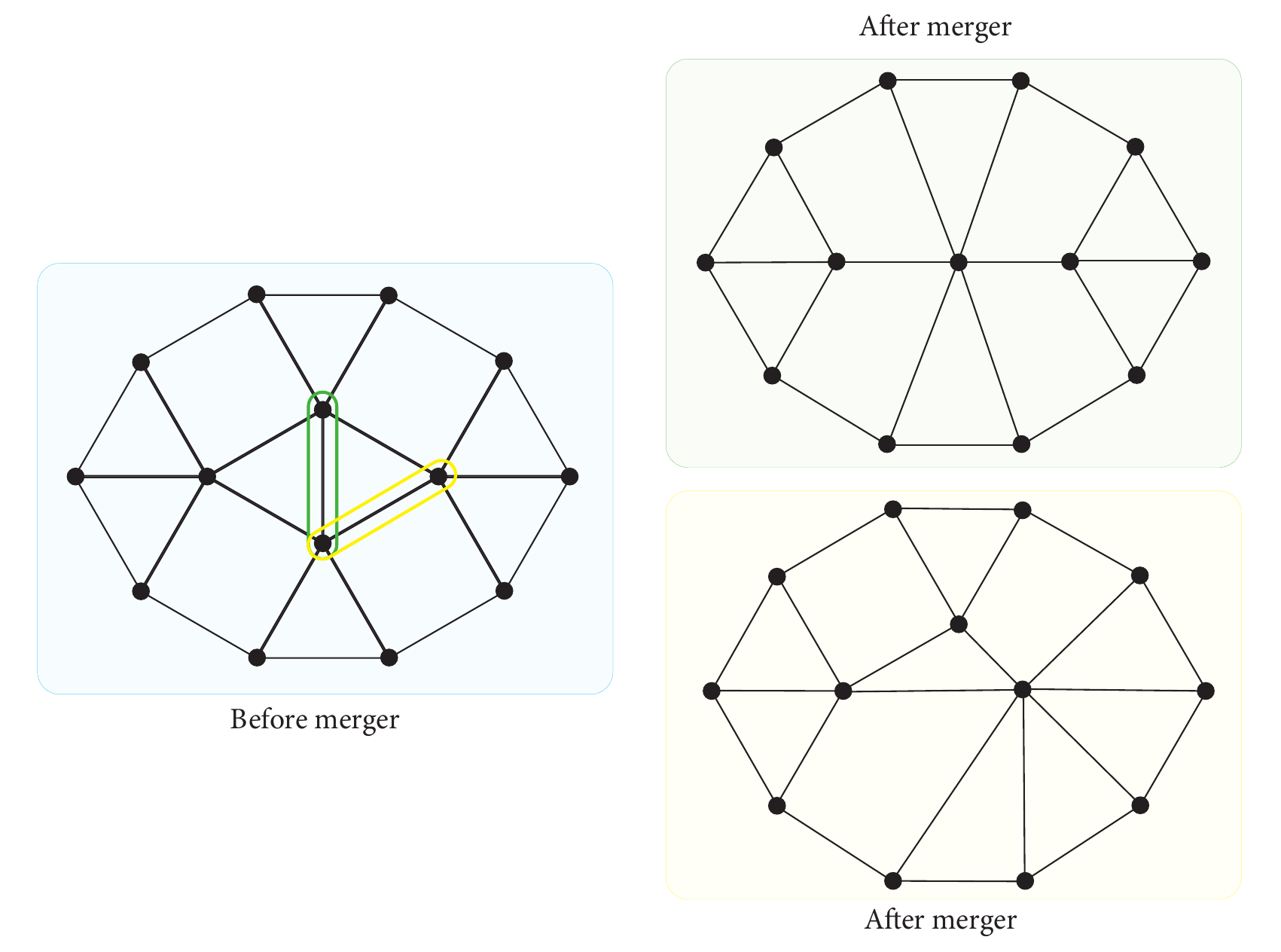}
		\caption{Two merging process examples demonstrated on a planar graph.}
		\label{Fig_Merging_Process_Geographical_Network}
	\end{center}
\end{figure}

Although the municipality merging process introduced in Section \ref{Sec_Merging_Process} changed the topology significantly, the average degree of the DMN remained almost unchanged, $d_{av}[k] \approx 5$.  Figure \ref{Fig_Merging_Process_Geographical_Network} shows a planar graph with two examples of centrally positioned merging municipality nodes, each having a degree 5. 
On the right-hand side of the Figure \ref{Fig_Merging_Process_Geographical_Network} the different scenarios of the two merger examples are given.

Two adjacent nodes $i$ and $j$ can have either $|\mathcal{N}_{i}[k]\cap \mathcal{N}_{j}[k]| = 2$ or $|\mathcal{N}_{i}[k]\cap \mathcal{N}_{j}[k]|=3$ common neighbours. When nodes $i$ and $j$ are merged into one node, the number of nodes $N[k+1]$ and number of links $L[k+1]$ in the next year $k+1$ change as follows
\[
\begin{split}
	L[k+1] =& L[k] - |\mathcal{N}_{i}[k]\cap \mathcal{N}_{j}[k]| - 1 \\
	N[k+1] =& N [k] - 1.
\end{split}
\]
We perform the merging of two adjacent municipalities and provide the updated topology on the right-hand side of Figure \ref{Fig_Merging_Process_Geographical_Network}. The conservation law on the average degree $d_{av}[k+1]$, when nodes $i$ and $j$ are merged at the end of the year $k$, can be obtained by importing $d_{av}[k]=2\frac{L[k]}{N[k]}$ into the above equation
\begin{equation}\label{Eq_App_d_av}
	d_{av}[k+1] = 	\left(1+\frac{1}{N[k]-1}\right)\cdot d_{av}[k] - 2\frac{|\mathcal{N}_{i}[k]\cap \mathcal{N}_{j}[k]|-1}{N[k]-1}.
\end{equation}
We further consider a case when three municipalities $i$, $j$ and $m$ are merged into one, at the end of year $k$. The newly formed municipality in the following year $k+1$ is connected to another municipality if either municipality $i$, $j$ or $m$ was connected to that municipality in year $k$. Therefore, the degree of the newly formed municipality equals $|\mathcal{N}_i \cup \mathcal{N}_j \cup \mathcal{N}_m|$. To determine the number of removed links $L[k+1]-L[k]$ in the DMN due to the merger, we apply the inclusion-exclusion formula (see for example \cite[p.10]{PVM2014PerformanceAnalysis})  and obtain
\[
|\mathcal{N}_i \cup \mathcal{N}_j \cup \mathcal{N}_m| = |\mathcal{N}_i| + |\mathcal{N}_j| + |\mathcal{N}_m| - |\mathcal{N}_i \cap \mathcal{N}_j| - |\mathcal{N}_i \cap \mathcal{N}_m| - |\mathcal{N}_j \cap \mathcal{N}_m| + 
|\mathcal{N}_i \cap \mathcal{N}_j \cap \mathcal{N}_m|,
\]
from where we derive the number of nodes $N[k+1]$ and the number of links $L[k+1]$ in the year $k+1$
\[
\begin{split}
	L[k+1] =& L[k] - |\mathcal{N}_i \cap \mathcal{N}_j| - |\mathcal{N}_i \cap \mathcal{N}_m| - |\mathcal{N}_j \cap \mathcal{N}_m| - a_{ij}[k] - a_{im}[k] - a_{jm}[k] + 
    |\mathcal{N}_i \cap \mathcal{N}_j \cap \mathcal{N}_m| \\
	N[k+1] =& N [k] - 2,
\end{split}
\]
leading to the following conservation law of the average degree $d_{av}[k]$
\begin{equation}\label{Eq_App_d_av_3_merger}
\begin{split}
    d_{av}[k+1] =& 	\left(1+\frac{1}{N[k]-2}\right)\cdot d_{av}[k] -\\
    &2\frac{|\mathcal{N}_i \cap \mathcal{N}_j| + |\mathcal{N}_i \cap \mathcal{N}_m| + |\mathcal{N}_j \cap \mathcal{N}_m| + a_{ij}[k] + a_{im}[k] + a_{jm}[k] - 
    |\mathcal{N}_i \cap \mathcal{N}_j \cap \mathcal{N}_m|}{N[k]-2}.
\end{split}
\end{equation}
In case three municipalities merge into one municipality, the average degree $d_{av}[k]$ slightly decreases over time, which is the opposite effect of when two municipalities merge. During the period $1960-2000$, mergers involving more than two municipalities were common, causing a decreasing trend in the average degree $d_{av}[k]$, as visible in the lower part of Figure \ref{Fig_Nodes_number_f1}.

\section{Different Distributions}\label{App_Distribution_Models}

\subsection{Normal Distribution}\label{App_Normal_Distribution}

A Gaussian random variable $X=N\left(\mu,\sigma^2\right)$ is a continuous random variable with an extent over the entire real axis and is defined \cite{PVM2014PerformanceAnalysis} by the distribution function $F_{X}(x)=\Pr[X \leq x]$ as
\begin{equation}\label{Eq_Normal_Distribution_F}
	F_{X}(x) = \frac{1}{\sigma\cdot \sqrt{2\pi}}\int\limits_{-\infty}^{x}\exp\left(-\frac{\left(t-\mu\right)^{2}}{2\sigma^{2}}\right)dt,
\end{equation}
with the mean $E[X]=\mu$ and with the variance Var$[X]=\sigma^{2}$.
The corresponding probability density function $f_X(x)= \frac{d}{dx}\Pr[X\leq x]$ is
\begin{equation}\label{Eq_Normal_Distribution_f}
	f_{X}(x) = \frac{e^{-\frac{(x-\mu)^{2}}{2\sigma^{2}}}}{\sigma\cdot \sqrt{2\pi}}.
\end{equation}

\subsection{Lognormal Distribution}\label{App_Lognormal_Distribution}

A lognormal random variable is defined as $Y = e^{X}$, where $X=N\left(\mu,\sigma^2\right)$
is a Gaussian or normal random variable \cite{PVM2014PerformanceAnalysis}. The distribution function $F_{Y}(y)=\Pr \left[Y\le y\right] = \Pr\left[X\le\log y\right]$ follows from (\ref{Eq_Normal_Distribution_F}), for non-negative real values of $y$, as
\begin{equation}\label{eq_lognormal_df}
	F_{Y}(y)=\frac{1}{\sigma \sqrt{2\pi}}\int\limits_{-\infty}^{\log y}\exp \left[-\frac{\left(t-\mu\right)^{2}}{2\sigma^{2}}\right]dt,
\end{equation}
The corresponding probability density function $f_{Y}(y)=\frac{dF_{Y}(y)}{dy}$ of a lognormal random variable $Y$ follows by differentiation of (\ref{eq_lognormal_df}) as
\begin{equation}\label{eq_lognormal_pdf}
	f_{Y}(y)=\frac{\exp \left[-\frac{\left(\log y - \mu\right)^{2}}{2\sigma^{2}}\right]}{\sigma \cdot \sqrt{2\pi} \cdot y},
\end{equation}
The mean $E[Y]$ and the variance $\text{Var}(Y)$ can be computed \cite[Sec. 3.5.5]{PVM2014PerformanceAnalysis} as
\begin{equation}\label{lognormal_mean_variance}
	\begin{split}
		E[Y] =& e^{\left(\mu+\frac{\sigma^{2}}{2}\right)} \\ \text{Var}[Y] =&  \left(e^{\sigma^{2}}-1\right)\cdot e^{\left(2 \mu+\sigma^{2}\right)}.
	\end{split}
\end{equation}

\subsection{Logistic distribution}\label{App_Logistic_Distribution}
A logistic random variable $X$, also known as a Fermi-Dirac random variable \cite[Sec. 19.6.2]{PVM2014PerformanceAnalysis}, has the distribution function
\begin{equation}\label{Eq_Logistic_Distribution_F}
	F_{X}(x) = \frac{1}{1 + e^{-\frac{x-\mu}{s}}} = \frac{1}{2} + \frac{1}{2}\tanh\left(\frac{x-\mu}{2s}\right),
\end{equation}
The probability density function (pdf) of a logistic random variable $X$ again follows by differentiation of (\ref{Eq_Logistic_Distribution_F}) as
\begin{equation}\label{Eq_Logistic_Distribution_f}
	f_{X}(x) = \frac{1}{4s}\text{sech}^2\left(\frac{x-\mu}{2s}\right).
\end{equation}
It is more convenient to consider the normalized Fermi-Dirac random variable $Z=\frac{X-\mu}{s}$ that obeys\footnote{Indeed, after letting $z=-\frac{x-\mu}{s}$ in (\ref{Eq_Logistic_Distribution_F}), we have $\frac{1}{1+e^{-z}}=\Pr[X \leq \mu + s z]=\Pr[\frac{X-\mu}{s} \leq  z]$.} $F_Z(z)=\Pr[Z\leq z]=\frac{1}{1+e^{-z}} = 1-\frac{1}{1+e^{z}}$. The probability generating function (pgf) $\phi(w) = E\left[e^{-w Z}\right] = \int_{-\infty}^{\infty} e^{-w t} f_Z(t) dt$ is the double-sided Laplace transform \cite[p. 20]{PVM2014PerformanceAnalysis} of $f_Z(t)$ and equals
\[
\varphi\left(  w\right)  =\int_{-\infty}^{\infty}e^{-wt}\frac{d}{dt}\frac
{1}{1+e^{-t}}dt
\]
Partial integration leads to%
\[
\varphi\left(  w\right)  =\left.  \frac{e^{-wt}}{1+e^{-t}}\right\vert
_{-\infty}^{\infty}+w\int_{-\infty}^{\infty}\frac{e^{-wt}}{1+e^{-t}}dt
\]
and the first term vanishes, provided that $0<\operatorname{Re}(w)<1$ holds, with $\operatorname{Re}(w)$ denoting the real part of $w$. Let $u=e^{-t}$ and
$t=-\log u$, then%
\[
\frac{\varphi\left(  w\right)  }{w}=\int_{0}^{\infty}\frac{u^{w-1}}{1+u}du
\]
One of the Beta function integrals, $B\left(  x,y\right)  =\int_{0}^{\infty
}\frac{t^{x-1}}{\left(  1+t\right)  ^{x+y}}dt=\frac{\Gamma\left(  x\right)
\Gamma\left(  y\right)  }{\Gamma\left(  x+y\right)  }$, valid for
$\operatorname{Re}\left(  x\right)  >0$ and $\operatorname{Re}\left(
y\right)  >0$, shows that, for $0<\operatorname{Re}(w)<1$,%
\[
\varphi\left(  w\right)  =\Gamma\left(  w\right)  \Gamma\left(  1-w\right)
=\frac{\pi w}{\sin\pi w}%
\]
where the last equality is the reflection formula of the Gamma function $\Gamma(w)$, valid for all complex numbers $w$.
The pgf $\varphi\left(  w\right)  =E\left[  e^{-wZ}\right] =\sum_{n=0}^{\infty}\frac{\left(  -1\right)  ^{n}}{n!}E\left[  Z^{n}\right]
w^{n} $ contains
all moments, whereas the Taylor series of $\frac{\pi w}{\sin\pi w}$ around $w=0$
equals%
\[
\frac{\pi w}{\sin\pi w}=1+\sum_{n=1}^{\infty}\left(  2^{1-2n}-1\right)
\left(  2\pi\right)  ^{2n}B_{2n}\,\frac{(-1)^{n}w^{2n}}{(2n)!}%
\]
where $B_{n}$ is the $n$-th Bernoulli number. By equating the
corresponding powers of $w$ in $\frac{\pi w}{\sin\pi w}=E\left[  e^{-wZ}\right]$, we find all even moments, for $n>0$%
\[
E\left[  Z^{2n}\right]  =\left(  2^{1-2n}-1\right)  \left(  2\pi\right)
^{2n}(-1)^{n}B_{2n}\,
\]
and while all odd $E\left[  Z^{2n+1}\right]  =0$ for $n\geq0$. Since $Z=\frac{X-\mu}{s}$ is
a normalized random variable, the mean $E\left[  Z\right]  =0$ and thus the mean $E[X]= \mu$. The
variance Var$\left[  Z\right]  =E\left[  \left(  Z-E\left[  Z\right]  \right)
^{2}\right]  =E\left[  Z^{2}\right]  =\frac{\pi^{2}}{3}$, because $B_{2}%
=\frac{1}{6}$. Hence,%
\[
E\left[  \left(  \frac{X-\mu}{s}\right)  ^{2}\right]  =\frac{1}{s^{2}%
}\text{Var}\left[  X\right]  =\frac{\pi^{2}}{3}%
\]
resulting in Var$\left[  X\right]  =\sigma^{2}=\frac{\pi^{2}s^{2}}{3}$.

\subsection{Log-logistic distribution}\label{App_Log_Logistic_Distribution}
A log-logistic random variable, defined by $Y = e^{X}$ where $X$ is a logistic random variable with mean $\mu$ and variance $\sigma$, has the probability function
\begin{equation}\label{eq_logistic_Y_F}
	F_{Y}(y) = \frac{1}{1 + e^{-\frac{\log(y)-\mu}{s}}} = \frac{1}{2} + \frac{1}{2}\tanh\left(\frac{\log(y)-\mu}{2s}\right),
\end{equation}
with the corresponding probability density function
\begin{equation}\label{eq_loglogistic_pdf}
	f_{Y}(y) = \frac{1}{4s\cdot y}\text{sech}^2\left(\frac{\log y-\mu}{2s}\right).
\end{equation}

We concentrate first on the normalized log-logistic random variable.
The probability distribution of a log-logistic random variable $Y=e^{X}$,
which is always positive, is%
\[
\Pr\left[  Y\leq y\right]  =\Pr\left[  e^{X}\leq y\right]  =\Pr\left[
X\leq\log y\right]  =\Pr\left[  \frac{X-\mu}{s}\leq\frac{\log y-\mu}%
{s}\right]
\]
which, in terms of the normalized logistic random variable $Z$ with $\log
z=\frac{\log y-\mu}{s}$ (and thus $z=\left(  ye^{-\mu}\right)  ^{\frac{1}{s}}$
and $z>0$) is%
\[
\Pr\left[  Z\leq\log z\right]  =\frac{1}{1+e^{-\log z}}=\frac{z}{z+1}%
=1-\frac{1}{z+1}%
\]
and the pdf is%
\[
f_{Z}\left(  z\right)  =\frac{1}{\left(  z+1\right)  ^{2}}%
\]
The moments%
\[
E\left[  Z^{k}\right]  =\int_{0}^{\infty}\frac{t^{k}}{\left(  t+1\right)
^{2}}dt=\Gamma\left(  k+1\right)  \Gamma\left(  1-k\right)  =\frac{\pi k}{\sin\pi k}%
\]
do not exist for integers $k\geq1$. 

With a little more effort, we compute the moments directly for a log-logistic random variable $Y$,%
\[
E\left[  Y^{k}\right]  =\frac{1}{4s}\int_{0}^{\infty}\frac{t^{k-1}}{\cosh
^{2}\left(  \frac{\log t-\mu}{s}\right)  }dt
\]
We modify this integral by a series of substitutions. First, let $u=\log t$,
then%
\[
E\left[  Y^{k}\right]  =\frac{1}{4s}\int_{-\infty}^{\infty}\frac{e^{ku}}%
{\cosh^{2}\left(  \frac{u-\mu}{s}\right)  }du
\]
followed by the substitution $w=\frac{u-\mu}{s}$ yields%
\[
E\left[  Y^{k}\right]  =\frac{e^{k\mu}}{4}\int_{-\infty}^{\infty}\frac
{e^{ksw}}{\cosh^{2}\left(  w\right)  }dw
\]
illustrating that the integral exists provided $-1<\frac{ks}{2}<1$, thus, the
integer $k<\frac{2}{s}$. Next, let $p=e^{w}$, then%
\[
E\left[  Y^{k}\right]  =e^{k\mu}\int_{0}^{\infty}\frac{p^{ks+1}}{\left(
p^{2}+1\right)  ^{2}}dp
\]
A last substitution $t=p^{2}$ reveals again the above Beta function integral%
\begin{align*}
E\left[  Y^{k}\right]    & =\frac{1}{2}e^{k\mu}\int_{0}^{\infty}\frac
{t^{\frac{ks}{2}}}{\left(  t+1\right)  ^{2}}dp\\
& =\frac{1}{2}e^{k\mu}\Gamma\left(  \frac{ks}{2}+1\right)  \Gamma\left(
1-\frac{ks}{2}\right)  =\frac{1}{2}e^{k\mu}\frac{\pi\frac{ks}{2}}{\sin\pi
\frac{ks}{2}}%
\end{align*}
Hence, provided that $-\frac{2}{s}<\operatorname{Re}\alpha<\frac{2}{s}$ where
$\alpha$ is now a complex number, the $\alpha-$moments of the log-logistic random variable exists%
\[
E\left[  Y^{\alpha}\right]  =\frac{1}{2}e^{\alpha\mu}\frac{\frac{\pi\alpha
s}{2}}{\sin\frac{\pi\alpha s}{2}}.%
\]
The mean $\text{E}[Y]$ and the variance $\text{E}[Y^{2}]-(\text{E}[Y])^2$ are 
\begin{equation}\label{Eq_logistic_Y_moments}
	\begin{cases}
		E[Y] = \frac{1}{4}\cdot e^{\mu}\cdot \frac{\pi s}{\sin(\frac{\pi s}{2})} & \text{If} \, s < 2 \\ 
		\text{Var}[Y] =  \frac{1}{4}\cdot e^{2 \mu}\cdot \left(\frac{2\pi s}{\sin(\pi s)} - \frac{1}{4}\cdot \frac{\pi^{2}s^{2}}{\sin^{2}(\frac{\pi s}{2})}\right)		 & \text{If} \, s < 1. 
	\end{cases}
\end{equation}

\subsection{Tail distributions}\label{Sec_Tail_distributions}
The probability density function $f_{Z[k]}(z)$ of the logistic distribution in (\ref{Eq_Logistic_Distribution_f}) can be transformed as follows
\begin{equation}\label{Eq_Tail_Logistic_Dist}
f_{Z[k]}(z) = \frac{1}{4s}\cdot \frac{e^{\frac{z-\mu}{s}}}{\left(e^{\frac{z-\mu}{s}} + 1\right)^{2}}.
\end{equation}
We introduce the logarithm of the probability density function as $L_{l}(z) = \log(f_{Z[k]}(z))$ and obtain from (\ref{Eq_Tail_Logistic_Dist})
\[
L_{l}(z) = -\log(4s) + \frac{z-\mu}{s} - 2\log\left(e^{\frac{z-\mu}{s}} + 1\right).
\]
Since we are interested in tail distribution, it holds $e^{\frac{z-\mu}{s}} >> 1$, allowing us to introduce the approximation $e^{\frac{z-\mu}{s}} + 1 \approx e^{\frac{z-\mu}{s}}$, simplifying the above equation as follows
\[
L_{l}(z) = -\log(4s) - \frac{z-\mu}{s}.
\]
Finally, by importing $z = \log p$, we obtain
\begin{equation}\label{Eq_Logistic_Tail}
    \log(f_{Z[k]}(z)) = \frac{\mu}{s} -\log(4s) - \frac{1}{s}\log(p),
\end{equation}
informing us that the probability density function $f_{Z[k]}(z)$ of the logistic distribution in (\ref{Eq_Logistic_Distribution_f}), decays linearly on a double logarithmic scale, for $z >> \mu$. Therefore, the subset of the largest municipalities in population follows a power-law distribution, as discussed in Section \ref{Sec_Power_Law}.

Further, we define the logarithm of the probability density function $L_n(z) = \log(f_{Z[k]}(z))$ of a  normal distribution in (\ref{Eq_Normal_Distribution_f}) and obtain
\[
L_n(z) = -\log\left(\sigma \sqrt{2\pi}\right) - \frac{\left(z-\mu\right)^2}{2\pi^2}.
\]
The above equation further transforms after importing $z=\log p$
\begin{equation}\label{Eq_Normal_Tail}
    \log(f_{Z[k]}(z)) =  -\log\left(\sigma \sqrt{2\pi}\right) - \frac{\left(\log p-\mu\right)^2}{2\pi^2},
\end{equation}
teaching us that the probability density function $f_{Z[k]}(z)$ of the normal distribution in (\ref{Eq_Normal_Distribution_f}) decreases on a double logarithmic scale as a square function of the population $p$. 
Tail distribution in (\ref{Eq_Normal_Tail}) better fits the area per Dutch municipality, given the constraint that the sum of area per municipality $\sum_{i=1}^{N[k]} s_i[k]$ remains relatively constant over time.

\section{Goodness-of-fit tests}\label{App_Goodness_of_Fit_Tests}
\begin{figure}[!h]
	\begin{center}
		\includegraphics[angle=0, scale=0.795]{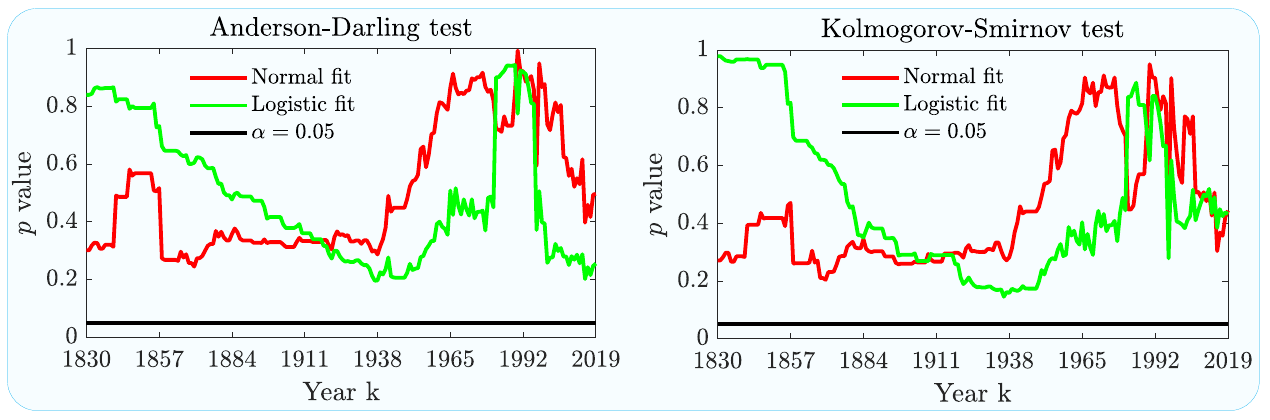}
		\caption{Anderson-Darling test (left-hand side) and Kolmogorov-Smirnov test (right-hand side) results of the normal distribution fit (red colour) and a logistic distribution fit (green colour) of the logarithm of the area $Y$ distribution in the period $1830 - 2019$.}
		\label{Fig_Surface_Hypothesis_Test}
	\end{center}
\end{figure}

The goodness-of-fit tests Anderson-Darling (AD) and Kolmogorov-Smirnov (KS) \cite[Ch. 14]{2005TestingHypotheses} are used, as shown in Figure \ref{Fig_Surface_Hypothesis_Test}, to determine the plausibility of the hypothesis that the logarithm of the area per Dutch municipality follows either a normal or a logistic distribution. Both tests provide a $p$ value that answers how likely the hypothesis holds. These tests are based on measuring the "distance" between the hypothesized distribution model and the measured distribution. Next, artificial datasets are created from the same distribution model, and the corresponding distances are computed. Finally, the $p$ value represents the ratio of the artificial distances larger than the measured distance from the empirical data. In the case computed $p$ value is close to $0$, the measured data does not agree with the hypothesized distribution, whereas $p$ close to 1 confirms the hypothesis.

\subsection{Area distribution}\label{App_Area_Goodness_of_Fit}
Both the AD and KS tests indicate that, from 1830 until $1918$, the logarithm of the area of a typical Dutch municipality follows a logistic distribution rather than a normal distribution, while from $1918$ until $1990$, the opposite holds. The AD and the KS test do not favour any distribution consistently during the last three decades.

\subsection{Population distribution}\label{App_Population_Goodness_of_Fit}

\begin{figure}[!h]
	\begin{center}
		\includegraphics[angle=0, scale=0.77]{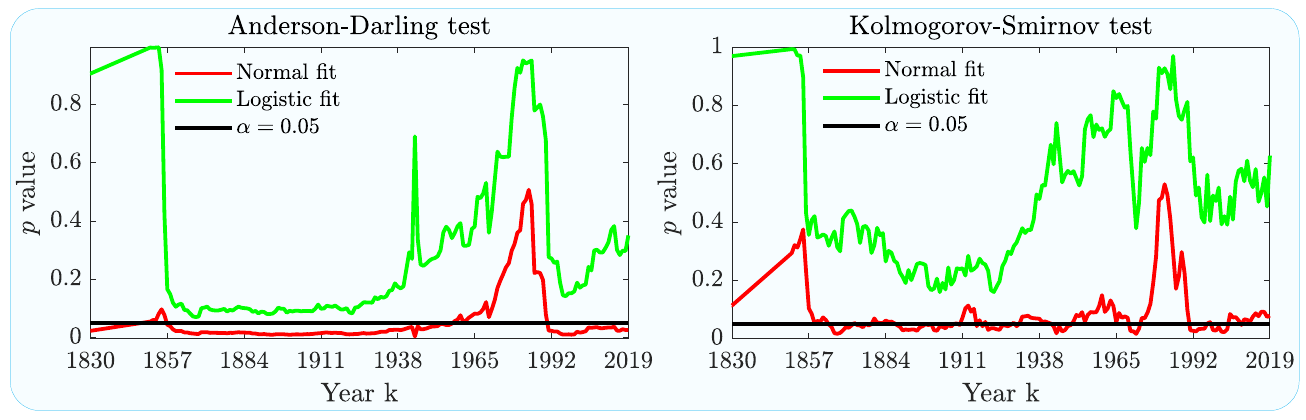}
		\caption{The $p$ value of the Anderson-Darling test (left-hand side) and Kolmogorov-Smirnov test (right-hand side) for the hypothesis that the logarithm of population vector $Z[k]$ follows the normal (red colour) and logistic (green colour) distribution in the period $1809 - 2019$.}
		\label{Fig_Population_LogL_vs_LogN_1}
	\end{center}
\end{figure}

The $p$ value of the AD and KS goodness-of-fit tests is shown for both the normal and logistic distribution of the logarithmic of the population in Figure \ref{Fig_Population_LogL_vs_LogN_1}. 
Over the entire period $(1809-2019)$, the $p$ value of the AD and the KS goodness-of-fit tests indicates that the logarithm of population $Z[k]$ per municipality follows the logistic distribution more closely than the normal distribution.

\section{Conservation laws of population and area}\label{App_Conservation_Laws}

\subsection{Area evolution}\label{App_Area_Evolution}
In this section, we analyse a merger case when $N_{a}[k] = |\mathcal{A}[k]|$ municipalities are abolished at the end of year $k$ and annexed by an existing municipality $\eta \in \mathcal{N}[k]$. The mean $y_{av}[k]$ of the $N[k]\times 1$ logarithm of area vector $y[k]$ evolves after the merger as follows
\begin{equation}\label{Eq_Area_Mean_Log_Year_k}
y_{av}[k] =  \frac{1}{N[k]}\sum\limits_{i\in \mathcal{N}[k]} y_{i}[k] = \frac{1}{N[k]}\log \left(\prod\limits_{i\in \mathcal{N}[k]}s_{i}[k]\right),
\end{equation}
while in the next year $k+1$, after abolishing $N_{a}[k]$ municipalities into a single new municipality $\eta$ with area $s_{\eta}[k+1] = s_{\eta}[k]+\sum_{j\in \mathcal{A}[k]}s_{j}[k]$ in year $k+1$, we obtain
\[
y_{av}[k+1] =  \frac{1}{N[k]-N_{a}[k]}\log  \left(\prod\limits_{j\in \mathcal{N}[k]\setminus \left(\eta \cup \mathcal{A}[k]\right)}s_{j}[k]\right) +  \frac{1}{N[k]-N_{a}[k]}\log \left(\sum\limits_{i\in \eta \cup \mathcal{A}[k]} s_{i}[k]\right).
\]
By adding and subtracting the term $\frac{1}{N[k]-N_{a}[k]}\sum_{j\in \eta \cup \mathcal{A}[k]} \log (s_{j}[k])$ from the above relation, we obtain
\[
y_{av}[k+1] = \frac{N[k]}{N[k]-N_{a}[k]}y_{av}[k] - \frac{1}{N[k]-N_{a}[k]}\sum\limits_{j\in \eta \cup \mathcal{A}[k]} \log (s_{j}[k]) + \frac{1}{N[k]-N_{a}[k]}\log \left(\sum\limits_{i\in \eta \cup \mathcal{A}[k]} s_{i}[k]\right),
\]
from where the governing equation for the average of logarithm of the area vector $y_{av}[k]$ is as follows
\begin{equation}\label{Eq_Area_Mean_Law}
y_{av}[k+1] = y_{av}[k] + \frac{N_{a}[k]}{N[k]-N_{a}[k]}\cdot y_{av}[k] + \frac{1}{N[k]-N_{a}[k]}\log \left(\frac{\sum\limits_{i\in \eta \cup \mathcal{A}[k]}s_{i}[k]}{\prod\limits_{j\in \eta \cup \mathcal{A}[k]}s_{j}[k]}\right).
\end{equation}

\subsection{Population evolution}\label{App_Pop_Evolution}

Based on the revealed rank size distribution of population per municipality, presented in Figure \ref{Fig_Population_Rank_Size_1}, municipality $i$ population in year $k$ can be approximated as
\begin{equation}\label{Eq_App_Rank_Size_Mun_i}
	z_{i}[k] \approx -\beta[k]\cdot \log r_{i}[k] + z_{A}[k],	
\end{equation}
where $z_{A}[k]=\log(x_{A}[k])$ denotes the logarithm of Amsterdam population in year $k$. 
By assuming that the ranking of a municipality $i$ does not change $r_{i}[k+1] = r_{i}[k]$ between years $k$ and $k+1$, we obtain
\begin{equation}\label{Eq_Pop_Zipf_Law_year_2}
z_{i}[k+1]=-\beta[k+1]\cdot \log r_{i}[k] + \log \left(p_{A}[k+1]\right).
\end{equation}
By combining (\ref{Eq_Pop_Zipf_Law}) and (\ref{Eq_Pop_Zipf_Law_year_2}), the difference $z_{i}[k+1] - z_{i}[k]$ in logarithm of the population of the $i$-th municipality in two consecutive years $k$ and $k+1$ follows
\[
z_{i}[k+1] - z_{i}[k] = -\left(\beta[k+1]-\beta[k]\right)\cdot \log r_{i}[k] + \left(z_{A}[k+1] - z_{A}[k]\right).
\]
After importing (\ref{Eq_slope_linear_fit}), the relation above translates into
\[
\log \left(\frac{p_i[k+1]}{p_i[k]}\right) = -b_{1}\cdot \log r_{i}[k]+ \left(\log \left(p_{A}[k+1]\right) - \log \left(p_{A}[k]\right)\right).
\]
Finally, we obtain the population increase of municipality $i$ in year $k$
\[
\frac{p_i[k+1]}{p_i[k]} = \left(r_{i}[k]\right)^{-b_{1}}\cdot \frac{p_{A}[k+1]}{p_{A}[k]}.
\]
By assuming that each municipality follows the rank-size distribution in the equation above, we derive the mean $z_{av}[k]$ as follows
\begin{equation}\label{Eq_App_Rank_Size_Mean}
	z_{av}[k] = \frac{1}{N[k]}\cdot \sum_{i=1}^{N[k]} \left(-\beta[k]\cdot \log r_{i}[k] + z_{A}[k]\right) = -\beta[k]\cdot \log\left(N[k]!^{\frac{1}{N[k]}}\right) + z_{A}[k].
\end{equation}
After importing (\ref{Eq_App_Rank_Size_Mun_i}) and (\ref{Eq_App_Rank_Size_Mean}) into the definition of the variance $\text{Var}(z[k])$, we obtain
\[
\text{Var}(z[k]) = \frac{1}{N[k]}\cdot \sum_{i=1}^{N[k]} \left(z_{A}[k] + \beta[k]\cdot \log\left(N[k]!^{\frac{1}{N[k]}}\right) - \beta[k]\cdot \log r_{i}[k] - z_{A}[k]\right)^{2}.
\]
that further simplifies as follows
\begin{equation}\label{Eq_App_Var_z_via_Beta}
	\text{Var}(z[k]) = \beta^{2}[k]\cdot g(N[k]),
\end{equation}
where 
\[
g(N[k]) = \frac{1}{N[k]}\sum_{i=1}^{N[k]}\left(\log \frac{N[k]!^{\frac{1}{N[k]}}}{r_{i}}\right)^{2} = \frac{1}{N[k]}\sum_{i=1}^{N[k]}\left(\log \frac{N[k]!^{\frac{1}{N[k]}}}{i}\right)^{2},
\]
when the sum terms are ordered in descending order.
Equation (\ref{Eq_App_Var_z_via_Beta}) teaches us that, for a given $N[k]$, the variance $\text{Var}(z[k])$ is a square function of the rank-size distribution slope $\beta[k]$.

\subsection{Rank-size distribution versus power-law distribution}\label{Sec_App_Rank_Size_Power_Law}
Under the assumptions introduced in Section \ref{Sec_Pop_Ranking_Distribution}, we derive the following probability distribution function
\begin{equation}\label{Eq_App_PDF_Rank_Size}
	\text{Pr}(X[k]>x_{i}) = \frac{r_{i}}{N[k]}.
\end{equation}
From (\ref{Eq_Pop_Zipf_Law}), we obtain
\[
r_{i}[k] = \left(\frac{x_{i}[k]}{x_{A}[k]}\right)^{-\frac{1}{\beta[k]}},
\]
transforming further (\ref{Eq_App_PDF_Rank_Size})
\[
\text{Pr}(X[k]>x_{i}) = \frac{1}{N[k]}\cdot \left(\frac{x_{i}[k]}{x_{A}[k]}\right)^{-\frac{1}{\beta[k]}}.
\]
On the other side, probability distribution function of the power-law distribution is
\[
\text{Pr}(X[k]>x_{i}) = \left(\frac{x_{i}}{x_{min}}\right)^{-(\tau[k] -1)}.
\]
By comparing the last relation with (\ref{Eq_App_PDF_Rank_Size}), we obtain
\[
\frac{1}{\beta[k]} \approx \tau[k] - 1,
\]
leading to the dependece between the rank-size distribution slope $\beta[k]$ and the exponent of the power-law distribution $\tau[k]$
\[
\beta[k] = \frac{1}{\tau[k]-1}.
\]
 
\section{Properties of the Migration Model}\label{App_Migration_Model}

\begin{theorem}\label{Theorem_Migration_Model_Mean}
	The proposed migration model (\ref{Eq_Migration_Process_Law}) does not change the total population over time, but internally redistributes the population among neighbouring municipalities:
	\begin{equation}
		T[k] = T[0], k > 0
	\end{equation}
\end{theorem}
\textit{Proof} 
Total population in year $k$ is denoted as $T[k] = u^{T}\cdot p[k]$. The total population $T[k+1]$ in the following $k+1$ is as follows:
\[T[k+1] = u^{T}\cdot p[k+1].\] By implementing (\ref{Eq_Migration_Process_Law}) we obtain: \[T[k+1] = u^{T} \cdot \left(I + \alpha\cdot M^{T}[k] + \delta\cdot M[k] - \delta\cdot \text{diag}\left(M^{T}[k]\cdot u\right) - \alpha\cdot \text{diag}\left(M[k]\cdot u\right)\right)\cdot p[k].\] We further group terms of the previous equation
\begin{equation}\label{Total_Population_Migration_Model_Eq}
	\begin{aligned}
		T[k+1] &= T[k] \\
		&+ \left(\delta \cdot \left(M^{T}\cdot u\right)^{T} - \delta \cdot \left(M^{T}\cdot u\right)^{T} \right) \cdot p[k] \\
		&+ \left(\alpha \cdot \left(M\cdot u\right)^{T} - \alpha \cdot \left(M\cdot u\right)^{T} \right) \cdot p[k], \\
	\end{aligned}
\end{equation}
from where we conclude $T[k+1] = T[k]$ or  $T[k] = T[0]$, which completes the proof. $\hfill \square $

\begin{theorem}\label{Migration_Model_Steady_State}
	The proposed migration model, defined in (\ref{Eq_Migration_Process_Law}), is in a steady state if the following condition holds for each node $i \in \mathcal{N}[k]$:
	\begin{equation}
		\begin{matrix}
			p_i[k] = \frac{\sum\limits_{j \in \mathcal{N}_{i}^{+}[k]} \delta\cdot p_j[k] + \sum\limits_{m \in \mathcal{N}_{i}^{-}[k]} \alpha\cdot p_m[k]}{\alpha \cdot d_{i}^{+} + \delta\cdot d_{i}^{-}}
		\end{matrix}
	\end{equation}
	where $\mathcal{N}_{i}^{+}[k] = \{\, j \mid j \in \mathcal{N}_{i}[k]\,p_i[k] > p_j[k] \,\}$ defines a set of neighbours of node $i$, that have a larger population, while the set of smaller neighbours is given by $\mathcal{N}_{i}^{-}[k] = \{\, j \mid j \in \mathcal{N}_{i}[k],\,p_i[k] < p_j[k] \,\}$ with the complete set of neighbour nodes of node $i$ defined as $\mathcal{N}_{i}[k] = \{\, j \mid j \in \mathcal{N}[k],\, a_{ij}[k] = 1 \,\}$.
\end{theorem}
\textit{Proof} 
We transform the governing equation (\ref{Eq_Migration_Process_Law}) of the migration model into a node-level governing equation:
\begin{equation}\label{Eq_Migration_Model_Node_Level}
	\begin{aligned}
		p_i[k+1] &= p_i[k] \\
		&+ \sum\limits_{j \in \mathcal{N}_{i}^{+}[k]} \delta\cdot p_j[k] + \sum\limits_{m \in \mathcal{N}_{i}^{-}[k]} \alpha\cdot p_m[k] \\
		&- \sum\limits_{j \in \mathcal{N}_{i}^{+}[k]} \alpha\cdot p_i[k] - \sum\limits_{m \in \mathcal{N}_{i}^{-}[k]} \delta\cdot p_i[k]
	\end{aligned}
\end{equation}

We implement the steady state equality $p_i[k+1] = p_i[k]$ into (\ref{Eq_Migration_Model_Node_Level}) and obtain:
\begin{equation}\label{Migration_Model_Node_Level_Steady_State_Eq}
	\left(\sum\limits_{j \in \mathcal{N}_{i}^{+}[k]} \alpha + \sum\limits_{m \in \mathcal{N}_{i}^{-}[k]} \delta\right)\cdot p_i[k] = \sum\limits_{j \in \mathcal{N}_{i}^{+}[k]} \delta\cdot p_j[k] + \sum\limits_{m \in \mathcal{N}_{i}^{-}[k]} \alpha\cdot p_m[k],
\end{equation}
from where we conclude
\begin{equation}\label{Eq_Migration_Model_Node_Level_Steady_State_Solution}
	p_i[k] = \frac{\sum\limits_{j \in \mathcal{N}_{i}^{+}[k]} \delta\cdot p_j[k] + \sum\limits_{m \in \mathcal{N}_{i}^{-}[k]} \alpha\cdot p_m[k]}{\alpha \cdot d_{i}^{+} + \delta\cdot d_{i}^{-}},
\end{equation}
which completes the proof.$\hfill \square $

\section{List of Notations}\label{App_List_of_Notations}
Four tables with the list of used notations in the paper are provided below.
\begin{table}[!h]\label{Table_DMN}\caption{Notations used in the paper for the Dutch Municipality Network}
	\centering
	\begin{tabular}{c|l}
		\hline
		\textbf{Notation} &		\textbf{Explanation} \\
		\hline		
		$k$								&		Year $k$ \\
		$N[k]$							&		Number of active municipalities in year $k$ \\ 
		$N_{a}[k]$ 						&		Number of abolished municipalities in year $k$ \\ 
		$N_{n}[k]$ 						&		Number of newly established municipalities in year $k$ \\ 
		$\mathcal{N}[k]$ 				&		Set of active municipalities in year $k$ \\ 
		$\mathcal{A}[k]$ 				&		Set of abolished municipalities at the end of year $k$ \\ 
		$\mathcal{N}_{i}[k]$ 			&		Set of municipality $i$ neighbouring municipalities in year $k$ \\ 
		$\mathcal{N}_{i}^{+}[k]$ 		&		Set of municipality $i$ neighbours with larger population in year $k$ \\ 
		$\mathcal{N}_{i}^{-}[k]$		&		Set of municipality $i$ neighbours with smaller population in year $k$ \\ 
		$d_{i}[k]$						&		Degree of node $i$ in year $k$ \\ 
		$d_{i}^{+}[k]$						&		Number of municipality $i$ neighbours with larger population in year $k$ \\ 
		$d_{i}^{-}[k]$						&		Number of municipality $i$ neighbours with smaller population in year $k$ \\ 
		$d_{av}[k]$						&		Average degree of the DMN in year $k$ \\ 
		$L[k]$							&		Number of links in the DMN in year $k$ \\ 
		$\mathcal{L}[k]$				&		Set of links in the DMN in year $k$ \\ 
		$A[k]$							&		Adjacency matrix of the DMN in year $k$ \\ 
		$a_{ij}[k]$						&		$ij$-th element of the adjacency matrix $A[k]$ in year $k$ \\ 
	\end{tabular}
\end{table}

\begin{table}[!h]\label{Table_Pop}\caption{Notations used for the analysis of population dynamics}
	\centering
	\begin{tabular}{c|l}
		\hline
		\textbf{Notation} &		\textbf{Explanation} \\
		\hline		
		$p_i[k]$ 						&		Population of municipality $i$ in year $k$ \\ 
		$p_{A}[k]$ 						&		Population of Amsterdam in year $k$ \\ 
		$z_{i}[k]$ 						&		Logarithm of the population size of municipality $i$ in year $k$ \\ 
		$z_{A}[k]$ 						&		Logarithm of the population size of municipality Amsterdam in year $k$ \\ 
		$p[k]$	 						&		The $N[k]\times 1$ vector of the population per municipality in year $k$ \\ 
		$z[k]$	 						&		The $N[k]\times 1$ vector of logarithm of the population size per municipality in year $k$ \\ 
		$p_{av}[k]$	 						&		Average population per municipality in year $k$ \\ 
		$z_{av}[k]$	 						&		Average logarithm of the population per municipality in year $k$ \\ 
		$P[k]$	 &	Population random variable in year $k$ \\ 
		$Z[k]$  &	Logarithm of the population random variable in year $k$\\ 
		$T[k]$ 							&	 	Total population of The Netherlands in year $k$ \\ 
		$M[k]$							&		The $N[k]\times N[k]$ migration matrix of the DMN in year $k$ \\ 
		$m_{ij}[k]$						&		$ij$-th element of the migration matrix $M[k]$ in year $k$ \\ 
		$\alpha[k]$						&	 	Forward migration rate in year $k$ \\ 
		$\delta[k]$						&	 	Backward migration rate in year $k$ \\ 
		$c_{1}[k]$						&	 	Estimated slope of the population increase in year $k$ \\ 
		$c_{2}[k]$						&	 	Estimated additive constant of the population increase in year $k$ \\ 
	\end{tabular}
\end{table}

\begin{table}[!h]\label{Table_Area}\caption{Notations used for the area and the merging process}
	\centering
	\begin{tabular}{c|l}
		\hline
		\textbf{Notation} &		\textbf{Explanation} \\
		\hline		
		$s_{i}[k]$ 						&		Area size of municipality $i$ in year $k$ \\ 
		$y_{i}[k]$ 						&		Logarithm of the area size of municipality $i$ in year $k$ \\ 
		$s[k]$	 						&		The $N[k]\times 1$ vector of area size per municipality in year $k$ \\ 
		$y[k]$	 						&		The $N[k]\times 1$ vector of logarithm of the area size per municipality in year $k$ \\ 
		$s_{av}[k]$	 						&		Average area size per municipality in year $k$ \\ 
		$y_{av}[k]$	 						&		Average logarithm of the area size per municipality in year $k$ \\ 
		$S[k]$	 						&		Area random variable in year $k$ \\ 
		$Y[k]$	 						&	Logarithm of the area random variable in year $k$ \\
		$x_i[k]$						&	 	Abolishment Likelihood index of municipality $i$ in year $k$ \\ 
		$x[k]$							&	 	The $N[k]\times 1$ vector with an Abolishment Likelihood index per municipality in year $k$ \\  
	\end{tabular}
\end{table}

\begin{table}[!h]\label{Table_Dist_Merg}\caption{Notations used for distribution functions}
	\centering
	\begin{tabular}{c|l}
		\hline
		\textbf{Notation} &		\textbf{Explanation} \\
		\hline		
		$\mu_{n}[k]$				&	 Shape parameter of the normal distribution in year $k$ \\ 
		$\sigma_{n}[k]$				&	 	Scale parameter of the normal distribution in year $k$\\ 
		$\mu_{l}[k]$				&	 	Shape parameter of the logistic distribution in year $k$ \\ 
		$\sigma_{l}[k]$				&	 	Scale parameter of the logistic distribution in year $k$ \\ 
		$\beta[k]$						&	 	Population rank-size distribution slope in year $k$ \\ 
		$b_{1}[k]$						&	 	First parameter of a linear fit of $\beta[k]$ \\ 
		$b_{2}[k]$						&	 	Second parameter of a linear fit of $\beta[k]$ \\
		$\tau[k]$						&	 	Exponent of the power-law distribution in year $k$ \\ 
        $C[k]$						&	 	Normalisation constant of the power-law distribution in year $k$ \\ 
		$E[P]$						&	 	Expectation of the random variable $P$ \\ 
		$\text{Var}(P)$						&	 	Variance of the random variable $P$ \\ 
		$\text{Cov}(P)$						&	 	Covariance of the random variable $P$ \\ 
		$p$						&	 	$p$ value of a goodness-of-fit test \\ 
	\end{tabular}
\end{table}

\end{document}